\documentclass[twocolumn, twocolappendix]{aastex631}
\usepackage{apjfonts}

\newcommand\ttc{$^{13}$C}
\newcommand\eto{$^{18}$O}
\newcommand\sco{C\eto}
\newcommand\mco{\ttc O}
\newcommand\lco{$^{12}$CO}
\newcommand\pls{$^+$}
\newcommand\hcop{HCO\pls}
\newcommand\dcop{DCO\pls}
\newcommand\htcop{H\ttc O\pls}
\newcommand\hmol{H$_2$}
\newcommand\nmol{N$_2$}
\newcommand\nthp{\nmol H\pls}
\newcommand\ntdp{\nmol D\pls}
\newcommand\nht{NH$_3$}

\newcommand\scot{\sco\ (3--2)}
\newcommand\mcot{\mco\ (3--2)}
\newcommand\lcot{\lco\ (3--2)}
\newcommand\hcopt{\hcop\ (4--3)}
\newcommand\htcopt{\htcop\ (4--3)}

\newcommand\hso{\textit{Herschel}}
\newcommand\gaia{\textit{Gaia}}
\newcommand\plck{\textit{Planck}}

\newcommand\sth{$\sigma_\mathrm{th}$}
\newcommand\snth{$\sigma_\mathrm{NT}$}

\newcommand\sobs{$\sigma_\mathrm{obs}$}
\newcommand\cs{$c_\mathrm{s}$}

\newcommand\per{$^{-1}$}
\newcommand\psq{$^{-2}$}
\newcommand\pcb{$^{-3}$}
\newcommand\solm{M$_\odot$}

\mathchardef\mhyphen="2D

\received{}
\revised{}
\accepted{}
\published{}
\submitjournal{\apj}

\shorttitle{Role of Filamentary Structure in the Dense Core Formation}
\shortauthors{Kim et al.}

\begin{document}

\title{Role of Filamentary Structures in the Formation of Two Dense Cores, L1544 and L694-2}

\correspondingauthor{Chang Won Lee}
\email{cwl@kasi.re.kr}

\author[0000-0001-9333-5608]{Shinyoung Kim}
\affiliation{Korea Astronomy and Space Science Institute, 776 Daedeok-daero Yuseong-gu, Daejeon 34055, Republic of Korea}
\affiliation{University of Science and Technology, 217 Gajeong-ro Yuseong-gu, Daejeon 34113, Republic of Korea}

\author[0000-0002-3179-6334]{Chang Won Lee}
\affiliation{Korea Astronomy and Space Science Institute, 776 Daedeok-daero Yuseong-gu, Daejeon 34055, Republic of Korea}
\affiliation{University of Science and Technology, 217 Gajeong-ro Yuseong-gu, Daejeon 34113, Republic of Korea}

\author[0000-0002-2569-1253]{Mario Tafalla}
\affiliation{Observatorio Astron\'{o}mico Nacional (IGN), Alfonso XII 3, Madrid 28014, Spain}

\author{Maheswar Gophinathan}
\affiliation{Indian Institute of Astrophysics, 2nd Block, Koramangala, Bengaluru, Karnataka 560034, India}

\author[0000-0003-1481-7911]{Paola Caselli}
\affiliation{Max-Planck-Institut f\"{u}r Extraterrestrische Physik, Gie{\ss}enbachstrasse 1, D-85741 Garching bei M\"{u}nchen, Germany}

\author[0000-0002-2885-1806]{Philip C. Myers}
\affiliation{Center for Astrophysics | Harvard and Smithsonian (CfA), Cambridge, MA 02138, USA}

\author[0000-0003-0014-1527]{Eun Jung Chung}
\affiliation{Department of Astronomy and Space Science, Chungnam National University, 99 Daehak-ro, Yuseong-gu, Daejeon 34134, Republic of Korea}

\author[0000-0003-1275-5251]{Shanghuo Li}
\affiliation{Korea Astronomy and Space Science Institute, 776 Daedeok-daero Yuseong-gu, Daejeon 34055, Republic of Korea}

\begin{abstract}
We present mapping results of two prestellar cores, L1544 and L694-2, embedded in filamentary clouds in \scot, \mcot, \lcot, \hcopt, and \htcopt\ lines with the JCMT telescope to examine the role of the filamentary structures in the formation of dense cores in the clouds, with new distance estimates for L1544 ($175_{-3}^{+4}$ pc) and L694-2 ($203_{-7}^{+6}$ pc).
From these observations, we found that the non-thermal velocity dispersion of two prestellar cores and their surrounding clouds are smaller than or comparable to the sound speed.
This may indicate that the turbulence has been already dissipated for both filaments and cores during their formation time.
We also found a $\lambda/4$ shift between the periodic oscillations in the velocity and the column density distributions implying the possible presence of gravitational core-forming flow motion along the axis of the filament.
The mass accretion rates due to these flow motions are estimated to be 2--3 \solm\ Myr\per, being comparable to that for Serpens cloud but much smaller than those for the Hub filaments, cluster, or high mass forming filaments by 1 or 2 order of magnitudes.
From this study, we suggest that the filaments in our targets might be formed from the shock-compression of colliding clouds, and then the cores are formed by gravitational fragmentation of the filaments to evolve to the prestellar stage.
We conclude that the filamentary structures in the clouds play an important role in the entire process of formation of dense cores and their evolution.
\end{abstract}


\section{Introduction}

\begin{figure*}
    \centering
    \includegraphics[width=7.1in]{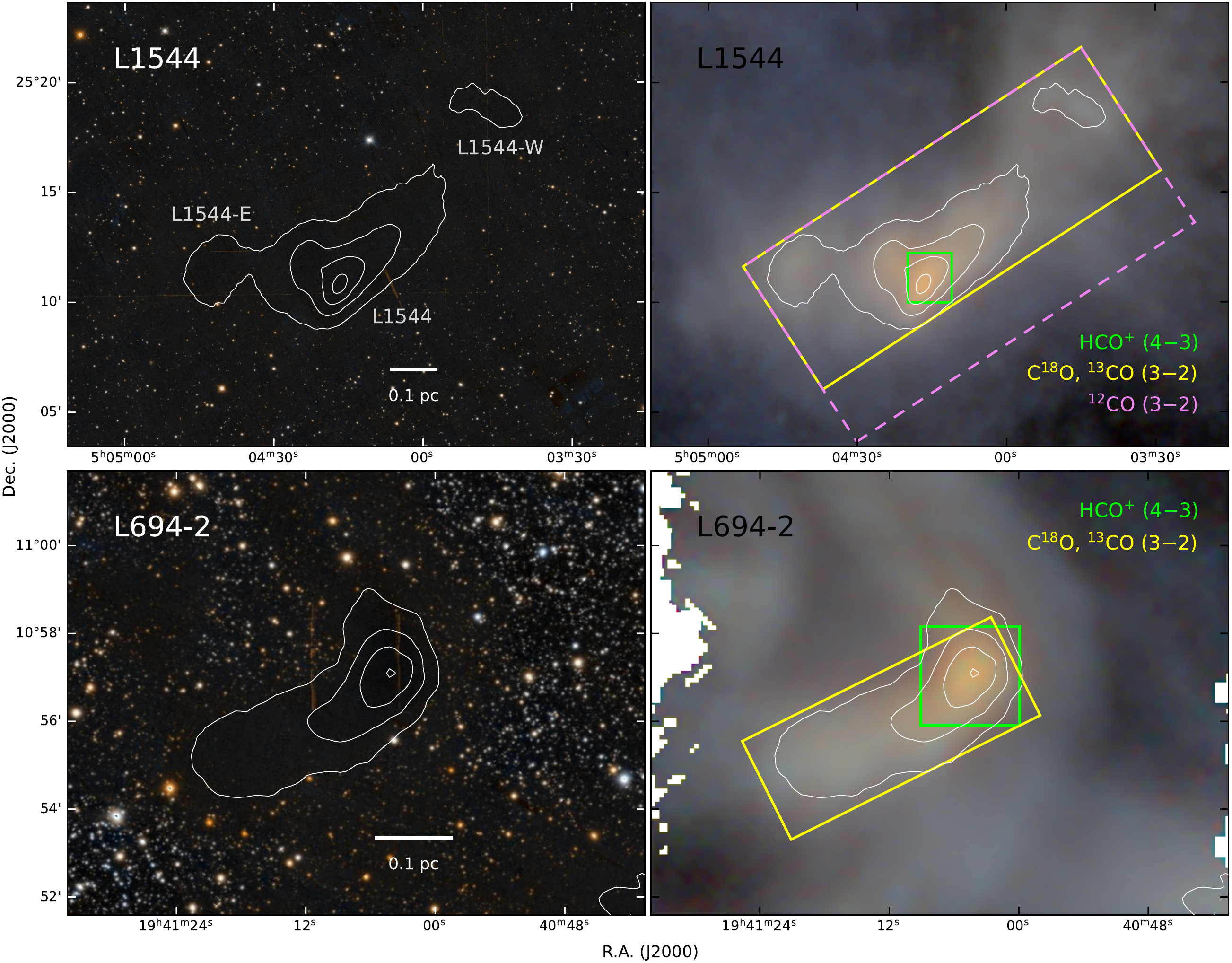}
    \caption{
    Target cores shown in optical (left panels, Pan-STARRS DR1 color from bands z and g) and far-IR (right panels, \hso\ SPIRE color from 250, 350, 500 \micron) images. 
    The contours of all panels show the \hmol\ column densities with levels of $2\times10^{21}$, $5\times10^{21}$, $10^{22}$, and $2\times10^{22}$ cm\psq\ derived from the SED fit for \hso\ dust continuum data (Figure \ref{fig:sed_result_h2cd_temp}).
    In the upper-right panel, three dense cores are distinguishable such as L1544-E, L1544, and L1544-W \citep{Tafalla_1998_ApJ_504_900} or L1544-2, L1544-1, and L1544-3 in other names, respectively \citep{Lee_1999_ApJS_123_233}. 
    In this study we only focus on `L1544' and its surrounding cloud.
    The colored rectangles of the right panels show the mapping areas for each molecular line; lime - \hcop\ and \htcopt, yellow - \sco, \mco, and \lcot, violet (dashed) - \lcot\ for L1544.
    }
    \label{fig:target_cores}
\end{figure*}

Dense cores embedded in less dense large clouds are the birth sites of stars. 
Since their characteristics represent the initial condition of star formation, the detailed processes of dense core formation and evolution are crucial issues in star formation.
A key challenge in the study of core formation is to explain the physical differences between the cores and their natal clouds. 
While dense cores have a typical size of 0.1 pc diameter and are often spherical in shape, the large-scale clouds in which these cores reside are of a few to tens of parsec in extent and have irregular shapes \citep[e.g.,][]{di_Francesco_2007_prpl.conf_17,Bergin_2007_ARA&A_45_339}.
The gas motions in low-mass cores are subsonic and coherent but are more turbulent in the parent cloud \citep{Myers_1983_ApJ_270_105,Goodman_1998_ApJ_504_223}.

A number of mechanisms, such as quasi-static contraction via ambipolar diffusion \citep{Shu_1987_ARA&A_25_23,Mouschovias_1991_ApJ_373_169} or turbulence dissipation by collisions of supersonic flows \citep{Padoan_2001_ApJ_553_227,Klessen_2005_ApJ_620_786}, have been proposed as possible mechanisms for the core formation at the stage prior to the formation of stars, but they do not entirely explain the observational results of the dense cores \citep{Ward-Thompson_2007_prpl.conf_33}. 
Thus the question, how the dense cores form in the clouds is still under debate.

Recent studies, including the representative \hso\ Gould Belt Survey \citep[HGBS,][]{Andre_2010_A&A_518_L102}, have shown that all molecular clouds consist of filamentary structures, and that the majority of gravitationally unstable cores on the verge of star formation, so-called ``prestellar cores \citep{Ward-Thompson_2007_prpl.conf_33},'' are in locations related to filaments \citep[e.g.,][]{Andre_2010_A&A_518_L102,Konyves_2015_A&A_584_A91}, suggesting that dense cores generally form in filamentary cloud structures.
In addition, spectroscopic observations of molecular lines have revealed that the large-scale filaments dominated by supersonic motions consist of sub-structures with a bundle of trans- or subsonic, velocity-coherent, small-scale filaments (a.k.a. `fibers,' e.g., \citealt{Hacar_2013_A&A_554_A55,Hacar_2018_A&A_610_A77,Li_2022_ApJ_926_165}).
Considering the sub-sonic and coherent characteristics of dense cores and intermediate size scale of such fibers between the large-scale filaments and cores, their positions in the hierarchical system of the filamentary structure seem to be at the same or lower level of small-scale sub-filaments.

A sequential top-down scenario for filament and core formation in such hierarchical filament(-fiber)-core system has been proposed by \citet{Tafalla_2015_A&A_574_A104}. 
In this scenario,
at the earlier stage, turbulent motions in the parent cloud produce a filamentary structure by turbulence dissipation (or a combination of turbulence and gravity) \citep[e.g.,][]{Padoan_2001_ApJ_553_227,Klessen_2005_ApJ_620_786,Hennebelle_2013_A&A_556_A153} and later the filament is fragmented into cores by gravitational instability \citep[e.g.,][]{Inutsuka_1997_ApJ_480_681} and/or magnetic fields \citep[e.g.,][]{Fiege_2000_MNRAS_311_105}. 
This scenario is supported by several observational evidences such as the presence of small-scale velocity-coherent subsonic filaments (or fibers), the distribution of cores spaced quasi-equally along the filaments, and the sinusoidal velocity oscillation shifted by $\lambda/4$ for the density fluctuation \citep[see also][]{Hacar_2011_A&A_533_A34,Hacar_2013_A&A_554_A55,Clarke_2016_MNRAS_458_319}.
However, such observational cases are quite limited, and usually the complex filament-fiber system makes it challenging to investigate the velocity structures of individual fibers.

The ``isolated'' prestellar cores, such as L1544 and L694-2, have been traditionally regarded as the ideal laboratories to study the physical, chemical, and kinematic properties of the evolved starless core and furthermore, to investigate the initial condition of star formation \citep[e.g.,][]{Lee_2001_ApJS_136_703,Crapsi_2005_ApJ_619_379,Bergin_2007_ARA&A_45_339,Keto_2015_MNRAS_446_3731}.
In contrast to the recent finding that the majority of gravitationally unstable cores are embedded within the filamentary clouds \citep[e.g.,][]{Andre_2010_A&A_518_L102,Konyves_2015_A&A_584_A91}, these isolated cores seem to be rather minority cases in the study of key issues in star formation under the filament paradigm in the sense that their location is well isolated and quite far away from large-scale filaments \citep[e.g.][]{Andre_2014_prpl.conf_27}.
Nevertheless, recent high-sensitivity \hso\ continuum data show that the clouds around the isolated dense cores are ``filamentary-shaped'' (e.g., Figure \ref{fig:target_cores}).
L1544 and L694-2 appear to be isolated filamentary clouds with one core each.
Therefore, if these systems are fully velocity coherent, they may represent ideal targets to investigate dense core formation in a filamentary environment with the least influence by other surrounding clouds and their star formation activities.

L1544 is the densest one of three optically selected dense cores \citep{Lee_1999_ApJS_123_233} in the L1544 dark cloud located in the eastern part of the Taurus molecular complex \citep{Benson_1989_ApJS_71_89}.
Its ``starless'' characteristics were known because of the lack of any far-infrared sources detected so far (e.g., \citet{Lee_1999_ApJS_123_233}, Figure \ref{fig:target_cores}).
L1544 core was identified as the first dense core where gas infalling motions are occurring over the core \citep{Tafalla_1998_ApJ_504_900}, and existence of these motions has been confirmed by many subsequent studies with various molecular line observations \citep[e.g.,][]{{Williams_1999_ApJ_513_L61,Ohashi_1999_ApJ_518_L41O,Lai_2000_ApJS_128_271, Caselli_1999_ApJ_523_L165,Caselli_2002_ApJ_565_331,Lee_2001_ApJS_136_703,Lee_2004_ApJS_153_523, Caselli_2012_ApJL_759_L37,Caselli_2017_A&A_603_L1}}.

Numerous observational studies of the physical and chemical properties of the L1544 core, such as a high central density of $\gtrsim 10^6$ cm\pcb\ \citep[e.g.,][]{Tafalla_2002_ApJ_569_815,Keto_2010_MNRAS_402_1625}, a severe molecular depletion in core center \citep[e.g.,][]{Caselli_1999_ApJ_523_L165,Caselli_2002_ApJ_565_331,Tafalla_2002_ApJ_569_815,Young_2004_ApJ_614_252,Kim_2020_ApJ_891_169}, and a significant deuterium fraction \citep[e.g.,][]{Caselli_1999_ApJ_523_L165,Caselli_2002_ApJ_565_344,Hirota_2003_ApJ_594_859,Crapsi_2005_ApJ_619_379,Crapsi_2007_A&A_470_221,Redaelli_2019_A&A_629_A15} suggest its highly evolved status.

As shown in Figure \ref{fig:target_cores}, the L1544 cloud contains three dense cores of L1544-E, L1544, and L1544-W.
The main core L1544 is the largest and densest one, but the other two cores appear to be not dense enough (for example, L1544-E (or L1544-2) was not detected with \nthp\ (1--0); \citealt{Lee_1999_ApJS_123_233}) and are located far from the filamentary cloud. 
Therefore, in this study, we focus on the main core L1544 only as it is the most evolved and believed to be the most appropriate target in examining the effect of the filamentary structure on the formation of dense core.

L694-2 was first identified in a search for optically selected cores in the L694 small dark cloud \citep{Lee_1999_ApJS_123_233}, and recognized as a source showing infall motions in molecular line observations by \citet{Lee_1999_ApJ_526_788}.
Since then, L694-2 is considered a classic case of the infall candidate just like L1544
\citep[e.g.,][]{Lee_2001_ApJS_136_703, Lee_2004_ApJS_153_523, Williams_2006_ApJ_636_952,Sohn_2007_ApJ_664_928,Keown_2016_ApJ_833_97}. 
Although L694-2 appears to be relatively less evolved compared to L1544 because of a slightly lower central density or a wider central flat region \citep{Crapsi_2005_ApJ_619_379,Kim_2020_ApJ_891_169}, both L1544 and L694-2 cores are believed to be the most evolved prestellar cores on the verge of initiating star formation.

In this paper we present new results from high-resolution mapping observations toward L1544 and L694-2 using the 15 m James Clerk Maxwell Telescope (JCMT) to investigate the kinematics of the dense gas in the cores and their surrounding filaments.
\scot\ line is usually optically thin and traces wider but fairly dense ($\gtrsim 10^4$ cm\pcb) regions of the filament \citep{Andre_2007_A&A_472_519,White_2015_MNRAS_447_1996}.
\lcot\ line is a good tracer of infalling motions because it traces less dense regions surrounding the core and has a high critical density ($>10^4$ cm\pcb) and a large optical depth \citep{Lee_2013_ApJ_777_50,Schneider_2015_A&A_578_A29}.
Separately, for the purpose of detecting the infalling motions near the innermost region, \hcopt\ and its isotopologue \htcopt\ lines are selected as optically thick and thin tracers, respectively \citep{Gregersen_1997_ApJ_484_256,Chira_2014_MNRAS_444_874}.
\hcop\ and its isotopologues are abundant enough to be detected over prestellar cores \citep{Redaelli_2019_A&A_629_A15,Li_2021_ApJL_912_L7}, and the $J=4$--3 transitions have high critical densities ($>10^6$ cm\pcb).

In the following sections, using these line observations with complementary dust continuum data obtained from the \hso\ science archive, we present the analysis of the physical structures and kinematics of the gas for two prestellar cores, L1544 and L694-2, embedded in filamentary clouds. 
Sections \ref{sec:obs} and \ref{sec:res} describe the JCMT observations, data reductions, the \hso\ data, and the results of our observations.
Section \ref{sec:dens} analyses density structures of the filamentary envelopes of two dense cores and their physical properties.
In Section \ref{sec:vels}, we examine the velocity structures of the filamentary envelopes by considering the Gaussian fit components of the line spectra.
Section \ref{sec:infl} discusses the radially contracting motions toward the filament and dense core by examining spatial distribution of the asymmetric line profiles.
We discuss our results and their implications on the dense core formation scenario in Section \ref{sec:disc}, and list our main conclusions in Section \ref{sec:conc}.

\section{Observations} \label{sec:obs}

\begin{deluxetable}{lcDcC}
\tablecaption{Observed Molecular Lines and Parameters\label{tab:lines}}
\tablewidth{0pt}
\tablehead{
\colhead{Line} & \colhead{$\nu_\mathrm{rest}$} & \multicolumn2c{$\delta \nu$} & \colhead{$\delta v$} & \colhead{$n_\mathrm{crit}$} \\
\colhead{} & \colhead{(GHz)} & \multicolumn2c{(kHz)} & \colhead{(km s\per)} & \colhead{(cm$^{-3}$)}
}
\decimals
\decimalcolnumbers
\startdata
\lcot   & 345.7959899 & 61   & 0.053 & 1.9\times 10^4 \\
\mcot   & 330.5879653 & 61   & 0.055 & 1.6\times 10^4 \\
\scot   & 329.3305525 & 61   & 0.056 & 1.6\times 10^4 \\
\hcopt  & 356.7342230 & 30.5 & 0.026 & 4.0\times 10^6 \\
\htcopt & 346.9983440 & 30.5 & 0.026 & 3.6\times 10^6 \\
\enddata
\tablecomments{(1) Observed line transitions. (2) Rest frequency of each transition referred from CDMS (Cologne Database for Molecular Spectroscopy, \citealt{Endres_2016_JMoSp_327_95}). (3) and (4) Observing spectral resolution in frequency and velocity. In this study, the velocity channels were resampled to 0.06 km s\per\ (for CO isotopologues) or 0.03 km s\per\ (for \hcop) for the channel matching between the data cubes. (5) Critical density of lines at 10 K. This was calculated using the equation of $A_u/\sum_{l<u} \gamma_{ul}$ where $A_u$ is the Einstein A coefficient of level $u$ and $\gamma_{ul}$ is the collision rate out of level $u$ into lower level $l$ at a temperature of 10 K.}
\end{deluxetable}

\begin{deluxetable*}{llCCccDCC}
\tablecaption{Mapping Areas and Sensitivities\label{tab:obs}}
\tablewidth{0pt}
\tablehead{
\colhead{Target} & \colhead{Line} & \multicolumn2c{Mapping area} & Mode & \multicolumn3c{$\sigma_\mathrm{rms}$} & \colhead{$SNR_\mathrm{peak}$} & \colhead{$N_\mathrm{H_2}^\mathrm{\,det}$} \\
\colhead{} & \colhead{} & \colhead{($''$)} & \colhead{(pc)} & & \colhead{(K)} & \multicolumn2c{(Jy)} & \colhead{} & \colhead{(cm\psq)}
}
\decimals
\decimalcolnumbers
\startdata
L1544 & \lcot & 1100\times 570 & 0.75\times 0.39 & R & 0.277 & 8.34 & 24 & \lesssim2.7\times 10^{19} \\
      & \mcot & 1100\times 400 & 0.75\times 0.27 & R & 0.157 & 4.73 & 18 & \phm{\lesssim}2.1\times 10^{20} \\
      & \scot & 1100\times 400 & 0.75\times 0.27 & R & 0.161 & 4.85 & 10 & \phm{\lesssim}2.5\times 10^{21} \\
      & \hcopt & 120\times 135 & 0.08\times 0.09 & G & 0.057 & 1.70 & 10 & \phm{\lesssim}6.9\times 10^{21} \\
      & \htcopt & 120\times 120 & 0.08\times 0.08 & G & 0.113 & 3.40 & \multicolumn2c{(not detected)} \\ 
\hline
L694-2 & \lcot & 380\times 150 & 0.37\times 0.15 & R & 0.348 & 10.46 & 17 & \phm{\lesssim}3.7\times 10^{19} \\
       & \mcot & 380\times 150 & 0.37\times 0.15 & R & 0.175 & 5.28 & 23 & \phm{\lesssim}3.7\times 10^{19} \\
       & \scot & 380\times 150 & 0.37\times 0.15 & R & 0.139 & 4.17 & 10 & \phm{\lesssim}1.7\times 10^{21} \\
       & \hcopt & 135\times 135 & 0.13\times 0.13 & G & 0.061 & 1.83 & 10 & \phm{\lesssim}5.6\times 10^{21} \\
       & \htcopt & 120\times 120 & 0.12\times 0.12 & G & 0.080 & 2.39 & \multicolumn2c{(not detected)} \\ 
\enddata
\tablecomments{(1) Source name. (2) Observed line transition. (3) and (4) Area of the mapping region in unit arcsec and pc, respectively. (5) Observing mode. `R' is to mean the `raster' mode which is the scan mapping with a basket-weave technique a.k.a. the On-The-Fly mapping, and `G' is to indicate the `grid' mode which is the mapping mode by the stare observations with the grid spacing. (6) and (7) Mapping sensitivity in K and Jy units with the velocity resolutions of 0.06 km s\per\ (for CO isotopologue lines) or 0.03 km s\per\ for \hcop\ line. The values of the brightness temperature (K) unit were measured in the antenna temperature ($T_\mathrm{A}^*$) scale, and the values of the spectral flux density (Jy) unit were converted using the equation $S\,\mathrm{[Jy]}=15.6\,T_\mathrm{A}^*\,\mathrm{[K]}/\eta_\mathrm{A}$ where the aperture efficiency $\eta_\mathrm{A}$ is 0.52 \citep{Rawlings_2019_JUM}. (8) Signal to noise ratio at the brightest position of each line. (9) The minimum value of $N_\mathrm{H_2}$ in the region detected with $SNR \sim 3$. The \lcot\ line for L1544 was strongly detected in every position and thus its $SNR$ for the minimum value of $N_\mathrm{H_2}$ was about 5.4.}
\end{deluxetable*}

\subsection{JCMT-HARP Observations}

Observations were carried out toward two dense cores, L1544 and L694-2, with the 15 m JCMT from 2016 July 12 to October 27 (project ID: M16AP025) and from 2017 February 4 to September 16 (M17AP061) under weather conditions with the precipitable water vapor (PWV) between 0.83 and 1.58 mm and $\tau_\mathrm{225\;GHz} \sim 0.05$--$0.08$\footnote{The conversion from $\tau_\mathrm{225\;GHz}$ to PWV is given in \citet{Dempsey_2013_MNRAS_430_2534}: $\tau_\mathrm{225\;GHz} = 0.04\times PWV+0.017$}. 

As for the front-end and the back-end for our observations, we used the HARP (Heterodyne Array Receiver Program, \citealt{Buckle_2009_MNRAS_399_1026}) and ACSIS (the Auto Correlation Spectral Imaging System). 
HARP was designed to consist of $4\times4$ mixers (so called receptors) in separation of 30$\arcsec$ which make it easier to scan large areas in an efficient way. 
However, during our observations two mixers were not operational, and thus 14 mixers in total were used.
The HPBW (half-power beam width) at 345 GHz is $\sim 14\arcsec$, varying 13$\farcs$5--14$\farcs$7 depending on the observed frequency.
The ACSIS correlator was set to have our required spectral resolutions of $\sim 0.03$--0.06 km s\per.

The main molecular lines used for our observations are \lcot, \mcot, \scot, \hcopt, and \htcopt. 
While we made a position switching mode observation for the background subtraction, we employed two different mapping modes of `raster' and `grid', for our main targets.
The large-scale and relatively low-density filamentary cloud regions were observed using CO isotopologue lines, \lcot, \mcot, and \scot, in the `raster' mapping mode with two spectral windows providing 4096 channels of 61 kHz for a 250 MHz bandwidth, while the central high-density regions of the cores were observed using \hcopt\ and \htcopt\ line in the `grid' mapping mode\footnote{The data cube consisting of $8\times8$ pixels with $15\arcsec$ spacing was obtained by the stare observations with offsets of $(0, 0)$, $(-15\arcsec, 0)$, $(-15\arcsec, 15\arcsec)$, and $(0, -15\arcsec)$ using the $4\times4$ array with a beam spacing of $30\arcsec$.} with a single spectral window providing 8192 channels of 30.5 kHz for the same bandwidth. 
The molecular lines, their parameters such as the rest frequencies and critical densities, and the achieved frequency and velocity resolutions in our observations are listed in Table \ref{tab:lines}.
The mapped area, the sensitivity achieved for each target, and the corresponding mode employed for the mapping observations are summarized in Table \ref{tab:obs}.

The raw data were reduced with the ORAC-DR pipeline software \citep{Jenness_2015_MNRAS_453_73} with a recipe of `REDUCE\_SCIENCE\_NARROWLINE' to produce data cubes of a velocity range $\pm10$ km s\per\ centered at the systemic velocity ($V_\mathrm{LSR}$) of the source.
The reduced data cubes were re-gridded to match each other's pixel grids with a pixel size of $15''\times 15''$ using \texttt{hcongrid} in \texttt{FITS\_tools}\footnote{https://github.com/keflavich/FITS\_tools}, and resampled to have a channel width of 0.06 km s\per\ for the CO isotopologue lines or 0.03 km s\per\ for the \hcop\ line using \texttt{CubicSpline} in \texttt{SciPy} \citep{Virtanen_2020_NatMe_17_261}. 
Antenna temperatures($T_\mathrm{A}^*$) originally given in the data cubes were converted to the main beam temperatures ($T_\mathrm{MB}$) using the main beam efficiency ($\eta_\mathrm{MB}$) of 0.64 \citep{Rawlings_2019_JUM}.

\subsection{\hso\ Archival Data}

To complement the JCMT observations, we used far-infrared dust continuum data observed with the PACS and SPIRE instrument at 160, 250, 350, and 500 \micron\ from the \hso\ Space Observatory.
The data were taken from the \hso\ Science Archive, and their OBSIDs (PI) are 134220484 \citep[P. Andr{\'e},][]{Herschel_gould_belt_survey} for L1544 and 1342230846 \citep[S. Schnee,][]{Herschel_high_level_images} for L694-2.
In case of L694-2, there is no PACS observation (160 \micron\ data), but this absence of PACS data is found to have negligible effect on the SED fitting result (see Appendix \ref{sec:app-sed} for details).
The L1544 data were obtained as a part of the \hso\ Gould Belt Survey \citep[HGBS,][]{Andre_2010_A&A_518_L102}.
The data used here were already zero-point corrected for extended emission based on the cross-calibration with \plck\ HFI-545 and HFI-857 all-sky maps \citep{Valtchanov_2017_SPIRE}.
The angular resolutions of 250, 350, and 500 \micron\ images are $18\farcs4$, $25\farcs2$, and $36\farcs7$, respectively.
The \hso\ images were also re-gridded to match our JCMT observations' pixel grid by the same method and tool for the line data cube (see above section).

\hspace{10pt}

\section{Results} \label{sec:res}

\begin{figure*}
    \centering
    \includegraphics[width=7.1in]{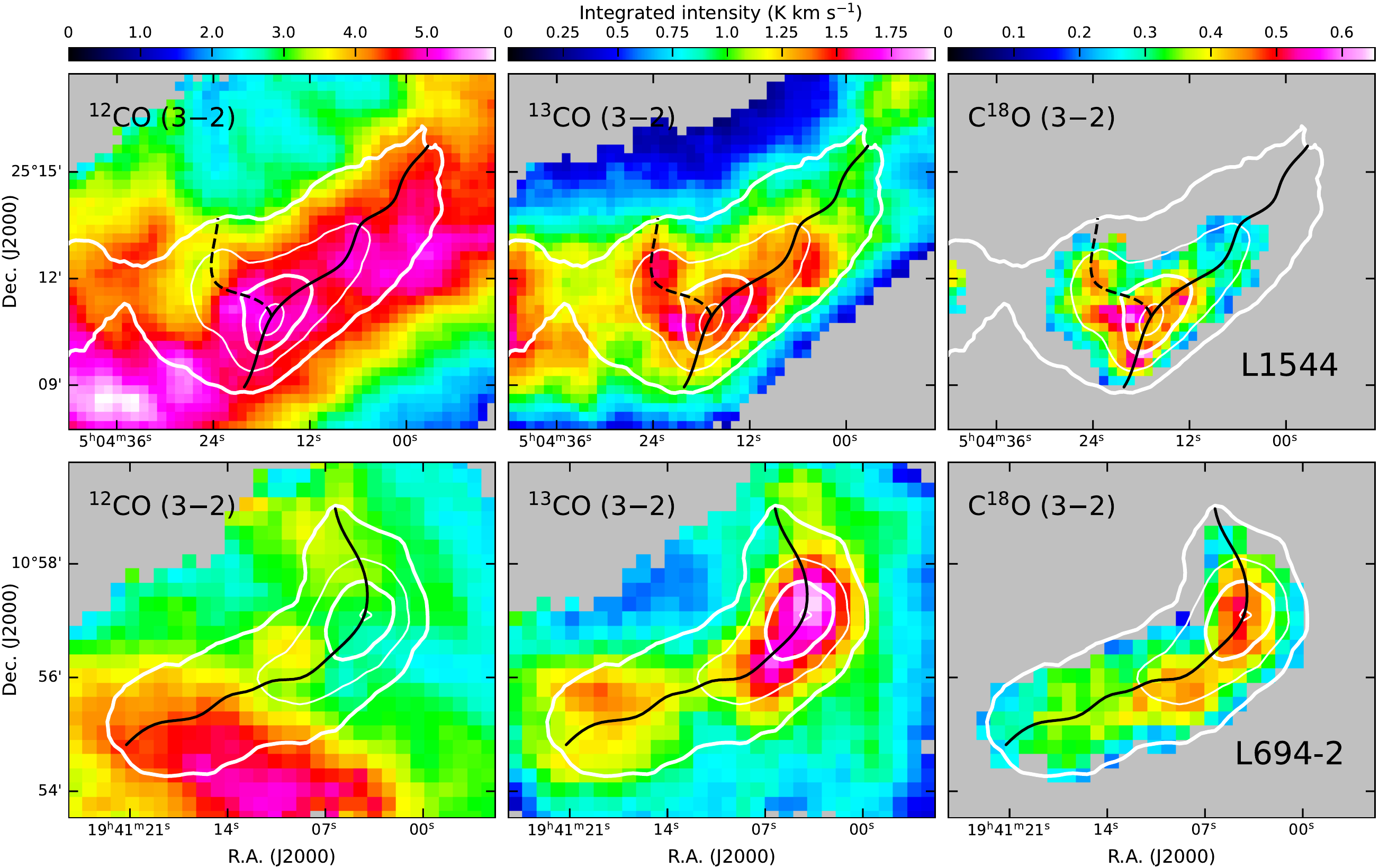}
    \caption{
    Integrated intensity maps (in color) of \lco\ (left panels), \mco\ (center panels), and \sco\ (right panels) emissions for L1544 (upper panels) and L694-2 (lower panels). 
    The white contours show \hmol\ column densities with levels of $2\times10^{21}$, $5\times10^{21}$, $10^{22}$, and $2\times10^{22}$ cm\psq\ derived from the SED fit for \hso\ dust continuum data (Figure \ref{fig:sed_result_h2cd_temp}). 
    The black solid and dashed lines show the skeletons of filamentary structures (Figure \ref{fig:find_skeletons}).
    }
    \label{fig:co_maps}
\end{figure*}

\begin{figure}
    \centering
    \includegraphics[width=3.39in]{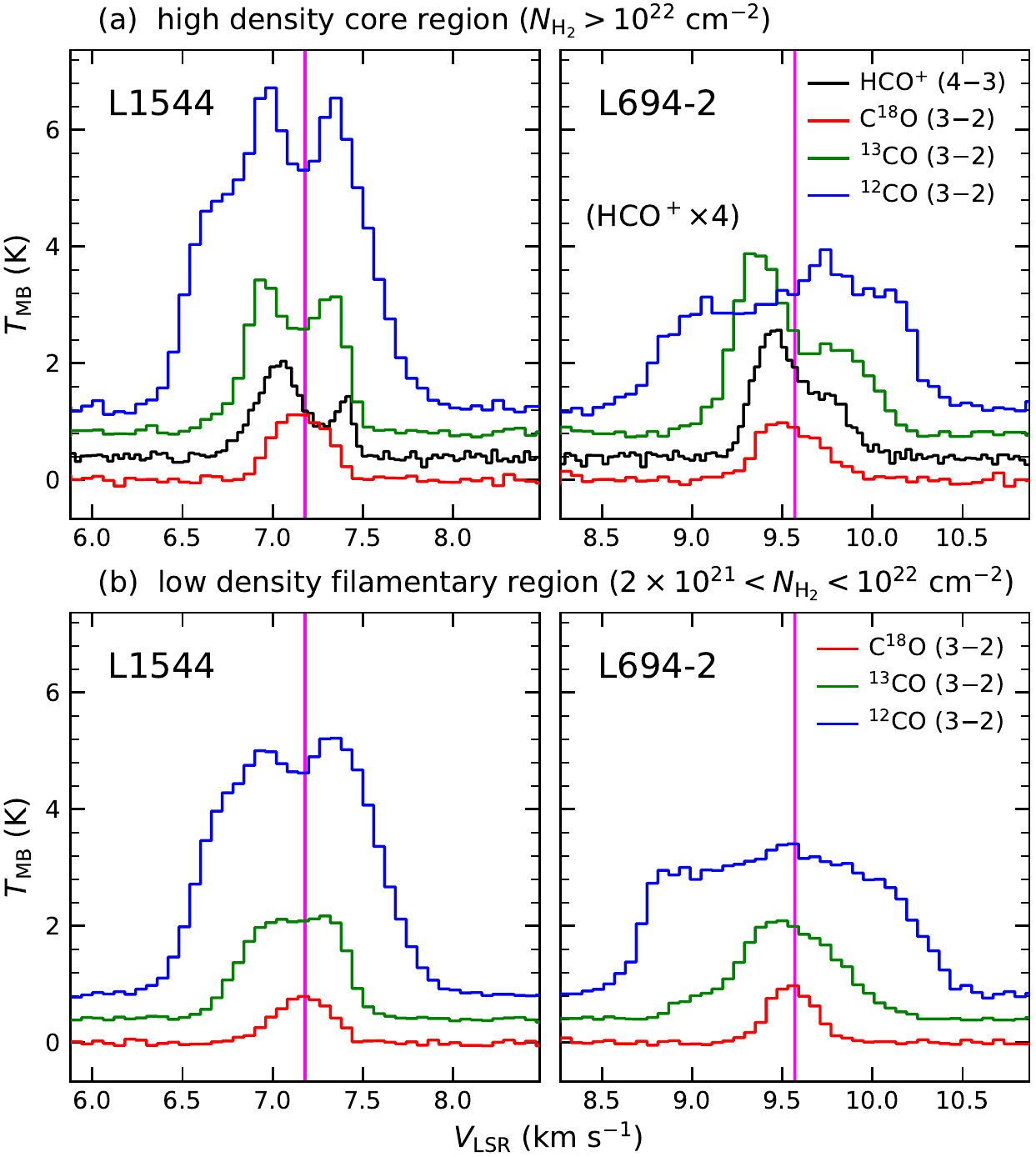}
    \caption{
    Line profiles of \lcot\ (blue), \mcot\ (green), \scot\ (red), and \hcopt\ (black) emissions for L1544 (left panels) and L694-2 (right panels). 
    The profiles were obtained by averaging the spectra for (a) the high density core region ($>10^{22}$ cm\psq) and (b) the entire filament region ($>2\times 10^{21}$ cm\psq), respectively.
    The magenta lines show the systemic velocities of target cores derived in \citet{Crapsi_2005_ApJ_619_379} using the optically thin tracer, $\mathrm{N_2D^+}$ (2--1).
    }
    \label{fig:co_line_profile}
\end{figure}

\begin{figure}
    \centering
    \includegraphics[width=3.39in]{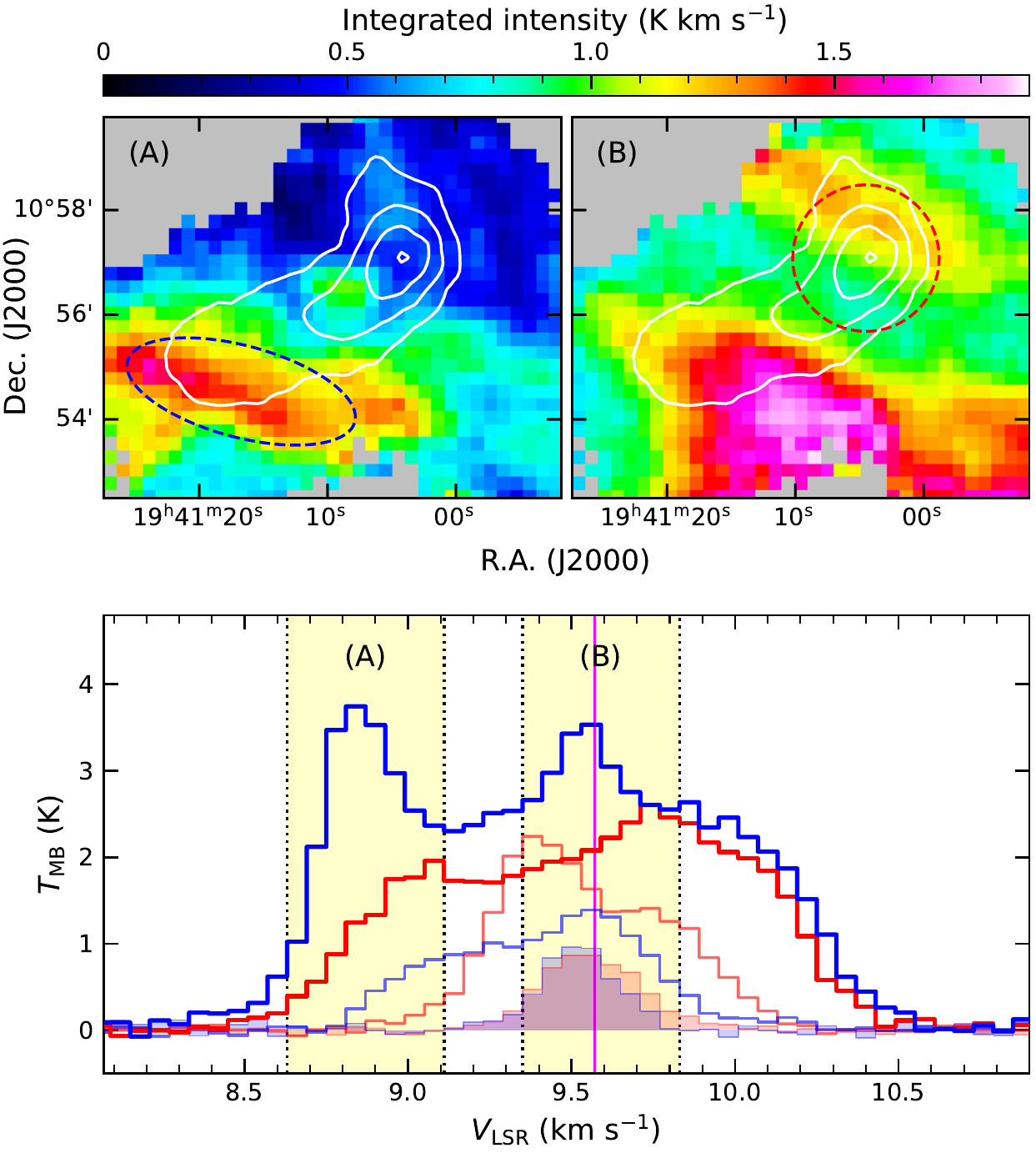}
    \caption{
    Emission distributions (upper panels) and line profiles (lower panel) of CO emissions for two different velocity components in L694-2. 
    The integrated intensity maps were made with a velocity width of 0.48 km s\per\ and a velocity center of $\sim8.8$ km s\per\ for (A) and $\sim9.6$ km s\per\ for (B), which are also displayed in the lower panel with the yellow tone and dotted lines. 
    The line profiles plotted with blue and red histograms represent the average of the spectra in the regions enclosed with dashed lines of the same colors. Their thick, thin, and filled profiles are for \lco, \mco, and \sco, respectively.
    The magenta line shows the systemic velocity of L694-2.
    }
    \label{fig:co_blue_shifted}
\end{figure}

\subsection{Parsec-scale CO emission} \label{sec:res-coe}

Figures \ref{fig:co_maps} and \ref{fig:co_line_profile} show the intensity maps and the average profiles for 3--2 transitional lines of CO isotopologues, \lco, \mco, and, \sco, detected in our JCMT-HARP observations. Our observing lines have rather high critical densities of 1.6--$1.9\times 10^4$ cm$^{-3}$ (Table \ref{tab:lines}). However, the filamentary structures around both L1544 and L694-2 (or envelopes that are extended widely around the target cores) are found to be well detected over the region traced by the \hso\ continuum emission.

$\mathbf{C^{18}O}$. The \scot\ emission was detected in regions with \hmol\ column densities $\gtrsim 2\times 10^{21}$ cm\psq\ (estimated from the dust continuum). 
The integrated intensity distribution of \sco\ generally follows the \hmol\ column density distribution well in the filamentary region, although the peak of the \sco\ emission is slightly offset from the peak position of the \hmol\ column density distribution in the high-density core region especially in the case of L1544.
Most of the line profiles also show a single Gaussian shape and their velocity positions are consistent with the systemic velocity of each core determined from the previous observation using optically thin tracer, $\mathrm{N_2 D^+}$ (2--1) \citep{Crapsi_2005_ApJ_619_379}.
Although there may be the effects of slightly high optical depth and/or depletion in \scot\ in the small central region with high-densities, \scot\ is considered as an optimal tracer for the kinematics of the filamentary structure around the prestellar core.

$\mathbf{^{13}CO}$. The emission of \mcot\ was detected down to the region where the \hmol\ column density was $\sim 2\times 10^{20}$ cm\psq\ for L1544 and $\sim 4\times 10^{19}$ cm\psq\ for L694-2. 
Similar to the case of \sco, \mco\ also traces the high column density region well, but the area over which the line was detected is much wider than that of the \sco\, covering almost the entire filamentary structure surrounding the core.
The \mco\ line profiles outside the core region have mostly single-peaked line shapes. 
Hence, it is easy to measure the velocity of the gas here by making a single component Gaussian fit even though the peaks are flat due to the high optical depth and do not exactly follow a Gaussian profile. 
One interesting point to note is that, in the high-density core region of both L1544 and L694-2, the \mco\ line profiles show strong self-absorption features (see Section \ref{sec:infl} and Figure \ref{fig:infall_profile_map}).
In L1544, most of the \mco\ lines have a double peak and a central absorption dip, while \sco\ lines show a single Gaussian profile whose peak is located between the two peaks of \mco. 
Although there are relatively more profiles with a brighter blue peak (so-called ``blue profiles''), about 40\% of profiles with comparable blue and red peaks or a brighter red peak are also distributed.
However, in the case of L694-2, the blue profiles were predominantly observed. 
The double-peaked profiles having a clear brighter blue peak or the skewed profiles with a brighter blue peak and a distinct red shoulder were widely found around this core.
Therefore, \mco\ line is also considered as an useful tracer for the kinematics of both the less dense part of filamentary structures and the high-density core region.

$\mathbf{^{12}CO}$. The \lcot\ lines were detected on the entire area covered in our observations with \hmol\ column densities down to 2--$3\times 10^{19}$ cm\psq\ in both the targets.
Overall the \lco\ spectra in L1544 and L694-2 seem to be highly affected by their high optical depths and thus self-absorbed. 
There are some differences in these self-absorbed features where \lco\ lines in L1544 are more likely double-peaked while those in L694-2 show flat-top shapes as shown in Figure \ref{fig:co_line_profile}. 
It is noted that the \lco\ emission distribution of L694-2 is significantly distinct from the density distribution.
This is attributed to the presence of two velocity components in L694-2 as shown in the bottom panel of Figure \ref{fig:co_blue_shifted}, one for the main core of L694-2 and the other for a less dense part whose systemic velocity is relatively blue-shifted with respect to that of the main core by $\rm \sim 0.8~km~s^{-1}$. 
The velocity maps of L694-2 shown in Figure \ref{fig:co_blue_shifted} indicate that the $\rm 8.8~km~s^{-1}$ component associated with less dense parts of the cloud is distributed from east to southeast, crossing the main $\rm 9.6~km~s^{-1}$component of the extended \lco\ emission.
The original purpose of the \lco\ observations was to trace low-density regions surrounding the filamentary structures where the dense cores are embedded, in order to identify any kinematical gas motion toward or along the filaments from low-density diffuse regions. 
However, contrary to our expectation, the distribution of \lco\ emission is quite different from those of \mco\ and \sco\ emission, and \hmol\ column density. 
This may be because the optical depths of the \lco\ lines are high throughout the core and also towards the low-density filamentary region. 
Indeed the optical depth can significantly reduce the effective critical density \citep{Shirley_2015_PASP_127_299}. 
Thus most of the \lco\ lines seem saturated, making it to trace only the surface of the filament.
Especially in the case of L694-2, the other additional velocity component seems to coexist with the main component. 
Therefore our \lcot\ line data appears to be not suitable for determining the kinematics properties of the filaments and the cores. 
Instead, the \mco\ lines are found to have sufficiently strong detection even towards the low-density parts of the filamentary structure with less saturation effects. 
This makes it a good tracer of the overall kinematics of the filaments, and their high-density regions with infall asymmetric profiles that can be used to trace the infalling motion of the outer part of core or filamentary region.

\subsection{Asymmetric \hcop\ emission}

The \hcop\ line, a well-known tracer of infall motion \citep[e.g.,][]{Gregersen_1997_ApJ_484_256,Chira_2014_MNRAS_444_874}, was selected to track the contracting motion in the innermost region of the core with the high critical density ($\sim 4\times10^{6}$ cm\pcb) of the 4--3 transition. 
As expected, \hcopt\ emission was significantly detected in both L1544 and L694-2 in an extended area with $N_\mathrm{H_2} \gtrsim 6\times10^{21}$ cm\psq. 
Especially the infall asymmetric profiles were significantly detected with a signal-to-noise ratio ($SNR$) $\gtrsim 10$ in the high density core regions ($>10^{22}$ cm\psq) of the two cores as shown in Figure \ref{fig:co_line_profile}, which may allow us to examine the characteristics of the gas kinematics in the cores well (Table \ref{tab:obs}). 
These infall profiles are found to be distributed over a radius of 0.05 pc from the core center (see Section \ref{sec:infl} and Figure \ref{fig:infall_profile_map}).

On the other hand, the \htcopt\ line which was adopted as an optically thin tracer of infall was not detected in both L1544 and L694-2. 
For the detection of \htcopt, a noise level of at least $\sim0.03$ K or lower is required as can be seen from Table \ref{tab:obs}.
The systemic velocities of the cores were alternatively estimated using \nthp\ (3--2) and \ntdp\ (3--2) data \citep{Crapsi_2005_ApJ_619_379}.

\section{Density Structures} \label{sec:dens}

\begin{figure}
    \centering
    \includegraphics[width=3.39in]{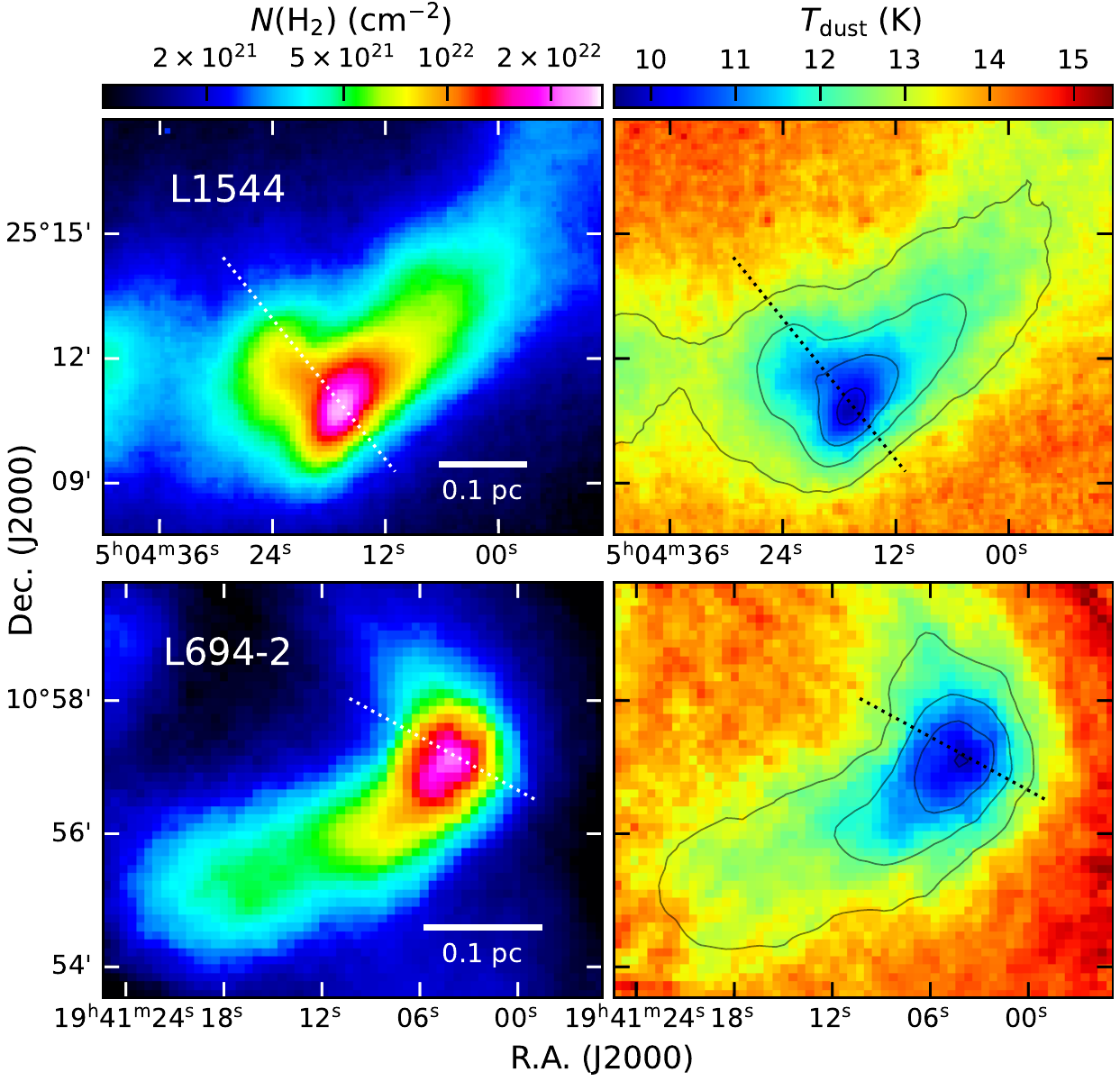}
    \caption{
    Distribution of column density ($N(\mathrm{H_2})$, left panels) and dust temperature ($T_\mathrm{dust}$, right panels) of L1544 (upper panels) and L694-2 (lower panels) derived from the SED fit for \hso\ dust continuum data.
    The dotted lines indicate the positions for which physical properties are displayed as their cut profiles in Figure \ref{fig:sed_result_subtract_effect}.
    }
    \label{fig:sed_result_h2cd_temp}
\end{figure}

In this section we analyse the physical structures of the filamentary envelopes of the two cores, especially focusing on the density structures as they are the basic ingredients to study the kinematics of the cloud cores and their filamentary envelopes. 

For this purpose we determined the \hmol\ column density distribution using the \hso\ continuum data. According to Section \ref{sec:res-coe}, main parts of two filaments are found to consist of a single velocity component. 
Note that L694-2 has one additional velocity component (``8.8 km s\per\ component''), but this component traces only the low density parts of the cloud (as it was not detected in \sco, see Figure \ref{fig:co_blue_shifted}) and may not have much affect on the physical structure of the main filament. 
Therefore we should be able to obtain the density structures of the filamentary envelopes using the continuum data. 
We derived the column density and the dust temperature maps by the iterative spectral energy distribution (SED) fitting of the \hso\ dust continuum data from 160 (250 for L694-2) to 500 \micron\ using the modified Planck function. 
In general, we followed the fitting procedure described in \citet{Kim_2020_ApJ_891_169}. 
However, there are slight differences compared to the procedure adopted by \citet{Kim_2020_ApJ_891_169} in the handling of the continuum data, especially in making the background subtraction and also in adopting optical depth in the equation for the dust emission. 
These are described in Appendix \ref{sec:app-sed}.

The resulting distributions of \hmol\ column density of L1544 and L694-2 including their dust temperature distribution are shown in Figure \ref{fig:sed_result_h2cd_temp}. 
The \hmol\ column density maps of the two targets reveal a high-density core and its surrounding filamentary structure of (sub-)parsec scale. 
The high-density core regions show an \hmol\ column density of $>10^{22}$ cm\psq\ and the dust temperature of 9.5--11 K, while the low-density filament regions show $10^{21}$--$10^{22}$ cm\psq\ and 11--13 K. 
These increasing density and decreasing temperature towards the central dense region in the filaments are typical of starless cores and their envelopes \citep[e.g.,][]{Galli_2002_A&A_394_275,Crapsi_2007_A&A_470_221,Bergin_2007_ARA&A_45_339}.
Note that the estimated \hmol\ column densities in the high-density core region are lower than those deduced from mm-continuum emission \citep[e.g.,][]{Crapsi_2005_ApJ_619_379}; however, the far-IR continuum, like \hso\ data, is more suitable to trace the dust emissions from the region with a lower density (and necessarily higher temperature), such as a filamentary envelope of core.

\begin{figure}
    \centering
    \includegraphics[width=3.39in]{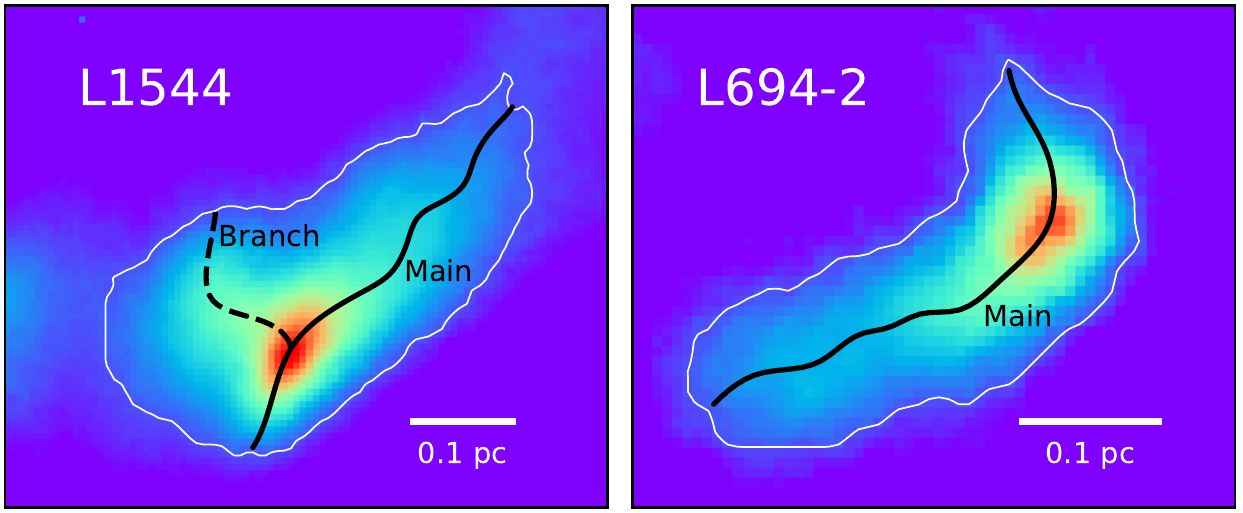}
    \caption{
    Skeletons of the filamentary structure surrounding L1544 (left panels) and L694-2 (right panels) identified using the \texttt{FilFinder} algorithm \citep{Koch_2015_MNRAS_452_3435}.
    The white contour on each panel indicates the filament boundary of an \hmol\ column density level defined at $2\times 10^{21}$ cm\psq\ (see text for detail).}
    \label{fig:find_skeletons}
\end{figure}

\subsection{Filamentary Structure} \label{sec:dens-fil}

We now find the skeletons of filamentary structures in L1544 and L694-2 using \hmol\ column density distribution, giving the central axis of filamentary structure or the ridge of density distribution (Figure \ref{fig:find_skeletons}). 
The skeletons were identified using the \texttt{FilFinder}\footnote{https://github.com/e-koch/FilFinder} algorithm \citep{Koch_2015_MNRAS_452_3435}.
As the boundary level of the filament which is the \texttt{FilFinder} parameter for the global threshold, we set an \hmol\ column density of $2\times 10^{21}$ cm\psq\ which closely matches the $3\sigma$ detection level of the \sco\ emission.
Figure \ref{fig:find_skeletons} shows that L694-2 has a single long skeleton while L1544 has one single main skeleton along the main axis of the filament and one small branch skeleton to the NE which is emerging out of the core.

\subsection{Target Distance}

In order to derive physical quantities like the length or mass of the filaments, accurate measurement of the distances of the targets is important.
The distance of L1544 is usually considered as 140 pc with the assumption that the cloud is associated with the Taurus molecular cloud \citep{Elias_1978_ApJ_224_857}. 
The cloud L694-2 is considered to be either at a distance of 250 pc by assuming it to be at a same distance to that of B335, which is located within a few degrees from L694-2 and shows a similar LSR velocity \citep{Lee_2001_ApJS_136_703}, or at a distance of $230\pm30$ pc estimated based on the Wolf diagram analysis for the L694 dark cloud containing L694-2 \citep{Kawamura_2001_PASJ_53_1097}. 
Thus it is apparent that the distances to both L1544 and L694-2 are not well constrained.

With the \gaia\ DR2 astrometric data, more reliable measurements of distance of molecular clouds have been recently made \citep[e.g.,][]{Zucker_2019_ApJ_879_125, Yan_2019_ApJ_885_19}. 
However, such reliable estimates of distance are not available for L1544 and L694-2. 
Therefore, for our study we made an attempt to determine the distances of L1544 and L694-2 using \gaia\ DR2 astrometric data \citep{Gaia_source_cat_dr2} and Pan-STARRS1 stellar photometry.
The $g-r$ and $r-i$ colours obtained from the Pan-STARRS1 catalog are used to segregate M-type dwarfs that are lying projected on our target clouds. The extinctions of the M dwarfs are determined by dereddening the M-dwarfs assuming a normal interstellar extinction law as described in Appendix \ref{sec:app-distance}. 
The distances to the identified M-dwarfs are obtained from the \gaia\ DR2 parallax measurements. 
We then plotted the extinctions of the M-dwarfs as a function of their distance and determined the distances to the cloud where the extinctions values showed an abruptly increase due to the presence of the cloud. 
The new distances determined in this way are $175_{-3}^{+4}$ pc for L1544 and $203_{-7}^{+6}$ pc for L694-2 which are used for our subsequent analysis. More detailed explanation on the procedure used to get the distance is given in Appendix \ref{sec:app-distance}.

\begin{deluxetable}{llCCC}
    \tablecaption{Physical Properties of Identified Filamentary Structures\label{tab:fil-prop}}
    \tablewidth{0pt}
    \tablehead{
    \colhead{Target} & \colhead{Fil.} & \colhead{$L$} & \colhead{$M$} & \colhead{$M_\mathrm{lin}$} \\
    \colhead{} & \colhead{} & \colhead{(pc)} & \colhead{(\solm)} & \colhead{(\solm\ pc\per)}
    }
    \decimalcolnumbers
    \startdata
    L1544                & Main   & 0.45 & 7.1\pm0.3 & 15.7\pm0.6 \\
    \multicolumn1c{$''$} & Branch & 0.19 & 3.1\pm0.1 & 16.3\pm0.7 \\
    L694-2               & Main   & 0.41 & 5.0\pm0.3 & 12.1\pm0.8 \\
    \enddata
    \tablecomments{(1--2) Target and filament name. (3) Projected filament length. (4) Filament mass within the filament boundary level of $2\times10^{21}$ cm\psq. (5) Filament mass per unit length (or line mass). Projection effect is not considered.}
\end{deluxetable}

\begin{figure*}
    \centering
    \includegraphics[width=7.2in]{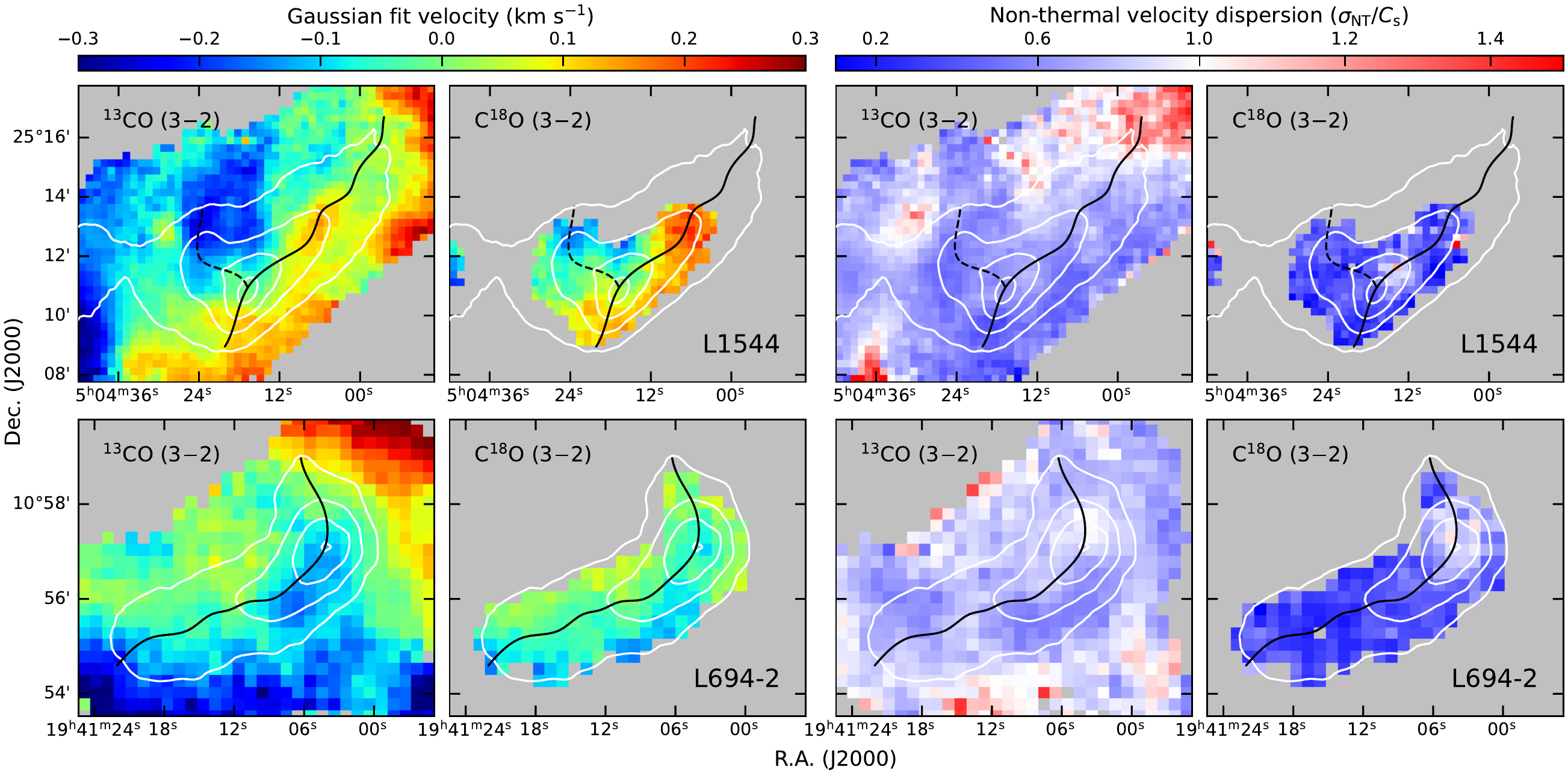}
    \caption{Velocity distributions of L1544 (upper panels) and L694-2 (lower panels) obtained from \mco\ and \sco\ line profiles.
    \textit{Left panels}: The distributions of Gaussian fit velocity displayed as the velocity difference from the median velocities (7.12 km s\per) for L1544 and (9.58 km s\per) for L694-2. \textit{Right panels}: The distributions of non-thermal velocity dispersion expressed by the sonic Mach number ($\sigma_\mathrm{NT}/c_\mathrm{s}$).
    The white contours and the black lines are as described in Figure \ref{fig:co_maps}.}
    \label{fig:gauss_velocity}
\end{figure*}

\subsection{Filament Properties}

The new distances of the target clouds enabled us to derive more reliably quantities that characterize the physical properties of the identified filaments. 
These quantities are listed in Table \ref{tab:fil-prop}. 
The filament length given in the table is defined as the (projected) length of the skeleton. 
The mass of the filament is obtained by making a summation of the \hmol\ column densities lying within the contour level of $2\times10^{21}$ cm\psq\ which is considered as the lowest column density in the filament.
For L1544, the $2\times10^{21}$ cm\psq\ contour was not closed due to the existence of another nearby dense core named L1544-E as shown in Figure \ref{fig:target_cores}, and thus the filament boundary between L1544 and L1544-E was manually drawn by following the minimum values ($\sim 2.2\times 10^{21}$ cm\psq) between column densities of the two cores (Figure \ref{fig:find_skeletons}).
The line mass of filament, defined as the filament mass divided by the filament length, is given in the table with no correction for its projection effect on the sky.

The mass and size scales of the filaments identified in this study are very similar to those of the filaments found in the L1517 cloud \citep{Hacar_2011_A&A_533_A34} and comparable with those of the velocity-coherent structures, `fibers,' which constitute the large-scale filaments in the L1495/B213 region, too \citep{Hacar_2013_A&A_554_A55}.
These filaments are found to have physical quantities of $L\sim 0.2$--0.7 pc and $M_\mathrm{lin}\sim 10$--20 \solm\ pc\per.
Compared to the mass per unit length ($\sim 16$ \solm\ pc\per) for an isothermal cylinder at 10 K expected from the models of \citet{Stodokiewicz_1963_AcA_13_30} and \citet{Ostriker_1964_ApJ_140_1056}, the line masses ($M_\mathrm{lin}$) of L1544 and L694-2 filaments are comparable to or less than this critical value, even though these filaments contain the prestellar cores.
We note that the $M_\mathrm{lin}$ might be underestimated by taking a large value of the filament length.
In fact most of the mass is concentrated toward the core, while the filament has a long tail of diffuse emission.
Therefore the global $M_\mathrm{lin}$ may not be the most suitable parameter in characterising the overall stability of the filamentary clouds.
This can be well examined when the $M_\mathrm{lin}$ is estimated by dividing the main filament into two parts as a core part and a tail part, $M_\mathrm{lin}^\mathrm{core}$ and $M_\mathrm{lin}^\mathrm{tail}$ are 20.7 and 9.8 \solm\ pc\per\ in L1544 and 16.2 and 8.1 \solm\ pc\per\ in L694-2, respectively.

\section{Velocity Structures} \label{sec:vels}

The previous section discusses the density structures by deriving the skeletons of the column density distributions of the two clouds. 
Now in this section the velocity structures along those density structures are discussed.
For this purpose we obtained the velocity centroid and the velocity dispersion for the \mco\ and \sco\ spectra from their Gaussian fits.
The velocity distributions that we obtain from these fits are displayed in the Figure \ref{fig:gauss_velocity}.
As mentioned in Section \ref{sec:res-coe}, most of the \sco\ spectra have single Gaussian component with narrow width, while a significant number of the \mco\ spectra often deviate from the Gaussian shape, particularly in the high column density regions where \sco\ is detected. 
At the same time the \mco\ spectral shapes get close to a single Gaussian function in the low column density regions where \sco\ is not detected. 
Therefore, for the discussion on the velocity structures of our targets along the filaments we used the velocity information obtained from the \sco\ spectra wherever it is detected and also from the \mco\ spectra in the low column density region where \sco\ is not detected.

Overall velocity distributions of two targets obtained using \mco\ and \sco\ line data are shown in Figure \ref{fig:gauss_velocity}. 
A distinct feature in the distribution of the centroid velocities is that there is a clear velocity variation between the filament and its surrounding in both targets. 
We found large velocity differences of $\sim 0.6$ km s\per\ in L1544 and L694-2. 
The other interesting thing we found is that the non-thermal velocity dispersion is somehow increased at the outer part of the filament. 
We will discuss implications of these features later in Section \ref{sec:disc} regarding the filament formation mechanism. 
In the velocity distributions, we also found significant variations in the local velocity and velocity dispersion along the filaments, especially around the dense cores. 
This section will focus on these variations and their possible role in the core formation process.

\begin{figure*}
    \centering
    \includegraphics[width=7.1in]{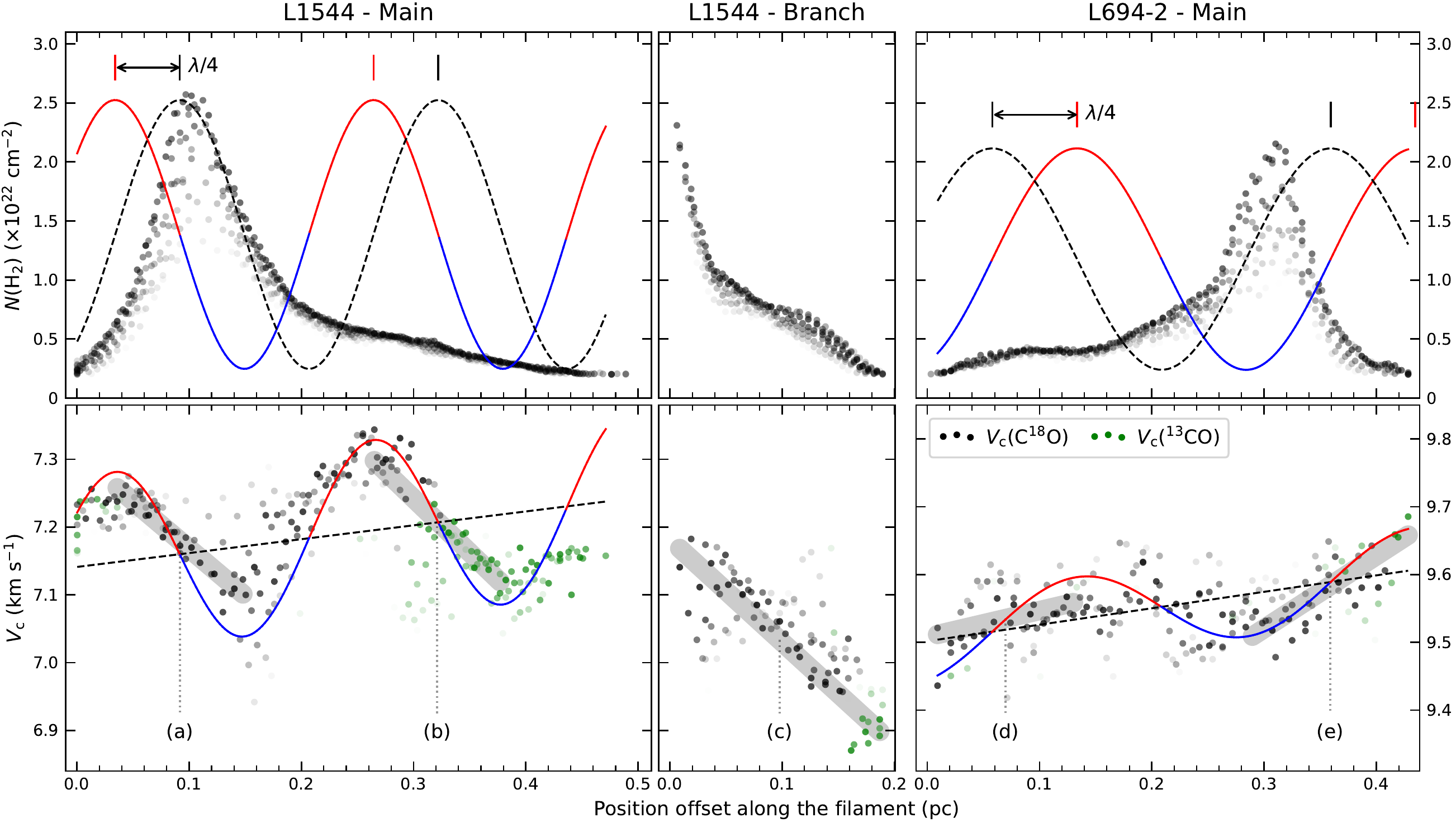}
    \caption{
    Distribution of velocity centroid and \hmol\ column density along the filamentary structures in L1544 and L694-2. 
    The x-axis is the offset distance in parsec measured along the skeleton from the southern end of the filament or branch. 
    \textit{In lower panels}: The black dots represent the \sco\ velocity centroid and green dots \mco\ values only for the pixels where \sco\ is not detected. 
    The data only within a radial distance of 0.15 pc from the skeleton are displayed (the width limit for the plots is 0.3 pc). 
    The red and blue curve is the best sinusoidal fit to the velocity centroid. 
    The black dashed line is a linear gradient in the sinusoidal fit. 
    The grey thick lines are to show the identified local velocity gradients of converging flow motions along the filament listed in Table \ref{tab:axial-flow-motion}. 
    \textit{In upper panels}: The black dots represent the \hmol\ column density. 
    The red and blue curve is a sinusoidal curve with the same frequency and phase values of the best-fit parameters for the velocity centroid, and the black dashed curve is a $\lambda /4$ shifted one. 
    The tone of dots is drawn to decrease with radial distance from the skeleton. 
    From (a) to (e) are to indicate the systemic velocities at the points of dense cores where its converging flow motions are measured.
    }
    \label{fig:velocity_profile}
\end{figure*}

\begin{deluxetable*}{cCCCCC}
    \tablecaption{Mass Accretion Along the Filaments\label{tab:axial-flow-motion}}
    \tablehead{
    \colhead{\#} & \colhead{$L_\mathrm{fil}$} & \colhead{$M_\mathrm{fil}$} & \colhead{$\nabla V_\parallel$} & \colhead{$\dot{M}_\parallel$ } & \colhead{$\dot{M}_{\parallel,\mathrm{line}}$} \\
    \colhead{} & \colhead{(pc)} & \colhead{(M$_\odot$)} & \colhead{(km s\per\ pc\per)} & \colhead{(\solm\ Myr\per)} & \colhead{(\solm\ Myr\per\ pc$^{-1}$) }
    }
    \decimalcolnumbers
    \startdata
    \multicolumn6c{L1544 -- Main filament} \\
    a & 0.11 & 2.1\pm0.1 & 1.40\pm0.08 & 3.0\pm0.2 & 26.6\pm1.9 \\
    b & 0.11 & 0.8\pm0.1 & 1.59\pm0.08 & 1.3\pm0.1 & 11.4\pm0.8 \\
    \hline
    \multicolumn6c{L1544 -- Branch} \\
    c & 0.18 & 2.0\pm0.1 & 1.52\pm0.05 & 3.0\pm0.2 & 17.0\pm0.9 \\
    \hline
    \multicolumn6c{L694-2 -- Main filament} \\
    d & 0.12 & 0.7\pm0.1 & 0.39\pm0.06 & 0.3\pm0.1 & \phn2.2\pm0.4 \\ 
    e & 0.14 & 1.7\pm0.1 & 1.08\pm0.08 & 1.8\pm0.2 & 13.1\pm1.4 \\
    \enddata
    \tablecomments{(1) Position where the converging flows are measured in Figure \ref{fig:velocity_profile}. (2) Filament length along which the velocity gradient is measured. (3) Filament mass within $L_\mathrm{fil}$. (4) Velocity gradient measured along the filament. (5) Mass accretion rate of the flow motion along the filament by a simple cylindrical model with $\tan{\alpha} = 1$. (6) Mass accretion rate per unit length (pc) by a simple cylindrical model with $\tan{\alpha} = 1$.}
\end{deluxetable*}

\subsection{Velocity Variation Along the Filament} \label{sec:vels-cen}

The velocity variation along the filament can be quantitatively examined with the velocity centroid distribution as a function of distance along the filament's skeleton.
The distance of each pixel position along the filament was measured to be the position offset in parsec measured along the filament's skeleton from the starting point (southern end) of the skeleton. 
We should note that there are multiple pixel positions along the line perpendicular to the skeleton. 
They are assumed to have the same distance as the pixel position along the skeleton. 
In this way, the velocity variation along the filaments can be displayed as shown in the bottom panels of Figure \ref{fig:velocity_profile}. 
At the same time we also derive the \hmol\ column density distribution along the filaments as shown in the upper panel of Figure \ref{fig:velocity_profile} that can be compared with the velocity variation along the filaments. 
We note that in Figure \ref{fig:velocity_profile} the `branch' filament of L1544 is also indicated as a separate filament structure connecting to the dense core embedded in the filament. 
Pixels located somewhere between the branch and the main filament were assigned as members of the nearer structure.

The velocity variation and the column density distribution as a function of positions on the filaments measured using the procedure described above and shown in Figure \ref{fig:velocity_profile} can be used to examine the kinematic structures in a quantitative way. 
In general, the velocity centroids along the filaments slightly change within a range of 0.3 km s\per\ or less without a large dispersion, indicating that these filaments are velocity-coherent on the sub-parsec scale.
However, we also note that there are velocity variations at the dense core size scale, especially across the dense cores embedded in the filaments. 
In the case of L1544, such variations are clearly seen over the dense core of L1544 as well as the branch filament of L1544. 
The branch part of L1544 is in fact directly connected to the dense core of L1544 and thus somehow should be regarded as the boundary part of the dense cores. 
Similar velocity variation is also seen in L694-2, especially across dense cores of L694-2, although it is less clear than the case of L1544.
These gradients over the dense cores remind us of the gas flow toward the dense cores related to the core formation in the filament. 
If the gas flow is somehow associated with the core formation, two velocity peaks would appear with the red-shifted and the blue-shifted peaks across the dense cores while the velocity at the column density peak position would be close to the systemic velocity of the dense core. 
This will result in $\lambda /4$ shift between the velocity and the column density variations where $\lambda$ is meant to be the wavelength of the variations of the velocity and column density \citep[e.g.,][]{Hacar_2011_A&A_533_A34, Liu_2019_MNRAS_487_1259}.

To examine whether such a $\lambda/4$ shift appears in the main filaments of L1544 and L694-2, we performed a sinusoidal fit on the velocity variation using a sine function (Figure \ref{fig:velocity_profile}). 
The sinusoidal fit was only considered for the data within a radial distance of 0.05 pc from the skeleton and was carried out with the Orthogonal Distance Regression (ODR) using the python package \texttt{scipy.odr}\footnote{https://docs.scipy.org/doc/scipy/reference/odr.html}.
In the fitting process, the results with a wavelength of sine function shorter than 0.1 pc or longer than 2$L_\mathrm{fil}$ (the filament length) were excluded.
The L1544 filament shows clear velocity oscillation, which is well fitted to the sine function with a wavelength of $\sim0.23$ pc.
In comparison, the velocity oscillation in the L694-2 filament is not as clear as in the case of the L1544 filament, but best fitted with the sine function of a wavelength of $\sim 0.3$ pc (lower panel of Figure \ref{fig:velocity_profile}).
On the other hand, the density oscillation is not fully seen because each of L1544 and L694-2 filaments contains only one core.
Even though the density oscillation is limited at the scale of the size of the dense core, in the case of L1544, $\lambda/4$ shift between the velocity oscillation and column density peak is fairly well seen. In the case of L694-2, the phase shift between the velocity and the column density distributions seems slightly larger than the $\lambda/4$ shift.
Considering that the density distribution of the dense cores in L1544 and L694-2 filaments is not symmetric along the filaments, this deviation from the $\lambda/4$ shift between density and velocity oscillations seems not very significant.

Nevertheless, the velocity gradients around the dense cores are more intuitively identifiable, as seen on (a), (c), and (e) in the lower panels of Figure 8, being likely the signature of the core-forming flow motion from the filament.
The quantitative scale of the core-forming flow motion in filamentary structures can be important in examining the role of filaments in core formation.
Following the method of \citet{Kirk_2013_ApJ_766_115}, the mass accretion rate of the axial flow motion assuming a simple cylindrical model is given by
\begin{equation}
    \dot{M}_\parallel = \frac{\nabla V_\parallel M_\mathrm{fil}}{\tan{\alpha}}\;, 
\end{equation} 
where $\nabla V_\parallel$ is the velocity gradient measured along the filament, $M_\mathrm{fil}$ is the filament mass where the velocity gradient is measured, and $\alpha$ is the angle between the axis of the cylinder and plane of the sky.
Local velocity gradients were estimated at five sections indicated by the thick grey lines in Figure \ref{fig:velocity_profile} and listed in Table \ref{tab:axial-flow-motion}: two in the main filament of L1544, one in its branch, and two for the filament of L694-2.
The sections were separated by the crest to the trough part of the best fit sinusoidal function, and the data within a radial distance of 0.05 pc from the skeleton were used for the linear fit.
These velocity gradients were measured by the linear fit for the data set of the velocity centroids and position offsets for each filament section so that the slopes given in the fits correspond to the mean values of the velocity gradients for the sections.
In L1544, we measured a similar velocity gradient of about 1.5 km s\per\ pc\per\ among the core, the branch, and the region of low-density filament.
In the filament of L694-2, its velocity gradients are measured to be 1.1 km s\per\ pc\per\ near the core and 0.4 km s\per\ pc\per\ for its low-density part.
Assuming the inclination angle ($\alpha$) is $45^\circ$, the mass accretion rates near the core are estimated to be 3 \solm\ Myr\per\ for L1544 and $\sim2$ \solm\ Myr\per\ for L694-2, respectively.
The extreme case where the $\alpha$ is $0\degr$ or $90\degr$ can be excluded from the fact that both the filamentary structures do exist and the velocity oscillation along those are significantly measurable. 
We simply assume $\tan{\alpha} = 1$ in the later discussion, and our estimated velocity gradients and mass accretion rates are expected to be uncertain by a factor of 2 when the $\alpha$ changes between $30\degr$ and $60\degr$.

The mass accretion rate, 2--3 \solm\ Myr\per, of the core-forming flow motion is the first quantitative measurement obtained from the sinusoidal velocity oscillations around low-mass prestellar cores and might be the smallest scale among the reported cases of axial mass accretion in the filament so far.
The longitudinal accretion rates have been reported as $\sim 10$ \solm\ Myr\per\ for the Serpens filament \citep{Gong_2018_A&A_620_A62}, 15--35 \solm\ Myr\per\ in the IC 5146 \citep{Chung_2021_ApJ_919_3}, and $\sim 45$ \solm\ Myr\per\ for the southern filament of Serpens south protocluster \citep{Kirk_2013_ApJ_766_115}.
In addition, the filamentary flow motions found in high-mass star-forming regions, such as hub-filaments, IRDCs, or protoclusters, show 1 or 2 order of magnitude greater mass accretion rates \citep{Lee_2013_ApJ_772_100,Peretto_2014_A&A_561_A83,Yuan_2018_ApJ_852_12,Lu_2018_ApJ_855_9}.
Indeed, all previous cases showing significant differences in accretion scale are not based on the sinusoidal velocity oscillations with the size scale of low-mass core but the local or global axial velocity gradients of filamentary structures.

The mass accretion rates per unit length of the core-forming flow motions in L1544 and L694-2 are about 11--27 \solm\ Myr\per\ pc\per\ (Table \ref{tab:axial-flow-motion}).
We note that this amount of the accretion rate is comparable to that estimated from the Serpens filament \citep[$\sim20$ \solm\ Myr\per\ pc\per,][]{Gong_2018_A&A_620_A62}.

\begin{figure*}
    \centering
    \includegraphics[width=7.1in]{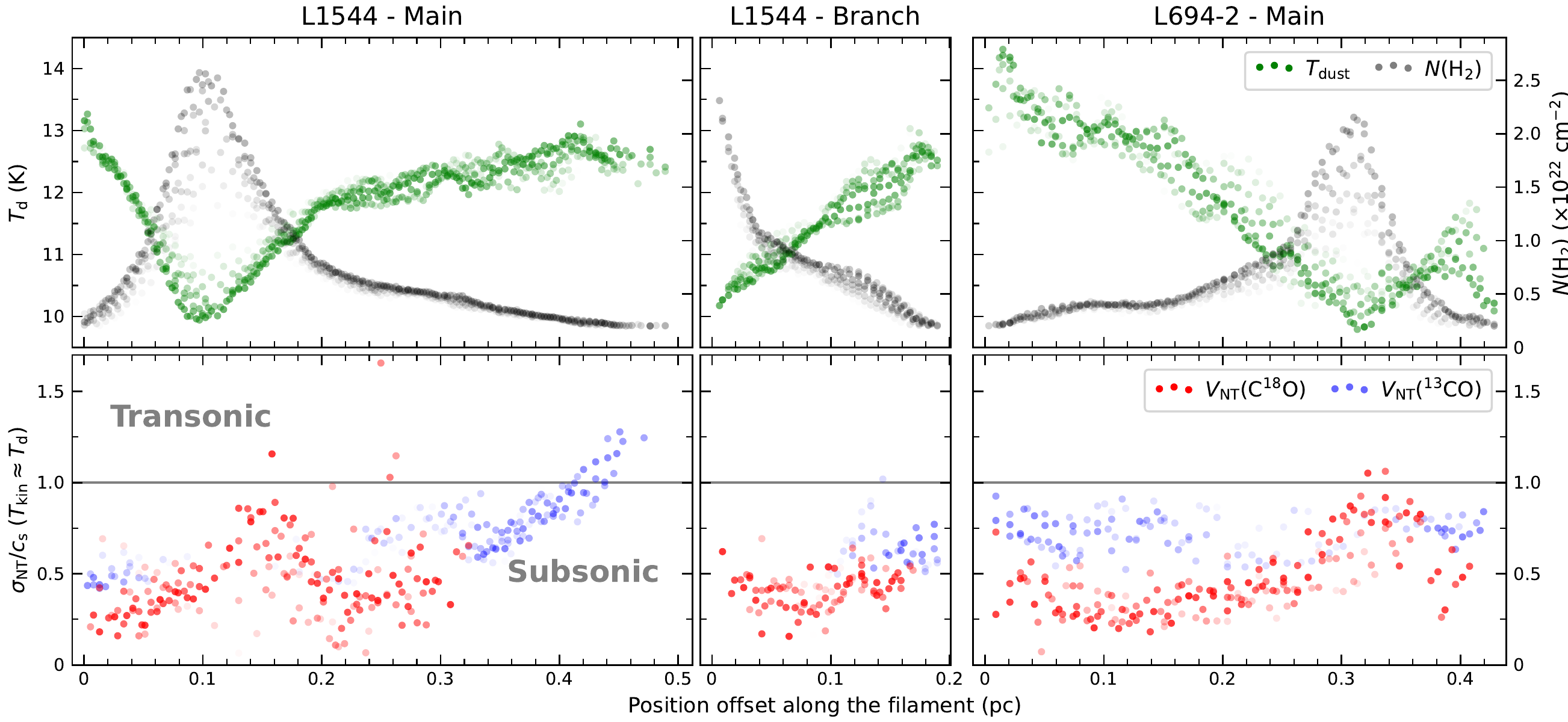}
    \caption{
    Distribution of non-thermal velocity dispersion as a function of position offset along the filamentary structures in L1544 and L694-2. 
    \textit{In lower panels}: The red and blue dots represent the non-thermal velocity dispersion in terms of the isothermal sound speed for \sco\ and \mco, respectively. The data points brighter than 3$\sigma$ level within a radial distance of 0.15 pc from the skeleton are plotted. The \mco\ data are corrected for the optical depth broadening and displayed for the positions only with \hmol\ column density lower than 10\% level of the peak value of \hmol\ column density (see Appendix \ref{sec:app-corr-width}). 
    The grey line divides between subsonic and transonic regimes. 
    \textit{In upper panels}: The black and green dots are to display the distributions of the \hmol\ column density and the dust temperature estimated using the \hso\ dust continuum data, respectively, in the regions of the core and filament cloud for two targets. 
    The tones of dots are drawn to decrease with radial distance from the skeleton.
    }
    \label{fig:velocity_dispersion}
\end{figure*}

\begin{figure}
    \centering
    \includegraphics[width=3.39in]{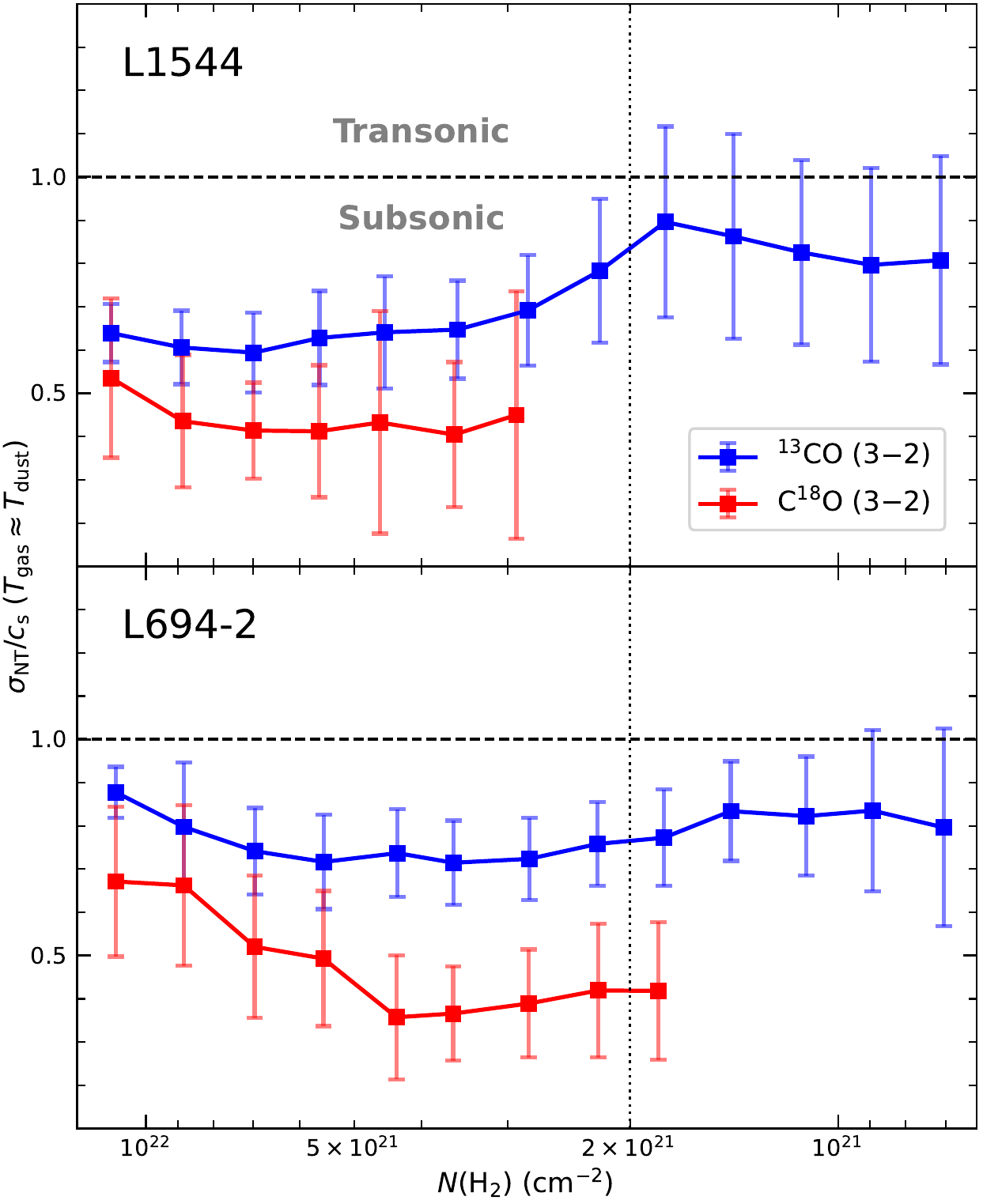}
    \caption{
    Non-thermal velocity dispersion (expressed by the sonic Mach number) obtained from \mco\ (blue) and \sco\ (red) spectra vs. \hmol\ column density in L1544 (upper panel) and L694-2 (lower panel). The dashed horizontal line is to divide between subsonic and transonic regimes, and the dotted vertical line indicates the filament boundary.
    }
    \label{fig:snth-vs-h2cd}
\end{figure}

\subsection{Velocity Dispersion} \label{sec:vels-disp}

The observed profile of a molecular line with a certain width is a result of the broadening effects mainly due to thermal and non-thermal motions of the molecular gas.
In this section, we examine the characteristics of the non-thermal motion of CO gas in the filamentary cloud. 
This non-thermal velocity dispersion (\snth) can be estimated by making a quadrature subtraction of the thermal velocity dispersion (\sth) from the observed velocity dispersion (\sobs):
\begin{equation}
    \sigma_\mathrm{NT} = \sqrt{\sigma_\mathrm{obs}^2-\sigma_\mathrm{th}^2}\;, 
\end{equation}
\begin{equation}
    \sigma_\mathrm{th} = \sqrt{\frac{k_\mathrm{B} T_\mathrm{kin}}{m_\mathrm{obs}}}\;, 
\end{equation}
where \sobs\ is the velocity dispersion (or Gaussian standard deviation) obtained by the Gaussian fit for observed \mco\ and \sco\ spectra, $k_\mathrm{B}$ is the Boltzmann constant, $T_\mathrm{kin}$ is the kinetic temperature, and $m_\mathrm{obs}$ is the mass of observed molecule.
An effect by the \snth\ can be expressed by the sonic Mach number ($\mathcal{M} = \sigma_\mathrm{NT}/c_\mathrm{s}$) which is the ratio of the nonthermal dispersion to the isothermal sound speed in a cloud, 
\begin{equation}
    c_\mathrm{s} = \sqrt{\frac{k T_\mathrm{kin}}{\mu_\mathrm{p} m_\mathrm{H}}}\;, 
\end{equation}
where $\mu_\mathrm{p}$ is the mean molecular weight per free particle and $m_\mathrm{H}$ is the mass of the hydrogen atom.
We used $\mu_\mathrm{p}$ of 2.37 obtained with the mass fraction of $\mathrm{H}:\mathrm{He}:\mathrm{Z} \approx 0.71:0.27:0.02$ \citep{Cox_2000_PhT_53_77,Kauffmann_2008_A&A_487_993}.

In this study, the optical depth broadening effects for the \sobs\ values are corrected (details for the correction are described in Appendix \ref{sec:app-corr-width}).
The \mco\ optical depths, $\tau_0(\mathrm{^{13}CO})$, are found to be in a range of about 1--10 in our target regions so that the optical depth broadening factors, $\beta_\tau(\mathrm{^{13}CO})$, are about 1.18--1.96. 
The velocity dispersion measured by the Gaussian fit ($\sigma_\mathrm{fit}$) was reduced to the observed velocity dispersion, \sobs, by dividing by these $\beta_\tau$ values.

As for the kinetic temperature ($T_\mathrm{kin}$), we adopted the dust temperature, $T_\mathrm{d}$, as its proxy.
The $T_\mathrm{d}$ all around two target regions are estimated to be in the range of 9.5--14 K. 
Then the corresponding \cs, and the \sth\ of \sco\ are calculated to be in the range of 0.18--0.22 km s\per\ and 0.05--0.06 km s\per, respectively. 
Note that the \snth\ estimation is very sensitive to the values of $T_\mathrm{kin}$ and $\tau$ determined.
For example, when $\sigma_\mathrm{fit} \approx 0.34$ km s\per\ and $\tau_0 \approx 5$, $\sigma_\mathrm{obs}$ becomes 0.2 km s\per if the optical depth broadening is taken into account, and then $\sigma_\mathrm{NT}/c_\mathrm{s}$ (hereafter Mach number, $\mathcal{M}$) can be quite differently estimated to be 1.06 at 9.5 K or 0.86 at 14 K.

The right panels of Figure \ref{fig:gauss_velocity} show that the non-thermal motions traced by \sco\ and \mco\ are more likely subsonic ($\mathcal{M} \le 1$) in the filament and core regions of both L1544 and L6940-2.
Such subsonic characteristics of both \sco\ and \mco\ are also revealed in the distribution of non-thermal velocity dispersion along the filamentary structure as shown in Figure \ref{fig:velocity_dispersion}.
This suggests that the prestellar cores, as well as the surrounding filamentary clouds, seem to be in a quiescent state where the non-thermal gas motions are smaller than or equivalent to the sound speed of the gas. 
This may be because the turbulence, one of the possible origins for the non-thermal motions, might have dissipated during the formation processes of these quiescent structures \citep[e.g.,][]{Myers_1998_ApJ_507_L157}.

However, although non-thermal motions in the cores and filamentary clouds are within the subsonic regime, it is interesting to note that the amplitudes of the non-thermal motions seem to increase toward the central region of the cores, being almost transonic (Figure \ref{fig:velocity_dispersion}).
Since \sco\ only traces the outer region of the core, unlike high-density tracers, this increase in non-thermal motions at the core region may be related to the gas motion in the less dense region moving toward the core region such as the core-forming flow discussed in the previous section or the radial infall to be discussed in the next section.
A similar increment of velocity dispersion toward the core has been also found in the SDC13 cores, which is suggested to be purely generated by gravity \citep{Williams_2018_A&A_613_A11}.
We also see a gradual increase of non-thermal motions to the transonic regime at the northwest end of the L1544 main filament, region traced by the \mco\ line. 
This is probably because the region is highly affected with the gas flow motions as shown in Figure \ref{fig:gauss_velocity} and \ref{fig:velocity_profile}.

In addition to the northwest part of L1544, the \snth(\mco) is slightly increased outside the filament compared to the inner part.
Figure \ref{fig:snth-vs-h2cd} shows the \snth\ variation of \mco\ and \sco\ with respect to the \hmol\ column density to examine the reliability of this increase.
The \snth\ data points plotted in Figure \ref{fig:snth-vs-h2cd} are averaged values along the isocontour of \hmol\ column density.
Although the averaged \snth\ lies in the subsonic regime over the entire density range, there are slight increases of \snth(\mco) from the inside to the outside of the filament boundary. 
In particular, the steep increase ($\Delta\sigma_\mathrm{NT}(\mathrm{^{13}CO}) \sim 0.25 c_\mathrm{s} \approx 0.06$ km s\per) in L1544 is significant.

\citet{Pineda_2010_ApJL_712_L116} have also found a sharp increase of the velocity dispersion in the B5 region of Perseus from subsonic to transonic across the filament boundary using a single tracer, \nht\ (1, 1) emission. 
In our target, such variation in \snth(\mco) between the inner and outer regions of the filament is less prominent than that shown in B5 region.

Note that recently \citet{Choudhury_2021_A&A_648_A114} found from \nht\ (1, 1) and (2, 2) observations of L1688 that a less sharp transition to coherence with $T_\mathrm{kin} \lesssim T_\mathrm{dust}$ (\hso-based) in the dense core and $T_\mathrm{kin} \gtrsim T_\mathrm{dust}$ in the cloud.
Thus, if this is the case for L1544, then the thermal contribution might be overestimated inside and underestimated outside, and therefore such an increase in \snth(\mco) toward outside the filament can be even less than shown here.

\begin{figure*}
    \centering
    \gridline{\fig{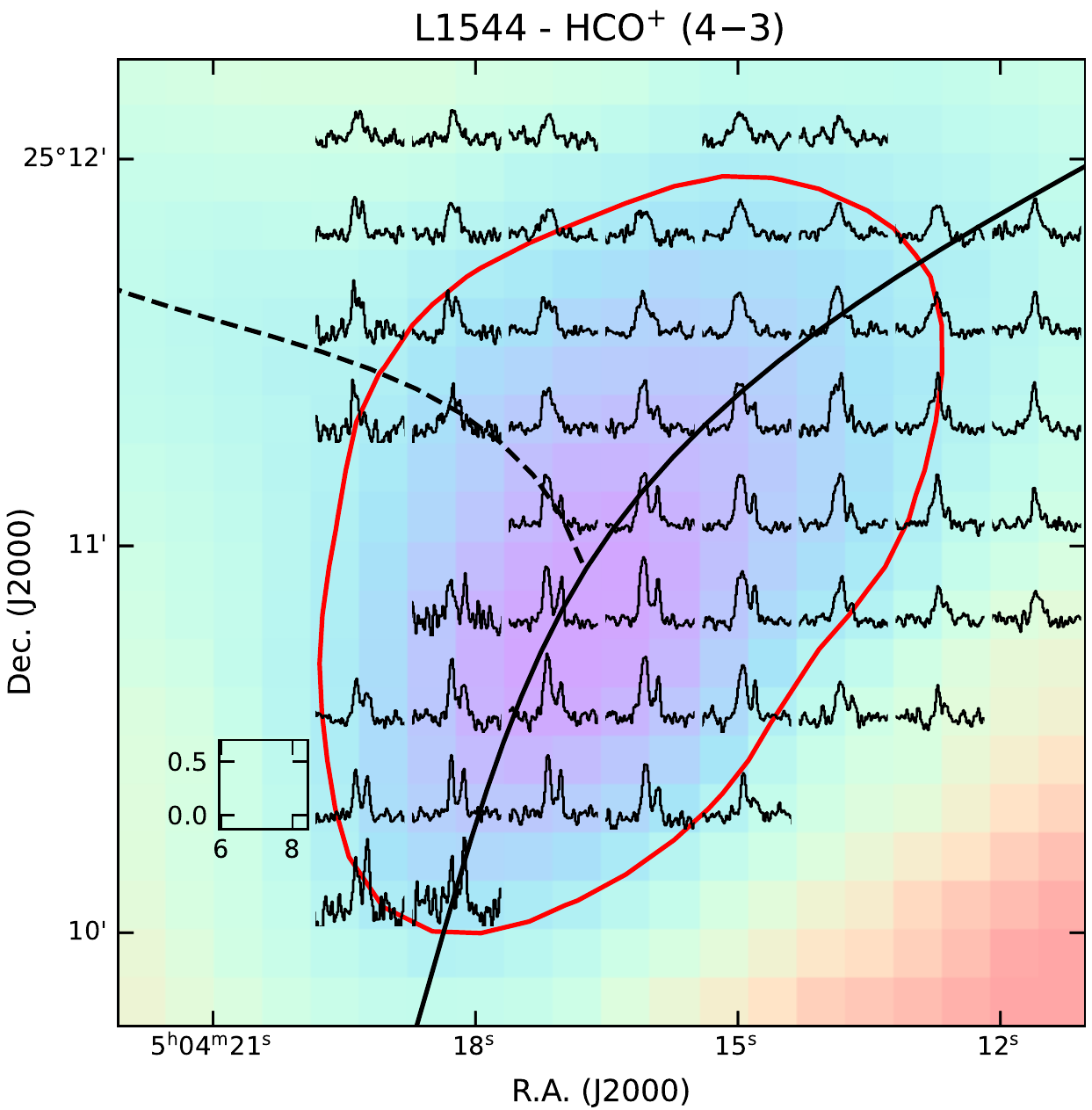}{3.39in}{}
              \fig{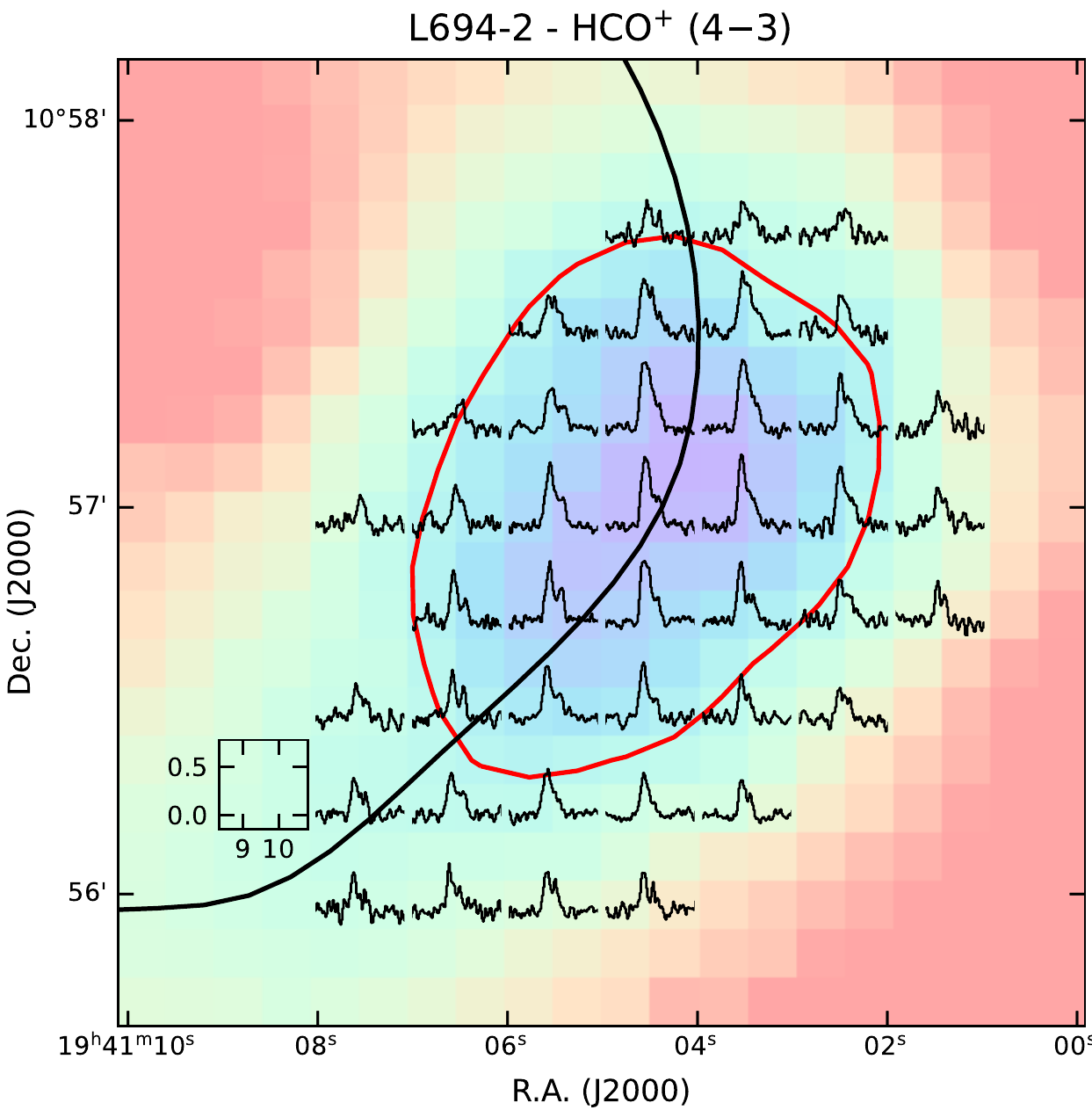}{3.39in}{}}
    
    \vspace{-0.5cm}
    \gridline{\fig{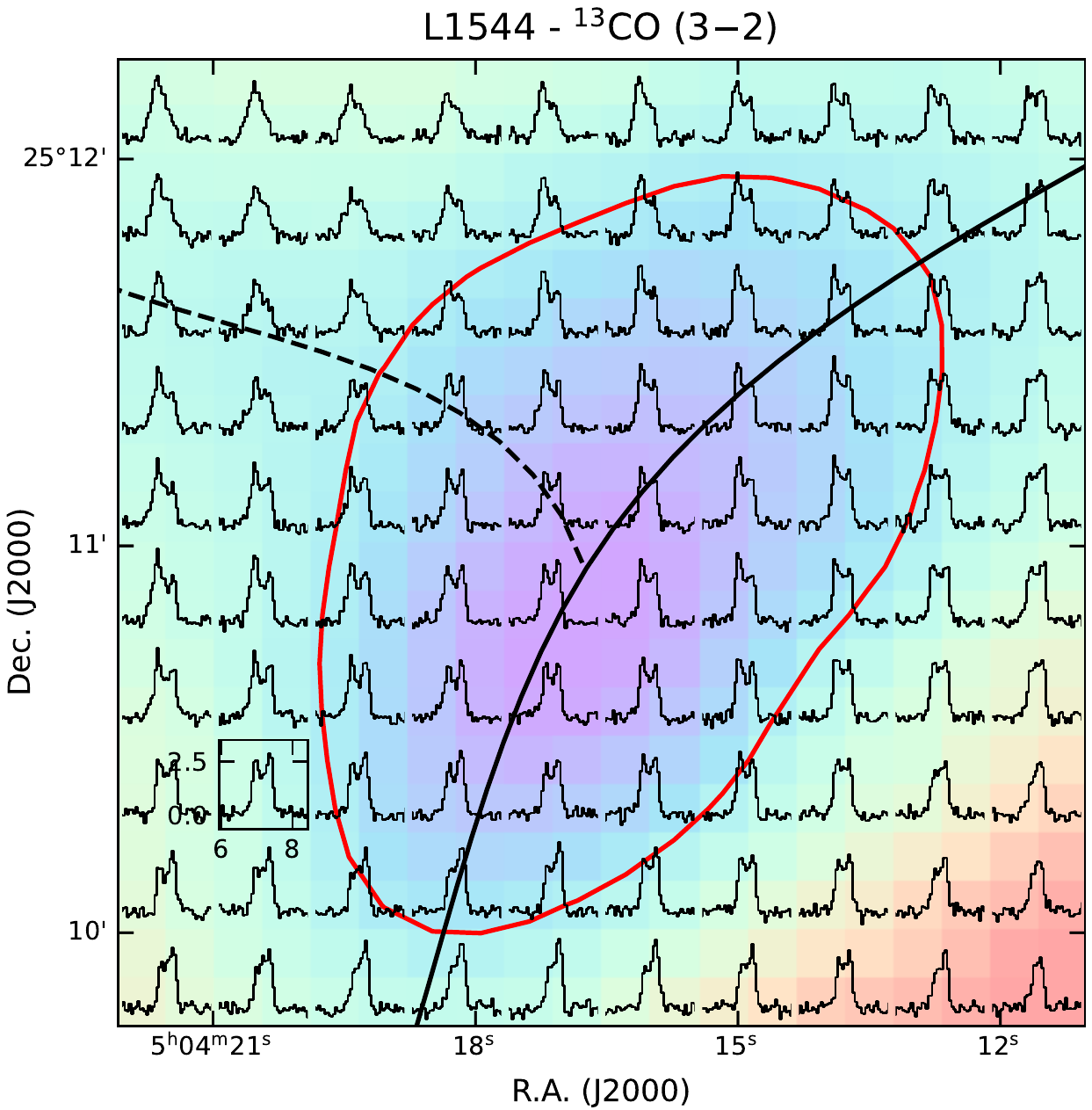}{3.39in}{}
              \fig{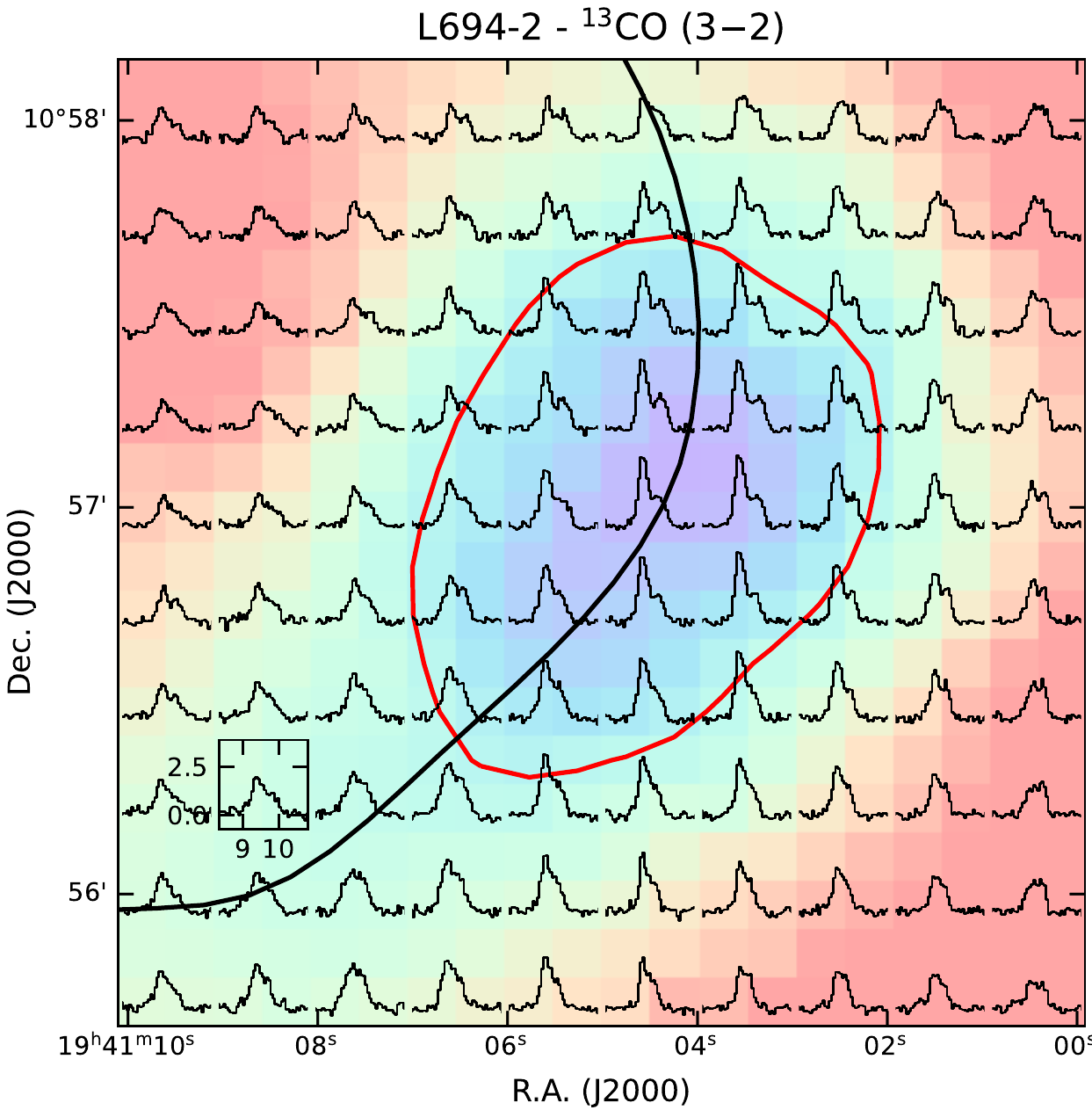}{3.39in}{}}
    
    \vspace{-0.5cm}
    \caption{
    Line profile maps of \hcopt\ and \mcot\ lines for L1544 and L694-2. 
    The background color indicates the \hmol\ column density distribution. 
    The red contour in each panel is to display the 40\% level of its peak value ($p_{40}$).
    The skeleton of filamentary structure is drawn in the black solid or dashed lines.
    }
    \label{fig:infall_profile_map}
\end{figure*}

\begin{figure}
    \centering
    \includegraphics[width=3.39in]{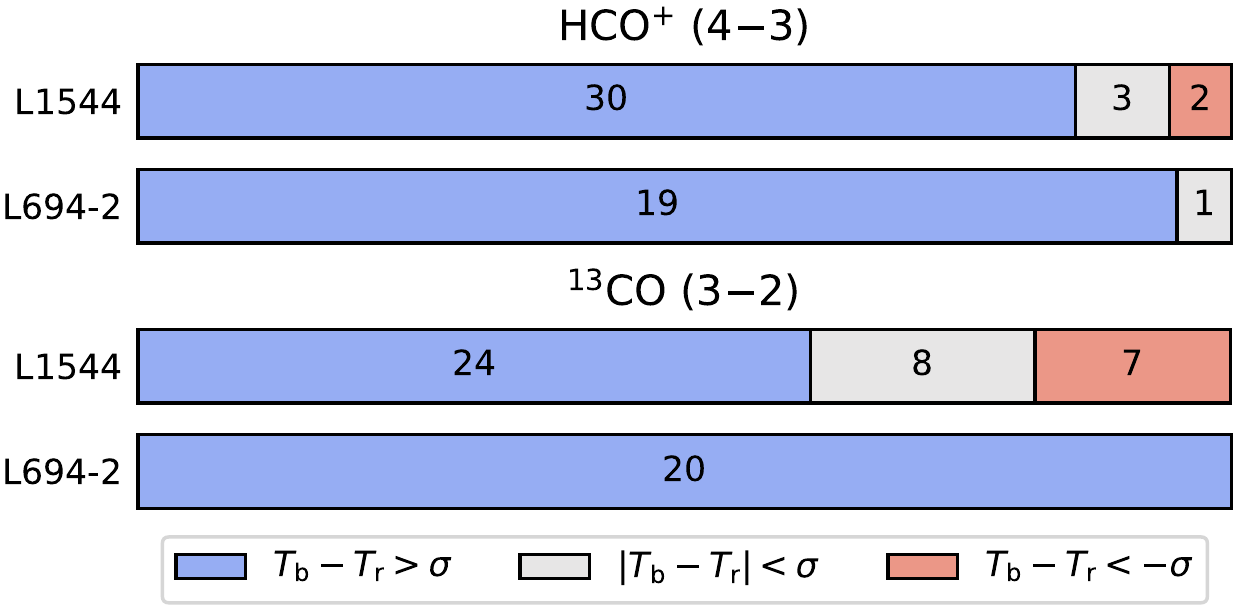}
    \caption{
    Number statistics of the blue and red profiles in \mco\ and \hcop\ lines within the 40\%  contour of the peak value of the \hmol\ column density  at the core center for two dense cores, L1544 and L694-2. 
    $T_\mathrm{b}$ and $T_\mathrm{r}$ are the brightness temperatures of the blue peak and the red peak (or shoulder) measured by double Gaussian fitting for each line profile, and $\sigma$ is the RMS noise level of each line profile.
    }
    \label{fig:blue_fraction}
\end{figure}

\begin{figure}
    \centering
    \includegraphics[width=3.39in]{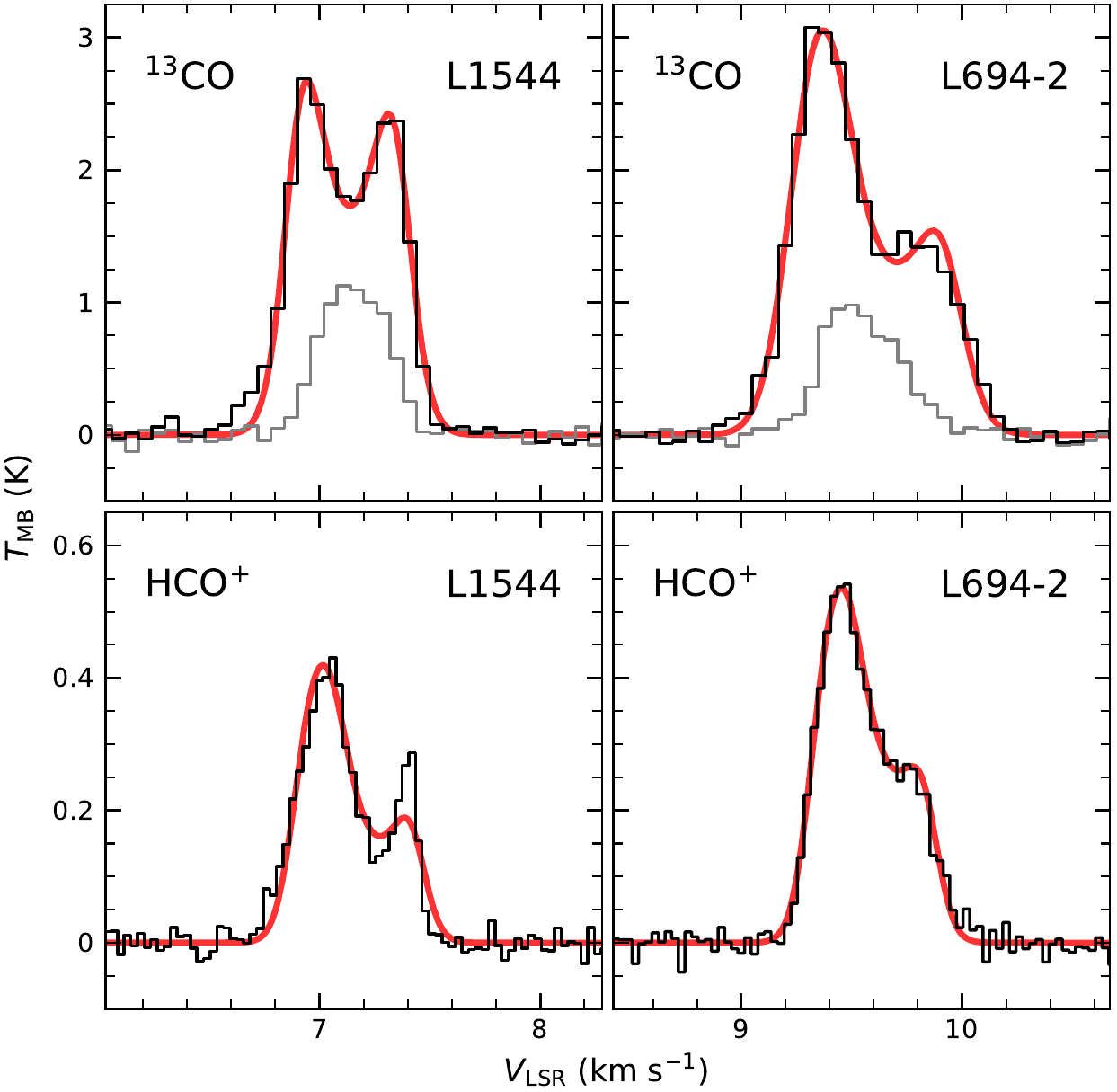}
    \caption{
    Infall asymmetric profiles of \mcot\ and \hcopt\ lines in L1544 (left panels) and L694-2 (right panels). 
    The line profiles given here in black (\mco\ and \hcop) and grey (\sco) were obtained by averaging the profiles within the 40\% level contour of the peak \hmol\ column density ($p_{40}$) of each dense core. 
    The red profiles are the best fitted ones using the Hill5 model \citep{DeVries_2005_ApJ_620_800} for each average profile.
    }
    \label{fig:infall_profiles_core}
\end{figure}

\begin{deluxetable*}{llCCCC}
    \tablecaption{Mass Accretion by Infall Motions in Core Region \label{tab:infl-core}}
    \tablewidth{0pt}
    \tablehead{
    \colhead{Target} & \colhead{Line} & \colhead{$V_\mathrm{in}$} & \colhead{$R_\mathrm{in}$} & \colhead{$M_\mathrm{in}$} & \colhead{$\dot{M}_\mathrm{in}$} \\
    \colhead{ } & \colhead{ } & \colhead{(km s\per)} & \colhead{(pc)} & \colhead{(\solm)} & \colhead{(\solm\ Myr\per)}
    }
    \decimalcolnumbers
    \startdata
    L1544    & \hcopt & 0.107\pm0.008 & 0.044 & 2.2\pm0.1 & 16\pm2 \\
    L694-2   & \hcopt & 0.113\pm0.015 & 0.038 & 1.4\pm0.1 & 13\pm3 \\
    L694-2   & \mcot  & 0.113\pm0.004 & 0.081\tablenotemark{a} & 1.4\pm0.1 & \phn6\pm1 \\
    \enddata
    \tablecomments{(1) Source name. (2) Transition. (3) Infall velocity derived by the Hill5 model fit. (4) Infall radius, $r_{40}$. (5) Mass calculated from the sum of \hmol\ column densities within the $p_{40}$ contour. (6) Mass accretion rate by infall motion (mass infall rate).}
    \tablenotetext{a}{The infall radius is assumed to be $r_{10}$ for \mco. See text for details.}
\end{deluxetable*}

\section{Infall Motions} \label{sec:infl}

In Section \ref{sec:vels-cen}, we found that a significant gas flow into the dense cores is in the form of oscillating motions along the filamentary envelopes.
Therefore it may be possible that such gas flow may affect the growth of the embedded cores. 
One way for checking this growth would be to look at the gas infalling motions around the dense cores.
In this section we examined the radially contracting motion towards the core or the filament by observing the infall signatures in the molecular lines, a feature of ``infall asymmetry'' in the self-absorbed spectra \citep[e.g.,][]{Lee_1999_ApJS_123_233, Lee_2001_ApJS_136_703}.

Figure \ref{fig:infall_profile_map} shows a host of infall asymmetries in the \hcopt\ and \mcot\ line profiles around the core region. 
We refer to these profiles showing infall asymmetry as ``blue profile'' in this paper.
The blue profile was identified by comparing the brightness temperatures of the blue peak ($T_\mathrm{b}$) and the red peak (or shoulder) ($T_\mathrm{r}$) measured by double Gaussian fitting for each line profiles.
Figure \ref{fig:blue_fraction} shows the statistics for the following three cases based on the RMS noise level ($\sigma$) of the line: (1) brighter blue peak than red peak ($T_\mathrm{b}-T_\mathrm{r}>\sigma$), (2) comparable blue and red peaks ($|T_\mathrm{b}-T_\mathrm{r}|<\sigma$), and (3) brighter red peak ($T_\mathrm{b}-T_\mathrm{r}<-\sigma$).
Most of the \hcopt\ and \mcot\ line profiles in L694-2 display blue asymmetry. 
In the case of L1544, 38\% of \mco\ and 14\% of \hcop\ spectra have shapes with comparable or brighter red peak, but the majority of the line profiles are classified as blue profiles.

\subsection{Infall Motions in Core Region} \label{sec:infl-core}

As mentioned above, both \mcot\ and \hcopt\ lines around the core region show strong self-absorption features, mostly seen as blue profile.
Considering that these two tracers have different critical densities by two orders of magnitudes, they likely trace different density regimes. 
Thus the dominance of the blue profiles in both \mco\ and \hcop\ lines around the core region may indicate that the gaseous material in both the surrounding envelope and the innermost region of the core is infalling in similar fashion onto the core center.
Note that this is consistent with recent findings from \citet{Ching_2022_Natur_601_49} for L1544. 
In particular, the agreement in infalling signatures in both molecular lines is remarkable in L694-2, implying that the core L694-2 is experiencing the contracting motions in its entire region.
On the other hand we note that the \mco\ line data in L1544 show some small portion of red asymmetric profiles (hereafter the red profile) indicating that there may be some other complicated motions in the low-density outer envelope of L1544 such as expansion, or oscillation, probably due to some possible inflowing motions along filaments.
Because of this complex pattern of the profiles in the envelope of L1544 core, we discuss infalling motions in the central region of L1544 core using \hcop\ lines only.

The line profiles of \hcop\ and \mco\ shown in Figure \ref{fig:infall_profiles_core} are the average of the profiles within the 40\% level ($p_{40}$) contour of the \hmol\ column density peak of each dense core (the red contours in Figure \ref{fig:infall_profile_map}).
We used the 40\% level contour in our analysis rather than a half-maximum contour generally used, as the 40\% level contour was found to be more useful in this study to include enough number of the blue profiles to cover the whole infalling central region and also to obtain an average profile with a sufficient $SNR$, which will enable us to make a better fit to the infall model. 
These average line profiles were used to analyze infall velocities and mass infall rates toward the cores.
The infall velocity of the average profile was determined by fitting the profiles with the Hill5 infall model \citep{DeVries_2005_ApJ_620_800}.
This Hill5 model is a simple radiative transfer model used for deriving infall speeds in a contracting molecular cloud by dealing with radiative transfer processes in two approaching layers whose excitation temperatures linearly increases toward the inner region.
Based on the error analysis for the Hill5 model, \citet{DeVries_2005_ApJ_620_800} suggested that a $SNR$ greater than $\sim 30$ is usually required to reliably determine the infall velocity from the asymmetric line profile shape.
In this study, we used a spectrum with a $SNR$ of at least 25 for fitting it with the Hill5 model.
The fitting was performed by Markov Chain Monte Carlo (MCMC) sampling using the \texttt{emcee}\footnote{https://github.com/dfm/emcee} python package \citep{Foreman-Mackey_2013_PASP_125_306} to ensure the reliability of the fitting result and accurately estimate the error.
The best-fit profiles are over-plotted on the averaged line profiles in Figure \ref{fig:infall_profiles_core}.
The resulting infall velocities from the fit are about 0.1 km s\per\ in all three cases as listed in Table \ref{tab:infl-core}.
Note that the infall velocity of 0.1 km s\per\ is very close to that found by \citet{Keto_2010_MNRAS_402_1625} in L1544, at a distance of about 1000 au, where the volume density is about $10^6$ cm\pcb.

These infall velocities can be used to derive the mass infall rate of gaseous material incoming toward the core center through the surface of the infall radius by an equation as follows \citep[e.g.,][]{Lee_2001_ApJS_136_703}:
\begin{equation}
    \label{eq:mass-infall-rate}
    \dot{M}_\mathrm{in} =4\pi R_\mathrm{in}^2 \rho V_\mathrm{in}\;, 
\end{equation}
where $R_\mathrm{in}$ is the radius of the region showing infall motion (hereafter, infall radius), $\rho$ is the density of the infalling shell, and $V_\mathrm{in}$ is the infall velocity.
In this approach, we assume that the infall motions are spherically symmetric toward the center of the cores.
Assuming the core has a spherical shape with uniform density, $\rho$, within the infall radius, its density is given by
\begin{equation}
     \rho = \frac{3M_\mathrm{in}}{4\pi R_\mathrm{in}^3}\;,
\end{equation} 
where ${M}_\mathrm{in}$ is the total mass in the region within $R_\mathrm{in}$, and thus, the mass infall rate can be expressed by 
\begin{equation}
     \dot{M}_\mathrm{in} = \frac{3M_\mathrm{in}}{R_\mathrm{in}} V_\mathrm{in}\;.
\end{equation}
The mass infall rates estimated from the \hcop\ line profile are $\sim16$ \solm\ Myr\per\ for L1544 and $\sim13$ \solm\ Myr\per\ for L694-2 (Table \ref{tab:infl-core}).
We have assumed $R_\mathrm{in}$ to be $r_{40}$, the radius of a circle embracing the same area as the area within the $p_{40}$ level contour. 
${M}_\mathrm{in}$ was obtained from the sum of $\rm H_2$ column densities inside the $p_{40}$ contour.
Since there is a flat density region in the prestellar core center to a radius of about 0.02--0.05 pc \citep{Kim_2020_ApJ_891_169}, this approximation is thought to be fairly reasonable with \hcop\ line data, but may not be so with \mco\ data which would trace the much less dense outer region.
The blue profiles in \mco\ are spatially more extended than the \hcop\ cases and are seen even down to the region of $\sim10$\% level of the peak column density value ($p_{10}$).
Thus, using $r_{10}$ as infall radius in the case of \mco\ data may be more reasonable.
In the outer envelope of L694-2, assuming $R_\mathrm{in}$ as $r_{10} \sim 0.08$ pc, the mass infall rate is estimated to be $6.0\pm0.8$ \solm\ Myr\per. 
This indicates that the mass infall rates inside (from \hcop\ data) are fairly larger than the mass infall rates outside (from \mco\ data).

Considering the velocity difference (or infall velocity of $\sim 0.1$ km s\per\ ) and the projected separation (or the infall radius of $\sim 0.04$ pc) between the opposite sides of the contracting cores inferred from \hcop\ data in both target cores, the velocity gradient of both core regions to the line of sight would be about 2.5 km s\per\ pc\per, being about a factor of 2--6 larger than the velocity gradient of the filaments at the core boundary (discussed in the Section \ref{sec:vels-cen}).
Moreover, the estimated mass infall rates of the contracting motion of both cores are a factor of 5--6 greater than the mass accretion rate of the axial flow motion around the core.
A core can grow initially by gaining mass from the axial flow motion along the filament. 
A sufficient accumulation of gaseous material around the core would lead to radial contraction of the dense core, helping the core to evolve. 
From the fact that L1544 and L694-2 are in a highly evolved state both chemically and kinematically \citep[e.g.,][]{Crapsi_2005_ApJ_619_379,Keto_2015_MNRAS_446_3731,Kim_2020_ApJ_891_169}, we can speculate that a significant amount of gaseous material has already been transported to these cores along the filament. 
In that sense, it is well explained why, in these targets, the amount of gas being accumulated by radial infall motions in the core region is presently more than from the axial flow motions along the filament at the core boundary.

\begin{figure}
    \centering
    \includegraphics[width=3.39in]{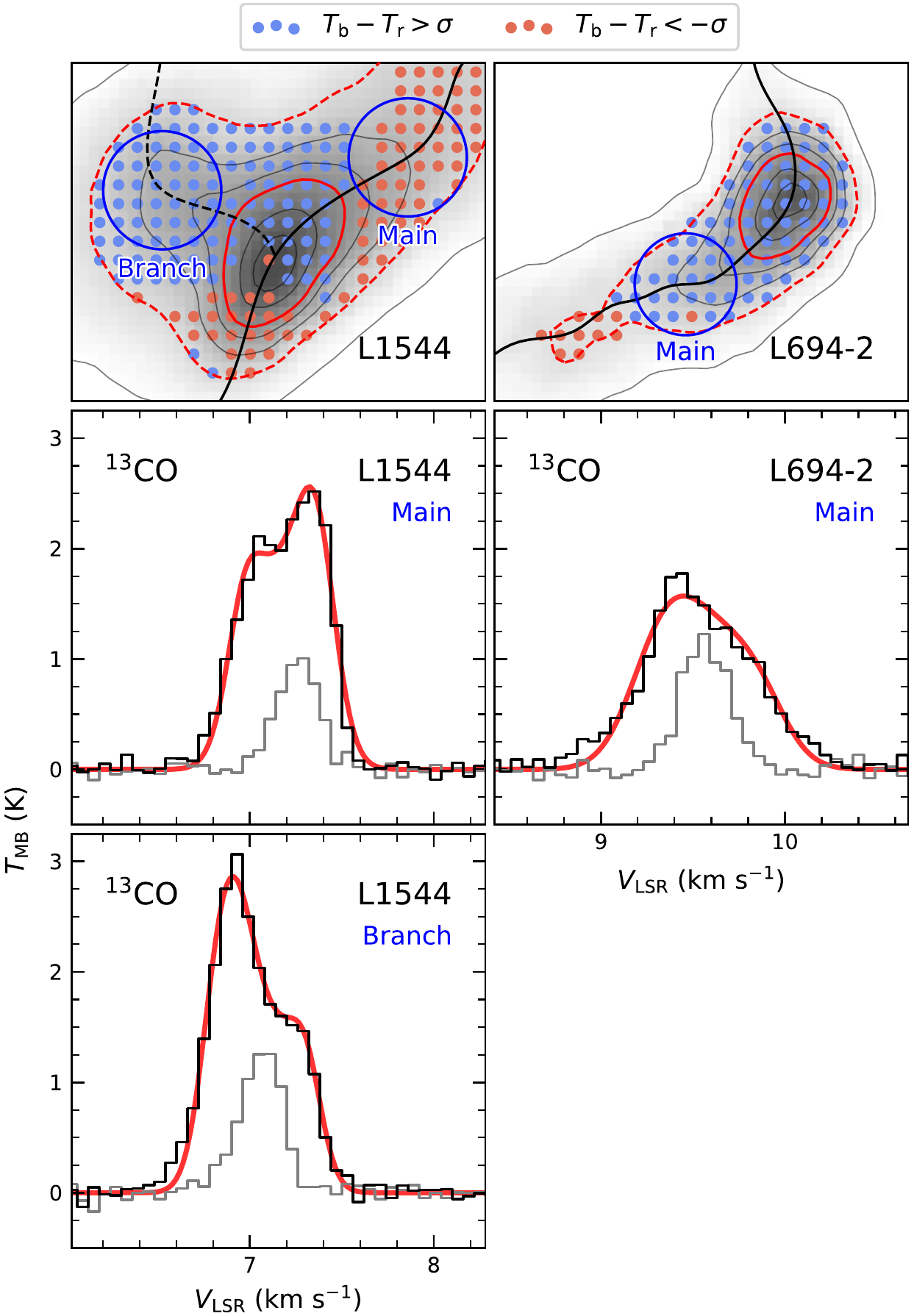}
    \caption{
    Line profiles of \mcot\ emission in low-density filament region for L1544 (left panels) and L694-2 (right panels).
    \textit{Top panels}: The blue and red dots indicate the pixel positions showing blue and red profiles of the \mcot\ line. 
    The background grey tone and the overlaid contours show the \hmol\ column density distributions. 
    The contours represent 80, 60, 40, 30, 20, and 10\% levels of the \hmol\ column peak density, $p_{40}$ and $p_{10}$ levels are highlighted with red solid and dashed contours (see Section \ref{sec:infl-core}). 
    The blue circles mark the regions for averaging the profiles to create the \mco\ line profiles in the lower panels. 
    \textit{Lower panels}: The black and grey line profiles are the averaged \mco\ (Figure \ref{fig:infall_profile_map_L1544_13CO} and \ref{fig:infall_profile_map_L6942_13CO}) and \sco\ spectra, respectively. 
    The red profiles show the best-fit results with the Hill5 model \citep{DeVries_2005_ApJ_620_800} for each averaged line profile.
    }
    \label{fig:infall_profiles_fil}
\end{figure}

\begin{deluxetable*}{llCCCCC}
    \tablecaption{Mass Accretion by Infall Motions in Filament Regions} \label{tab:infl-fil}
    \tablewidth{0pt}
    \tablehead{
        \colhead{Target} & \colhead{Fil.} & \colhead{$M_\mathrm{area}$} & \colhead{$\langle\sigma_\mathrm{tot}\rangle_\mathrm{area}$} & \colhead{$\alpha_\mathrm{vir}$} & \colhead{$V_\perp$} & \colhead{$\dot{M}_\perp$} \\
        \colhead{} & \colhead{} & \colhead{(\solm)} & \colhead{(km s\per)} & \colhead{} & \colhead{(km s\per)} & \colhead{(\solm\ Myr\per)}
    }
    \decimalcolnumbers
    \startdata
    L1544  & Main   & 0.73\pm0.03 & 0.22\pm0.05 & 1.9\pm0.8 & \cdot       & \cdot \\
           & Branch & 0.82\pm0.03 & 0.22\pm0.04 & 1.6\pm0.6 & 0.11\pm0.01 & 7\pm1 \\
    L694-2 & Main   & 0.64\pm0.04 & 0.22\pm0.03 & 2.1\pm0.6 & 0.13\pm0.01 & 6\pm1 \\
    \enddata
    \tablecomments{(1--2) Core and filament name. (3) Filament mass within the averaged area with radius of 0.04 pc for the line profiles in Figure \ref{fig:infall_profiles_fil}. (4) Mean total velocity dispersion from \sco\ observation. (5) Virial parameter, $\alpha_\mathrm{vir} = 3\sigma_\mathrm{tot}^2 R / G M$. (6) Infall velocity from the best fit for Hill5 model \citep{DeVries_2005_ApJ_620_800}. (7) Mass accretion rate onto the filament ($\tan \alpha \approx 45^\circ$).}
\end{deluxetable*}

\subsection{Infall motions in Filament Region} \label{sec:infl-fil}

The previous section examined the infall motions around the high-density core regions. 
We note that similar infall motions can be also seen in the filamentary regions (Figure \ref{fig:infall_profiles_fil}), analogous to the features reported in the low-density filamentary clouds by, for example, \citet{Kirk_2013_ApJ_766_115} and \citet{ Gong_2018_A&A_620_A62}.
In this case, the blue profiles of optically thick molecular lines were analyzed to infer the radial contraction towards the central axis of the filament.
Then, assuming that the filament is a simple linear cylinder \citep[see Figure 5 in][]{Kirk_2013_ApJ_766_115}, the mass accretion rate onto the filament in radial direction can be calculated by
\begin{equation}
    \dot{M}_\perp = \frac{2 V_\mathrm{in} M}{r \cos \alpha} \;,
\end{equation}
where $V_\mathrm{in}$ is the infall velocity, $M$ is the filament mass, $r$ is the half of the filament width, and $\alpha$ is the inclination to the plane of the sky.
The radial accretion rates onto the filament in the case of Serpens filaments were found to be quite larger than the longitudinal mass accretion rates along the filament by a factor of 4--7 \citep[]{Kirk_2013_ApJ_766_115, Gong_2018_A&A_620_A62}.

The \mco\ data in this study also show the blue profiles (extended to the $p_{10}$ level) indicative of the existence of inward motions in the filament region, not the core region.
Figure \ref{fig:infall_profiles_fil} shows the line profiles of \mco\ averaged over regions of the branch filaments of L1544 and the main filament of L694-2. They are seen as blue profiles.
Assuming that these blue profiles are due to the radial infall motions in the filament area in the same way as in \citet{Kirk_2013_ApJ_766_115}, the infall velocities obtained by the Hill5 model fit are 0.11--0.13 km s\per\ (corresponding to 0.15--0.18 km s\per\ for a filament with an inclination angle of $45^\circ$), and the radial mass accretion rates onto the filaments are estimated to be about 6--7 \solm\ Myr\per\ (Table \ref{tab:infl-fil}).
This is a factor of two greater than the axial mass accretion rates along the filaments that we calculated in section 4.2.

However, it is noted that we need to be careful in interpreting any asymmetric profiles from the periphery of the main filaments, especially if the filaments have large velocity gradient along them. 
This is because the observed profiles may have some contribution from the axial flow motions which could modify the shapes of the asymmetric profiles (more elaborated in Appendix \ref{sec:app-line-maps}). 
In fact, the line profiles in Figure \ref{fig:infall_profiles_fil} are averaged over a radius of 0.04 pc 
and thus the radial velocity difference due to the velocity gradient within the area used for the averaging is about 0.1--0.2 km s\per.
Therefore, it is possible that the blue profile shapes in the branch part in L1544 and main part in L694-2 could be significantly affected by the axial flows along the filaments. 
The red profile shown in the main filament part of L1544 can be explained in a similar context. 
The profile is probably highly affected by the red-shifted axial gas flow along the filament toward to the dense core region.

\section{Discussion} \label{sec:disc}

In this study, we found that the surrounding material associated with the two prestellar cores has a filamentary structure (Section \ref{sec:dens}), the gaseous material is being accreted toward the cores along the quiescent filaments (Section \ref{sec:vels}), and the inner and outer envelopes are radially collapsing toward the core center (Section \ref{sec:infl}).
The physical and kinematic characteristics of the filaments identified in L1544 and L694-2 are found to be consistent with those seen in filaments with embedded cores in the low-mass star-forming region \citep{Hacar_2011_A&A_533_A34}.
In particular, they have physical properties significantly similar to those of the small-scale, velocity-coherent filamentary structures, also called `fibers,' found by the decomposing techniques in turbulent large-scale filaments \citep[e.g.,][]{Hacar_2013_A&A_554_A55,Chung_2019_ApJ_877_114}.

\subsection{Usefulness of Our Targets as Dense Cores Embedded in Isolated and Velocity-coherent Filaments}

Decomposing velocity coherent filaments in a filamentary cloud seen in continuum emission, which often contains multiple velocity components, is highly tedious and also complex to do. 
Complication in the decomposition has something to do with `multi-Gaussian decomposing' and `friends-of-friends-in-velocity' method \citep{Hacar_2013_A&A_554_A55,Chung_2021_ApJ_919_3} or the Dendrogram technique \citep{Chung_2019_ApJ_877_114} which usually suffers from the ambiguities and uncertainties in getting the unique solution of decomposed components in the processes \citep[e.g.,][]{ZamoraAviles_2017_MNRAS_472_647}.
Therefore, in the isolated filaments and embedded dense cores, a spatial variation of their velocity structures can be easily identified without any help of such challenging decomposing techniques.
In that respect, L1544 and L694-2 could be suitable cases to study the role played by filamentary structures in the process of dense core formation.

\subsection{Origin of Subsonic Characteristic} \label{sec:disc-subs}

Prestellar/starless cores, including L1544 and L694-2, are generally subsonic \citep[][based on \nthp\ (1--0) observations]{Lee_1999_ApJ_526_788}, and the velocity-coherent filaments around our target cores are also mostly subsonic (see Section \ref{sec:vels-disp}).
The quiescent internal motions in these filamentary clouds with coherent velocities are somewhat different from those seen in many turbulent filaments where the dense cores are forming \citep[e.g.,][]{Chung_2019_ApJ_877_114,Chung_2021_ApJ_919_3}.
In the case of L1544 and L694-2, the subsonic characteristics of the dense cores seems to be inherited from the parent filaments.
One explanation for this could be that at the stage of the formation of the filament, the gas flows collide and their turbulent motions are dissipated to form a quiescent filament at the earlier stage where individual dense core forms \citep{Andre_2014_prpl.conf_27, Tafalla_2015_A&A_574_A104}.

\subsection{Core-forming Flow Motion along the filament} \label{sec:disc-cffm}

The existence of the core-forming flow motion along the filament axis for our targets has been inferred from the $\lambda/4$ shift in the periodic variation of density and velocity distributions along the filamentary clouds.
Compared to the clear variation in the density distribution, the sinusoidal oscillations of the velocity field were relatively less pronounced in the previously reported cases \citep{Hacar_2011_A&A_533_A34,Liu_2019_MNRAS_487_1259} and also in the case of L694-2 in our study.
However, the velocity oscillation found in the 0.5 pc-long main filament of L1544 was distinctly identified with double repetition of the red-shift and the blue-shift components (see Figure \ref{fig:velocity_profile}).
In the first wavelength of the sinusoidal variation of the velocity distribution it seems clear that there is a converging flow motion which can accumulate the gaseous material to form a central core. 
The formation of the dense core probably would produce more gravitational pull and then lead to more longitudinal contraction of the filamentary cloud to make the core to evolve to a more condensed stage.
On the other hand, it is noted that there is no corresponding density peak in the second wavelength of the velocity distribution in the northwest.

This is probably because the 1.3 \solm\ Myr\per\ mass accretion rate for this second part of the sinusoidal velocity pattern is significantly smaller than the 6.0 \solm\ Myr\per\ for the first part from the contributions of both main and branch filaments, and it may not be enough to form one more core at the second part of the sinusoidal in L1544. 
Other possible reason for the non-existence of the secondary density peak would be that there is not enough material to form the other core at the edge of the filamentary cloud.

Such periodic velocity variation and gas motion parallel to the axis in the uniform density cylinder due to gravitational instability has been predicted in theoretical studies \citep[e.g.,][]{Fiege_2000_MNRAS_311_85}.
\citet{Gritschneder_2017_ApJ_834_202} showed that the initial geometrical bend of a filamentary cloud could initiate the core formation and produce the sinusoidal velocity oscillation feature like the first wavelength case of the velocity distribution in L1544.
The initial geometrical perturbation can lead to a long-lived, stable oscillation until that filament becomes gravitationally unstable via a density enhancement.

\subsection{Radially Contracting Core}

In this study, we measured infall velocities of $\sim 0.11$ km s\per\ in the inner region of both L1544 and L694-2 cores from \hcopt\ observations, which are consistent with those measured from \nthp\ (1--0), \dcop\ (2--1) and (3--2), \hcop\ (3--2), o-H$_2$O (1$_{10}$--1$_{01}$), or o-\nht\ (1$_{0}$--0$_{0}$) observations in previous studies \citep{Williams_1999_ApJ_513_L61,Williams_2006_ApJ_636_952,Caselli_2012_ApJL_759_L37,Keown_2016_ApJ_833_97,Caselli_2017_A&A_603_L1}.
This suggests that, considering the high critical density of \hcopt\ line, the gas is contracting even in the innermost, very dense regions ($n_\mathrm{H_2}\gtrsim 10^6$ cm\pcb) of these cores.

For the outer regions of cores, the infall velocity is measured from the \mcot\ observations as similarly $\sim 0.11$ km s\per\ in L694-2. 
On the other hand, in case of L1544, the infall asymmetry (blue profile) is not dominant and thus the infall velocity is not well measurable in the outer region of the core, although there have been claims that there are infalling motions in L1544 cloud scales \citep[e.g.,][]{Tafalla_1998_ApJ_504_900,Ching_2022_Natur_601_49}.
This difference in the infall velocities of the outer region between L1544 and L694-2 cores is also consistent with the results obtained from CS (2--1) and (3--2) observations in \citet{Lee_2001_ApJS_136_703,Lee_2004_ApJS_153_523} showing the infall velocity of 0.02--0.03 km s\per\ for L1544 and 0.07--0.09 km s\per\ for L694-2.
The velocity gradient indicative of the core-forming flow around the core is clearer in L1544 than L694-2. 
In contrast, the infalling motions in both inner and outer regions of the core are more clearly identified in L694-2. The outer region of the L1544 core seems to have more complex motions, probably due to the core-forming flow motion around the L1544 core as discussed in Section \ref{sec:disc-cffm}.
For example, \citet{Lin_2022_arXiv_2205_09806} have found from the observations with CH$_3$OH lines a feature implying an asymmetric accretion of material at the dust peak of L1544.
Still, more observational studies are needed to determine the relationship between core-forming flow and contracting motion.

\subsection{Possible Formation Mechanisms of Dense Cores and Filaments}

Filament fragmentation by gravitational instability \citep[e.g.,][]{Stodokiewicz_1963_AcA_13_30,Inutsuka_1997_ApJ_480_681} is one of the essential core formation mechanisms well supported by observations, such as the existence of the filaments in gravitational equilibrium and the chain of dense cores along the filaments with quasi-equal spacing \citep[e.g.,][]{Tafalla_2015_A&A_574_A104}. 
The sinusoidal velocity variation along the filaments found in this study can also be a feature that presents the core formation by filament fragmentation.
In the final step of ``fray and fragment'' scenario \citep{Tafalla_2015_A&A_574_A104}, dense cores are formed out of velocity-coherent fibers by gravitational instability, and in the resulting core chain, the inter-core distance is estimated to be approximately 0.2--0.3 pc.
In the two clouds studied here, the filaments harbor only one core each, but the wavelength of velocity oscillation is about 0.2--0.3 pc (see Figure \ref{fig:velocity_profile}).
In our case, the initial filament mass seems not sufficient to make a series of multiple cores, but the velocity oscillation pattern is possibly a leftover signature of such fragmentation due to gravitational instability.

An alternative core-formation scenario is the shock-induced fragmentation, where turbulent flows collide to make shock-induced filamentary structures which in turn form dense cores from their fragmentation \citep{Klessen_2005_ApJ_620_786}.
This scenario predicts a sharp physical transition of gaseous material over the filament, for example, a velocity variation and an increase in the velocity dispersion across the boundary of the filament as shown by \citet{Pineda_2010_ApJL_712_L116}. 

We examined the \mco\ velocity centroid distribution for two target clouds in Figure \ref{fig:gauss_velocity}, finding a large-scale velocity difference across both target clouds (see Section \ref{sec:vels}).
In L1544, we also found a distinct increase of about 0.06 km s\per\ in \snth\ of \mco\ across the filament boundary (see Figure \ref{fig:snth-vs-h2cd}). 
On the other hand, it is noted that the \snth\ increase across the filament in case of L694-2 is not as significant as L1544 considering its uncertainty and also \snth\ lies in the subsonic regime over the whole filament region in both target cases as discussed in Section \ref{sec:vels-disp}. 
As discussed in Section \ref{sec:disc-subs}, such reduced \snth\ might be natural in these highly evolved systems in that turbulence dissipation can be sufficiently achieved.
Therefore, our results seem to neither strongly support nor rule out core formation by shock-induced fragmentation.

In summary, the filamentary structure seen in L1544 cloud might be the result of shock compression from the variations of the centroid velocity and the velocity dispersion across the filamentary cloud, and then the filament might contract gravitationally along the filament to form the cores. 
In case of L694-2, there is no clear evidence showing that the shock-induced compression has occurred. 
But it is likely that the filamentary structure in the L1694-2 envelope cloud has experienced gravitational fragmentation to form its dense core L694-2.

\section{Conclusion} \label{sec:conc}

We performed mapping observations in \scot, \mcot, \lcot, \hcopt, and \htcopt\ emission lines with the JCMT radio telescope toward L1544 and L694-2 cores and their surrounding filamentary envelopes.
By analyzing our molecular line observations and the \hso\ continuum data, we have investigated how the mass accretes onto the prestellar cores embedded in the pc-scale filamentary cloud in order to understand how the filamentary structures in the clouds play a critical role in the formation of the prestellar cores. 
\hmol\ column density and dust temperature were measured by the SED fitting procedure of the \hso\ dust continuum data.
Based on Gaussian fits to the \sco\ profiles (and also \mco\ profiles as an alternative method towards the low-density region where \sco\ is undetected), the velocity structure all along the skeleton of each filament was examined.
The main results of this study are summarized below:

\begin{enumerate}
    
    \item The \scot\ line is found to be an optimal tracer of the kinematics of the filamentary structure around the prestellar core.
    The \mcot\ line was detected over the entire filamentary cloud until the \hmol\ column density goes down to $2\times 10^{20}$ cm\psq, and its line profiles in the high-density region show a feature of strong self-absorption.
    The \lcot\ line seems to be a poor tracer of kinematics in the filaments and the cores due to its very different distribution from those of \mco\ and \sco\ as well as its saturated features in the spectra due to its high optical depth.
    The infall asymmetries of \hcopt\ line were significantly detected in the high-density core region.
    
    \item To obtain a better estimate of the physical parameters of L1544 and L694-2, we determined more reliable distances for these two targets by identifying the distances where an abrupt jump in the stellar extinction occurs due to existence of the clouds and cores using the \gaia\ DR2 parallax data and Pan-STARRS1 stellar photometry for M-dwarfs.
    The newly determined distances of L1544 and L694-2 are $175_{-3}^{+4}$ pc and $203_{-7}^{+6}$ pc, respectively.
    
    \item We identified $\sim0.5$ pc-long filamentary structures that harbour the cores in both L1544 and L694-2.
    L1544 has another small filament branched out from the core.
    The identified filaments in both L1544 and L694-2 have physical properties comparable with the individual filamentary structures (so-called ``fibers'') composing the large-scale filaments as complex bundles.
    We found that the prestellar cores as well as the surrounding filamentary clouds are in a quiescent state in the sense that their non-thermal velocity dispersion is smaller than or equivalent to the sound speed of the gas. 
    This may imply that the turbulence has already been dissipated during the forming processes of these quiescent structures.
    
    \item The filaments in our two targets are velocity-coherent on a (sub-)parsec scale, with velocity variations within a range of 0.3 km s\per\ in the dense core size scale.
    The sinusoidal velocity oscillations with a $\lambda/4$ shift with respect to the oscillations of the column density distribution are seen in L1544 and L694-2, possibly indicating the core-forming flow motion along the axis of the filament.
    The mass accretion rates of these longitudinal flow motions are estimated as 2--3 \solm\ Myr\per\ with an assumption of the filament inclination of $45^\circ$, being comparable to the mass accretion rate for Serpens cloud, but much smaller than those of the Hub filaments, cluster forming filaments, and high-mass star forming IRDCs by 1 or 2 orders of magnitude.

    \item Both \hcopt\ and \mcot\ lines in the high-density core regions show strong self-absorption features, mostly seen as blue profiles, indicating that the gaseous material in both the surrounding envelope and the inner region of cores is infalling onto the core center.
    The infall velocities are measured to be about 0.1 km s\per\ by the Hill5 model fit for both line profiles with the MCMC sampling technique.
    The mass infall rates are estimated as $\sim16$ \solm\ Myr\per\ for L1544 and $\sim13$ \solm\ Myr\per\ for L694-2 in the inner region of cores by \hcop\ spectra, and $\sim6$ \solm\ Myr\per\ in the outer envelope of L694-2 by \mco\ spectra. 
    This may imply that the mass infall rates are higher inside than outside.
    
    \item The \mcot\ lines show self-absorption features even in the filament region. 
    In the main filament of L694-2 and in the branch filament of L1544, the averaged \mco\ lines show typical infall asymmetric features.
    Assuming these are due to the radial contracting of filaments, the radial mass accretion rates onto the filaments are estimated to about 6--7 \solm\ Myr\per.
    However, as the core-forming flow motion parallel to the filament axis can also contribute on the shapes of such asymmetric line profiles, interpreting the asymmetric profiles in the filaments where there exist gas flow motions along the filaments requires caution.
    
    \item The filamentary clouds harboring two prestellar cores with coherent velocity structures are found to be an important laboratory to study the role of the filamentary structure of the clouds in the formation of dense cores.
    From this study we suggest that the filaments in our targets could be intermediate between turbulent cloud and quiescent core that probably fromed the shock-compression of colliding clouds, the cores could be then formed by gravitational fragmentation of the filaments, and the continuing mass accretion through the filaments might lead the cores to evolve to the prestellar dense core stage.
    We conclude that the filamentary structures in two clouds, L1544 and L694-2 consisting of dense cores, are likely playing a key role in the whole process of core formation and evolution.
    
\end{enumerate}

\begin{acknowledgments}
This work was supported by the Basic Science Research Program through the National Research Foundation of Korea (NRF) funded by the Ministry of Education, Science and Technology (NRF- 2019R1A2C1010851), and by the Korea Astronomy and Space Science Institute grant funded by the Korea government (MSIT; Project No. 2022-1-840-05).
S.K. is supported by National Research Council of Science \& Technology (NST) -- Korea Astronomy and Space Science Institute (KASI) Postdoctoral Research Fellowship for Young Scientists at KASI in South Korea.
S.K. is particularly thankful for the help from S.-S. Lee, J. Kim, K.-T. Kim, W. Kwon, and M.-Y. Lee during the dissertation evaluation.
M.T. acknowledges support from project PID2019-108765GB-I00 financed by MCIN/AEI/10.13039/501100011033.
E.J.C. is supported by Basic Science Research Program through the National Research Foundation of Korea (NRF) funded by the Ministry of Education (NRF-2022R1I1A1A01053862). 
The James Clerk Maxwell Telescope is operated by the East Asian Observatory on behalf of The National Astronomical Observatory of Japan; Academia Sinica Institute of Astronomy and Astrophysics; the Korea Astronomy and Space Science Institute; the National Astronomical Research Institute of Thailand; Center for Astronomical Mega-Science (as well as the National Key R\&D Program of China with No. 2017YFA0402700). Additional funding support is provided by the Science and Technology Facilities Council of the United Kingdom and participating universities and organizations in the United Kingdom and Canada.
The authors wish to recognize and acknowledge the very significant cultural role and reverence that the summit of Maunakea has always had within the indigenous Hawaiian community. 
We are most fortunate to have the opportunity to conduct observations from this mountain.
\end{acknowledgments}

\vspace{5mm}
\facilities{JCMT (HARP), Herschel (PACS, SPIRE)}
\software{Starlink \citep{Currie_2014_ASPC_485_391}, Astropy \citep{Astropy_2013_A&A_558_A33,Astropy_2018_AJ_156_123}, emcee \citep{Foreman-Mackey_2013_PASP_125_306}, FilFinder \citep{Koch_2015_MNRAS_452_3435}}

\appendix

\section{Estimation of \hmol\ column density and dust temperature} \label{sec:app-sed}

\begin{figure}
    \centering
    \includegraphics[width=3.39in]{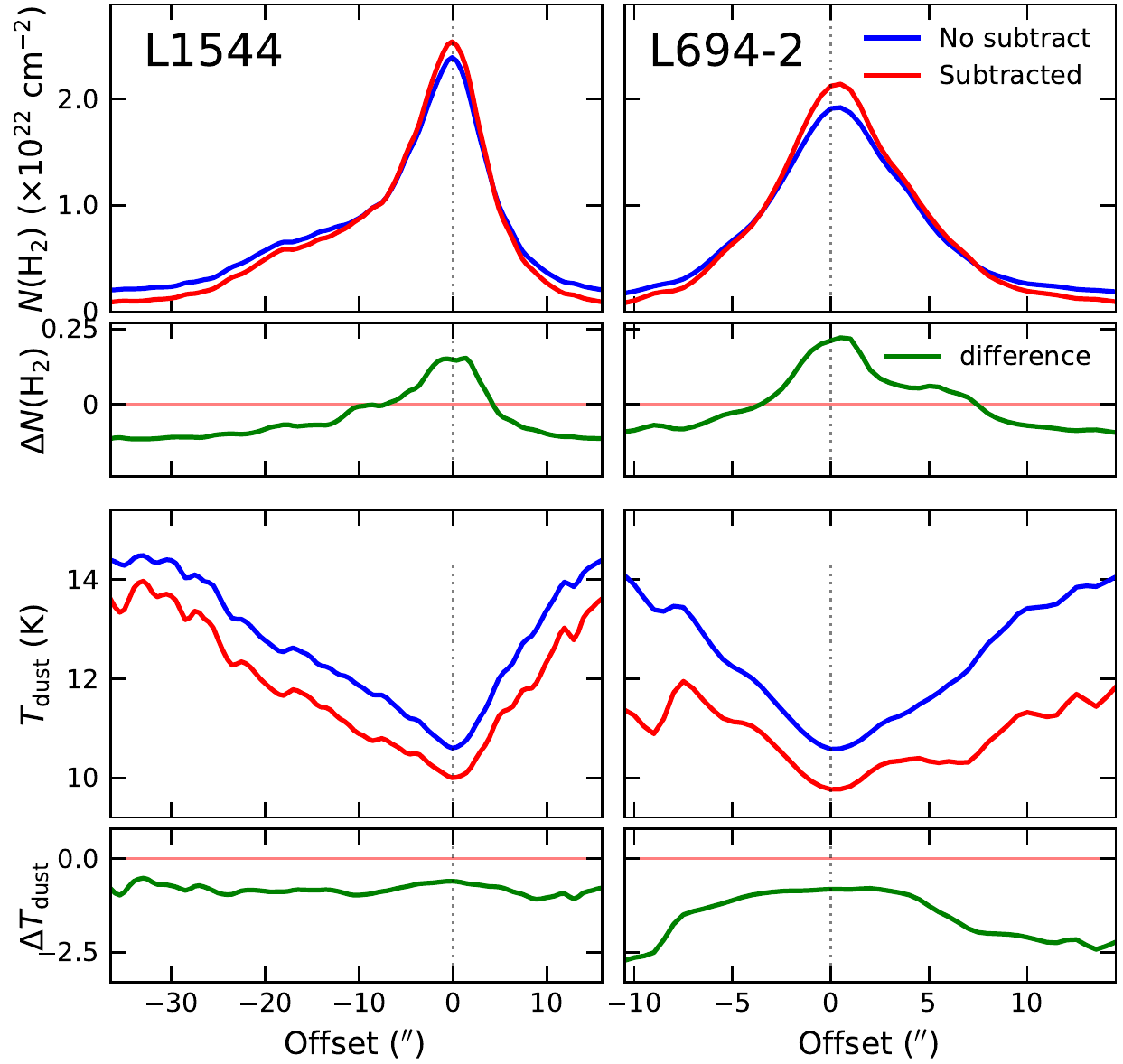}
    \caption{
    Effect of subtracting background emissions on L1544 (left panels) and L694-2 (right panels). 
    The upper panels show the comparison for the \hmol\ column density profile along the cut in Figure \ref{fig:sed_result_h2cd_temp} between before (blue) and after (red) subtraction and their difference (green). 
    The lower panels are the same, but for the dust temperature.
    }
    \label{fig:sed_result_subtract_effect}
\end{figure}

The \hmol\ column density and dust temperature were derived based on the SED fit using the \hso\ dust continuum images from 160 (250 for L694-2) to 500 \micron.
The observed dust emission can be expressed by
\begin{equation}
    I_\nu = B_\nu \left( T_\mathrm{d} \right) \left( 1-\mathrm{e}^{-\tau_\nu} \right) \;, 
\end{equation}
where $B_\nu (T)$ is the Planck function, $T_\mathrm{d}$ is the dust temperature, and ${\tau_{\nu}}$ is an optical depth at the observing frequency ${\nu}$, given as ${\tau_{\nu} = \mu_\mathrm{H_2} m_\mathrm{H} \kappa_\nu N_\mathrm{H_2}}$.
Here, $\mu_\mathrm{H_2}$ is the mean molecular weight per hydrogen molecule \citep[2.8, ][]{Kauffmann_2008_A&A_487_993}, $m_\mathrm{H}$ is the hydrogen atom mass, $\kappa_\nu$ is the dust opacity referred from OH5 \citep{Ossenkopf_1994_A&A_291_943}, and $N_\mathrm{H_2}$ is the \hmol\ column density.

The continuum image contains all of the emissions from the cloud components lying along the line of sight.
Thus, the peak of the SED curve of the dense core could be shifted toward a shorter wavelength due to the fact that the low-density molecular gas in the large-scale molecular cloud has a relatively high temperature.
Therefore, the background (or foreground, whatever it is) emission from such diffused gas should be subtracted before.
Assuming that the background level is the mean intensity of pixels for which \mco\ was not detected, we measured the background levels as 23, 54, 34, and 15 MJy sr\per\ for the \hso\ 160, 250, 350, and 500 \micron\ images, respectively.
The determined background levels are similar to the dust emission at the \hmol\ column density of $1.1\times 10^{21}$ cm\psq\ with dust temperature of 15 K.
After subtracting the background, we also masked some pixels that have values lower than the RMS noise level.
The masking level is 1.8, 2.6, 1.9, and 0.9 MJy sr\per\ for the 160, 250, 350, and 500 \micron\ images.

Figure \ref{fig:sed_result_subtract_effect} shows the difference in SED fit results between before and after background subtraction. 
Dust temperature was reduced by 1--2 K overall.
Based on the boundary where the \hmol\ column density is approximately $10^{22}$ cm\psq, the \hmol\ column density decreased as some emission was removed in the low-density region but inversely increased in the high-density region due to the reduced temperature.
As mentioned in Section \ref{sec:dens}, for the rest of the fitting procedure not described here, see \citet{Kim_2020_ApJ_891_169}.

In the case of L694-2, we used only SPIRE 250, 350, and 500 \micron\ data in the SED fit, because there is no PACS 160 \micron\ data in the \hso\ science archive. 
However, we found that the differences between the two SED fit results for L1544 with and without 160 \micron\ were quite negligible. 
Therefore, the absence of 160 \micron\ data for L694-2 did not affect this SED fit result.

\section{Distance Determination Based on \gaia\ DR2} \label{sec:app-distance}

\begin{figure*}
    \centering
    \gridline{\fig{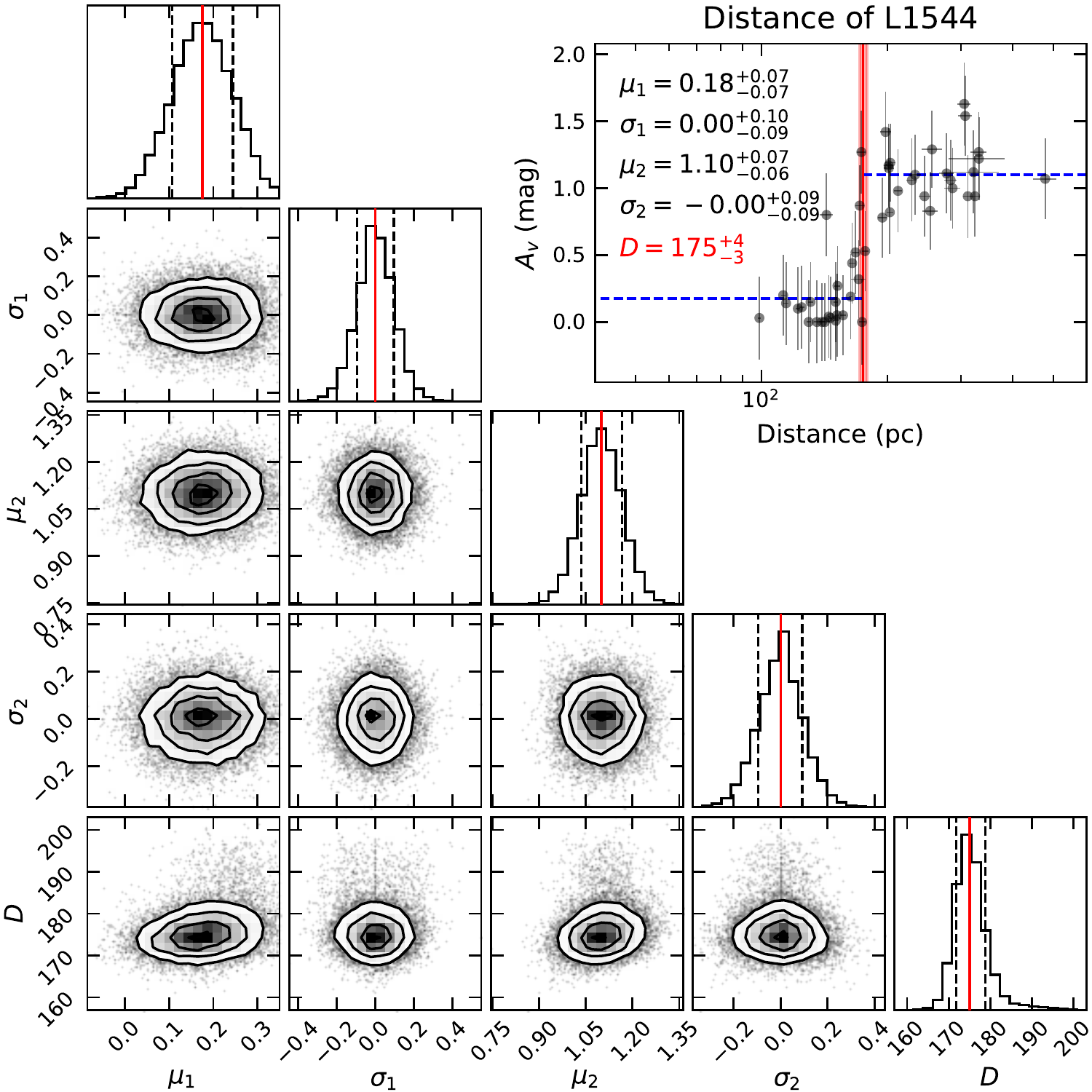}{3.39in}{}
              \fig{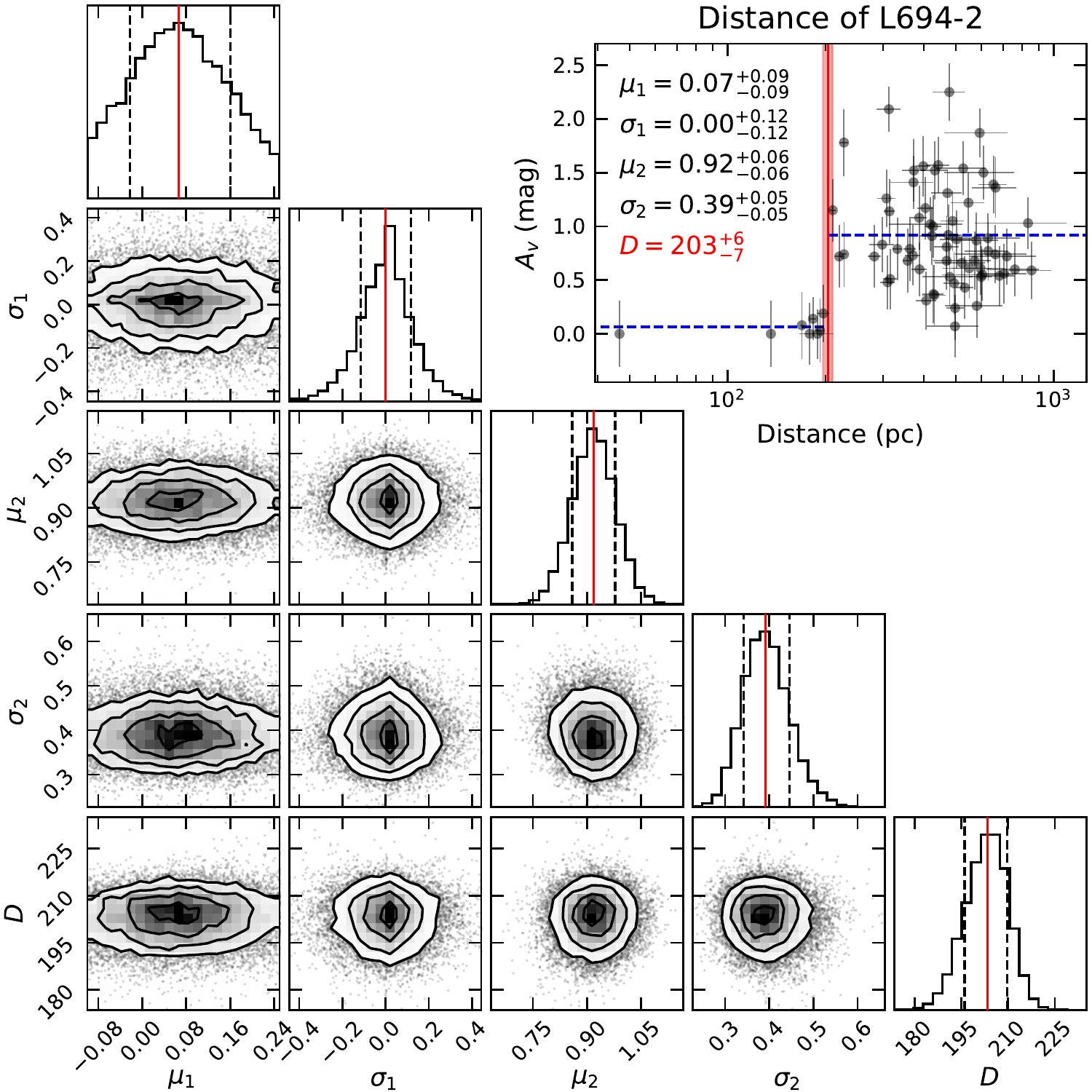}{3.39in}{}}
    
    \vspace{-5mm}
    \caption{
    Determination of distances of L1544 (left panels) and L694-2 (right panels). 
    The corner plots show the posterior probability distributions of parameters. 
    The red lines show the median values of the parameters, and the dashed lines indicates 16 and 84 percentiles of the parameters. 
    The best fit model and parameters show in the upper right plot of each target (the units of $\mu$ and $\sigma$ are `mag' and the unit of $D$ is `pc'). 
    The blue dashed lines show the extinction of foreground and background stars, and the red line and shaded area show the cloud distance and its uncertainty range.
    }
    \label{fig:dist_corner_plot}
\end{figure*}

We determined the distance of L1544 and L694-2 based on stellar photometry and \gaia\ DR2 parallax for the M-type dwarfs.
The procedures are roughly split into two parts: (1) making data set consisting of stellar optical extinction ($A_\mathrm{v}$) and parallax distance, and (2) determining cloud distance by identifying where a sharp increase in extinction occurs.

The first part of the procedure is referred to Maheswar et al. (2022, in preparation).
To separate the M-type dwarfs in the Pan-STARRS catalog \citep{Chambers_2016_arXiv161205560} into ON and OFF positions for the target cloud, we first set the cloud boundary to 22 MJy sr\per\ contour of \plck\ 857 GHz image.
The threshold was chosen arbitrarily to get the cloud region isolated from the ambient material.
The ON-cloud stars can be separated again into the foreground and background stars for the target cloud.
The background stars have relatively higher amounts of extinction than the foreground stars, and the extinction is sharply increased at the location of the target cloud.
A baseline to correct for the normal interstellar reddening was identified from the distribution of OFF-cloud stars in the (g, r, i) color-color diagram.
We dereddened ON-cloud stars using the baseline and obtained their extinction.
Their parallax distances were procured from the \gaia\ DR2.

The fitting method was referred to \citet{Yan_2019_ApJ_885_19}, who derived the distance with Bayesian analysis and Markov chain Monte Carlo (MCMC) sampling from the extinction-distance data set.
The likelihood model describing an extinction of a star is expressed by 
\begin{equation}
    p(A_{\mathrm{v}\,i}|\mu_\mathrm{f}, \sigma_\mathrm{f}, \mu_\mathrm{b}, \sigma_\mathrm{b}, D) = f_i \,PF_i + (1-f_i) \,PB_i \;,
\end{equation}
where $\mu_\mathrm{f}$ and $\sigma_\mathrm{f}$ are the extinction and its standard deviation of foreground stars, $\mu_\mathrm{b}$ and $\sigma_\mathrm{b}$ for background stars, and $D$ is the cloud distance.
$f_i$ and $(1-f_i)$ are the probability of the star being in the foreground and background, respectively.
$PF_i$ and $PB_i$ is the likelihood of measuring $A_{\mathrm{v}\,i}$ for a foreground star and a background star, respectively.

The MCMC sampling was performed by using \texttt{emcee} python package \citep{Foreman-Mackey_2013_PASP_125_306}, and the sampling result is shown in Figure \ref{fig:dist_corner_plot}.
The extinction change between foreground and background stars was about 1 mag in both targets.
The best fit values of cloud distance are $175_{-3}^{+4}$ pc for L1544 and $203_{-7}^{+6}$ pc for L694-2.

\section{Correction of Optical Depth Broadening} \label{sec:app-corr-width}

The observed velocity dispersion of optically thick line such as \mco\ should be interpreted as a combination of optical depth broadening and intrinsic velocity dispersion of the gas \citep{Hacar_2016_A&A_591_A104}.
The optical depth broadening, $\beta_\tau$, is analytically defined as \citep{Phillips_1979_ApJ_231_720}, 
\begin{equation}
    \beta_\tau = \frac{\Delta V_\mathrm{obs}}{\Delta V_\mathrm{int}} = \frac{1}{\sqrt{\ln{2}}} \left[\ln{\left(\frac{\tau_0}{\ln{\left(\displaystyle\frac{2}{\exp(\tau_0)+1}\right)}}\right)}\right]^{1/2}\;,
\end{equation}
where $\Delta V_\mathrm{int}$ is the intrinsic (or practically corrected) FWHM linewidth and $\tau_0$ is central optical depth.
To examine the optical depth broadening, the optical depth of \mco\ and \sco\ lines should be determined.
Using the isotopic ratio of \sco\ and \mco\ in the local ISM \citep{Wilson_1994_ARA&A_32_191}, the optical depth of \mco\ can be calculated by multiplying the optical depth of \sco\ and their isotopic ratio,
\begin{equation} \label{equ:tau-co}
    \tau_{13} \approx \tau_{18}\cdot X \approx 7.3 \tau_{18} \;,
\end{equation}
where $\tau_{13}$ and $\tau_{18}$ are optical depths of \mco\ and \sco, and $X$ is abundance ratio of \mco\ with respect to \sco.
The spectrum of a molecular line is represented by the radiative transfer equation \citep[e.g.,][]{Wilson_2013_TRA},
\begin{equation} \label{equ:rt-tmb}
    T_\mathrm{MB,\nu}=(J_\nu (T_\mathrm{ex})-J_\nu (T_\mathrm{bg}))\cdot (1-\exp(-\tau_\nu))\;,
\end{equation}
where $J_\nu(T_\mathrm{ex})$ and $J_\nu(T_\mathrm{bg})$ are the equivalent Rayleigh-Jeans excitation and background temperatures, and $\tau_\nu$ is optical depth at frequency $\nu$.
The spectral optical depth, $\tau_\nu$, with Gaussian distribution along the frequency is given as follows;
\begin{equation}
    \tau_\nu = \tau_0\cdot\exp\left(-(\nu-\nu_0)^2/2\sigma^2\right)\;,
\end{equation}
where $\tau_0$ is the peak optical depth at the central frequency $\nu_0$, and $\sigma$ is the intrinsic velocity dispersion.
From the Equation \ref{equ:tau-co} and \ref{equ:rt-tmb}, the line ratio of \mco\ and \sco\ can be approximated by
\begin{equation} \label{equ:line-ratio}
    \frac{T_\mathrm{MB}^{13}}{T_\mathrm{MB}^{18}} \approx \frac{1-\exp(-\tau_{13})}{1-\exp(-\tau_{18})} \approx \frac{1-\exp(-\tau_{13})}{1-\exp(-\tau_{13}/X)}\;,
\end{equation}
since the remaining term, 
\begin{equation}
    \frac{J_\nu^{13} (T_\mathrm{ex})-J_\nu^{13} (T_\mathrm{bg})}{J_\nu^{18} (T_\mathrm{ex})-J_\nu^{18} (T_\mathrm{bg})} \approx 1\;,
\end{equation}
only changes between 0.992--0.998 in the condition where $T_\mathrm{ex}$ is 5--20 K.

\begin{figure}
    \centering
    \includegraphics[width=3.39in]{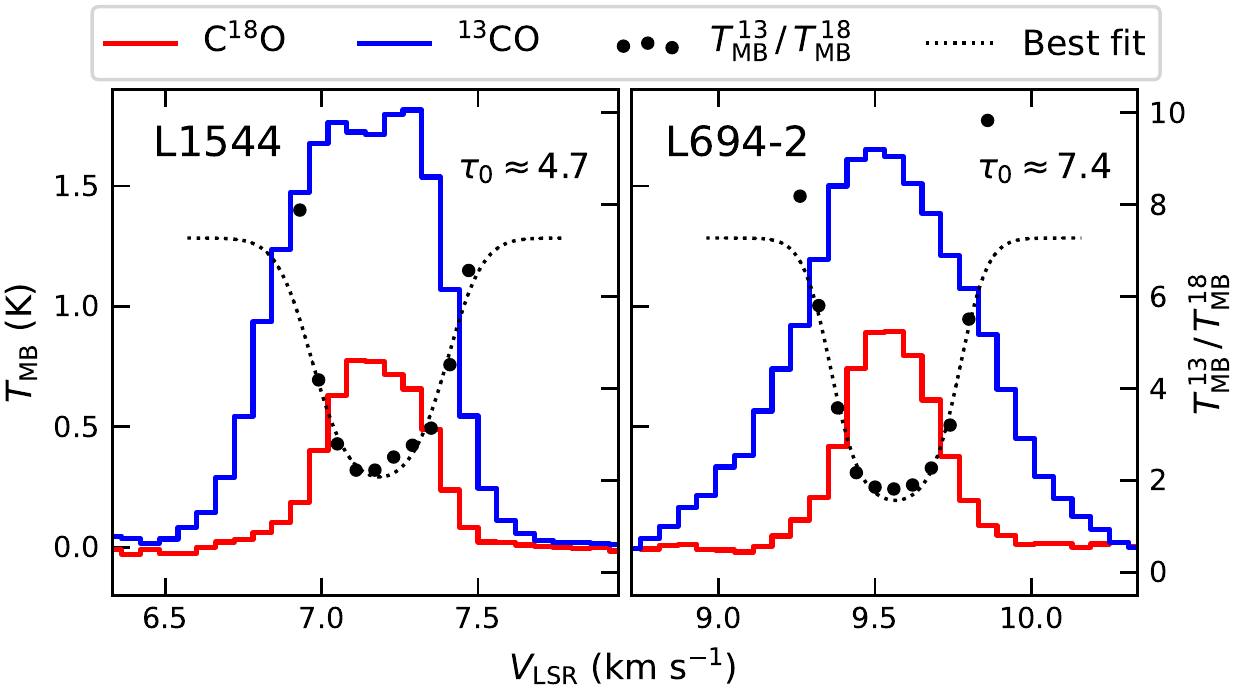}
    \caption{
    Median line profiles of \mco\ (blue) and \sco\ (red) emission and their ratio (black points) in L1544 (left panel) and L694-2 (right panel). 
    The dotted line is the best fit model of the line ratios to derive the peak optical depth $\tau_0$.
    }
    \label{fig:line_ratio_tau}
\end{figure}

\begin{figure}
    \centering
    \includegraphics[width=3.39in]{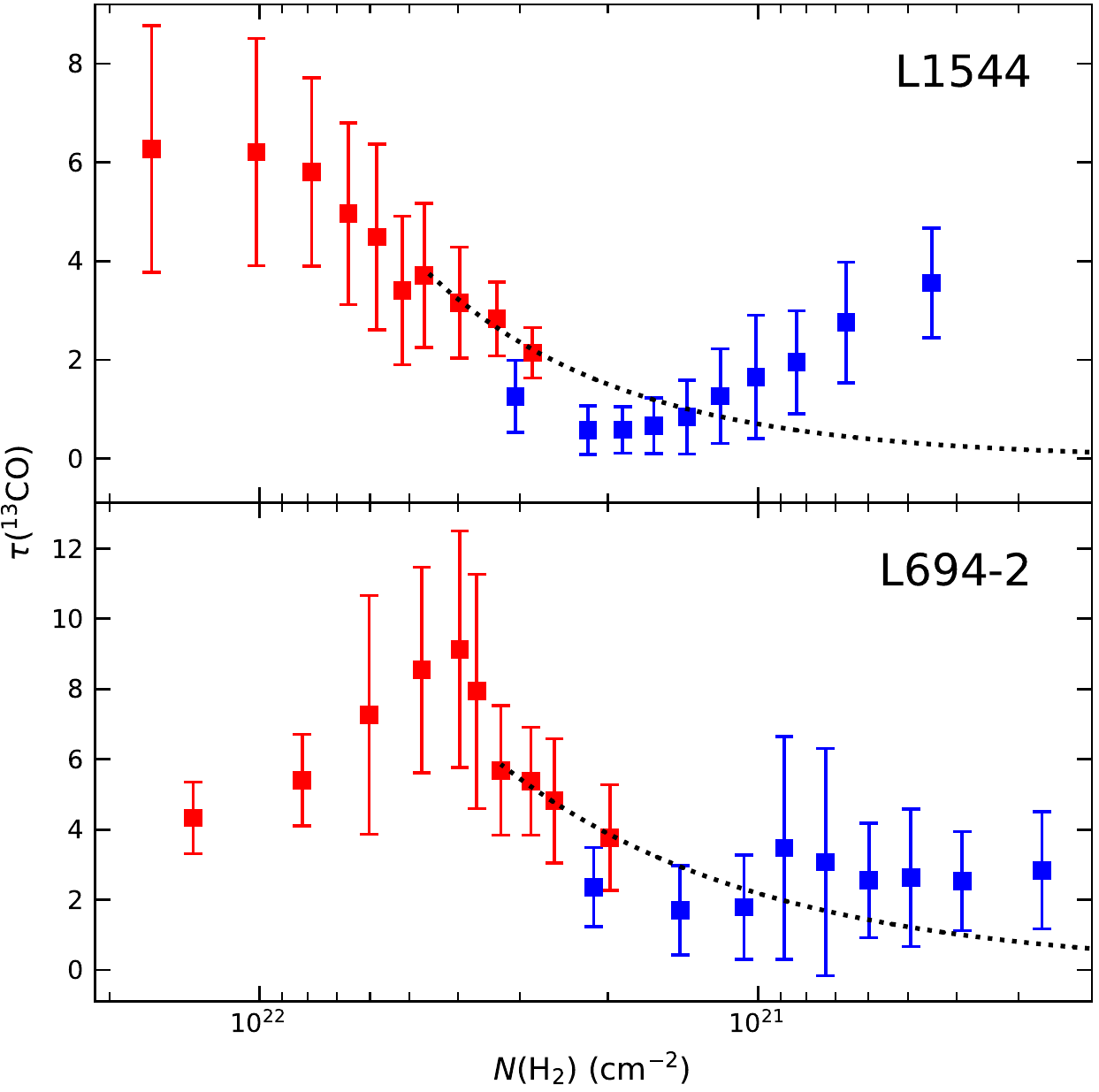}
    \caption{
    Variation of \mco\ optical depth with respect of \hmol\ column density. 
    The optical depths are estimated from the line ratio of \mco\ and \sco\ (red points). 
    For the regions where \sco\ was not detected, the optical depths were derived by assuming the \sco\ brightness as the $3\sigma$ level of \sco\ observation (blue points) or by extrapolating the relation between $\tau_{13}$ and $N(\mathrm{H_2})$ obtained near the filament boundary regions to those low column density regions (dotted lines).
    }
    \label{fig:h2cd_tau}
\end{figure}

Figure \ref{fig:line_ratio_tau} shows the median line profiles of \mco\ and \sco\ emission, and their ratio.
By fitting the model using Equation \ref{equ:line-ratio} to this line ratio, $\tau_0$ can be estimated.
The best fit values of $\tau_0$ for \mco\ are 4.7 for L1544 and 7.4 for L694-2.

The \mco\ optical depth ($\tau_{13}$) was determined by the method described above for all pixels where both \mco\ and \sco\ were detected.
Obtaining $\tau_{13}$ in the regions where \sco\ was not detected is not straightforward, as the line ratio in those regions can not be directly estimated.
There are two possible ways to estimate $\tau_{13}$ for those regions, as shown in Figure \ref{fig:h2cd_tau}: 
(1) by assuming the \sco\ line intensity to be  the $3\sigma$ noise level of the \sco\ observation (the blue data points in the Figure), or
(2) by extrapolating the relation between $\tau_{13}$ and $N(\mathrm{H_2})$ obtained near the filament boundary regions (the dotted lines in the Figure).
In the former case, the estimated $\tau_{13}$ rather increases at the lower column density regions because \mco\ line is getting weaker while \sco\ intensity is fixed as the $3\sigma$ level. 
This would be different from the real situation. 
Therefore we choose the latter method so that $\tau_{13}$ was estimated by extrapolating the relation between $\tau_{13}$ and $N(\mathrm{H_2})$ that were obtained at the high $N(\mathrm{H_2})$ regions ($2\times 10^{21} < N(\mathrm{H_2}) < 5\times 10^{21}$ cm\psq) to the low $N(\mathrm{H_2})$ regions where \sco\ was not detected.

\begin{figure}
    \centering
    \includegraphics[width=3.39in]{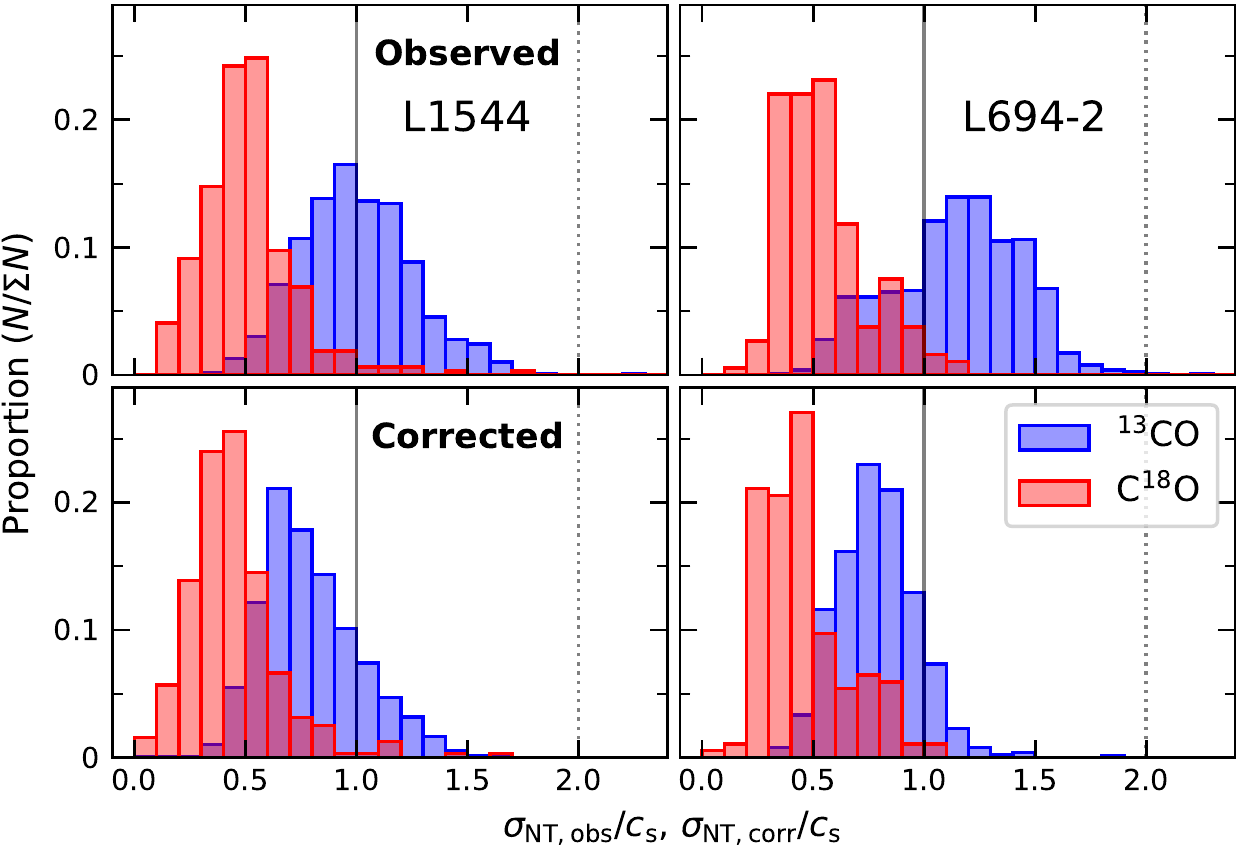}
    \caption{
    Histogram of non-thermal velocity dispersion expressed by Mach number for \mco\ (blue) and \sco\ (red) lines in L1544 (left panels) and L694-2 (right panels). 
    The upper panels show the distribution of observed non-thermal velocity dispersion, and the lower panels show the distribution of values corrected for optical depth broadening. 
    The solid and dotted lines indicate the dividing lines among sub-, trans-, and super-sonic regimes.
    }
    \label{fig:opacity_corr_vnth}
\end{figure}

Figure \ref{fig:opacity_corr_vnth} shows the results of correcting the optical depth broadening for the observed velocity dispersion.
The optical depth broadening contributes 46\% of the \mco\ linewidth at L1544 and 71\% at L694-2 in the core and filament region.
After correcting this, the non-thermal velocity dispersion of \mco\ was found to be mostly subsonic except for some regions outside the filament such as the northwestern end of the L1544 main filament (Figure \ref{fig:gauss_velocity} and \ref{fig:velocity_dispersion}).

\section{\mco\ Asymmetric Line Profiles} \label{sec:app-line-maps}

\begin{figure*}
    \centering
    \includegraphics[width=7.1in]{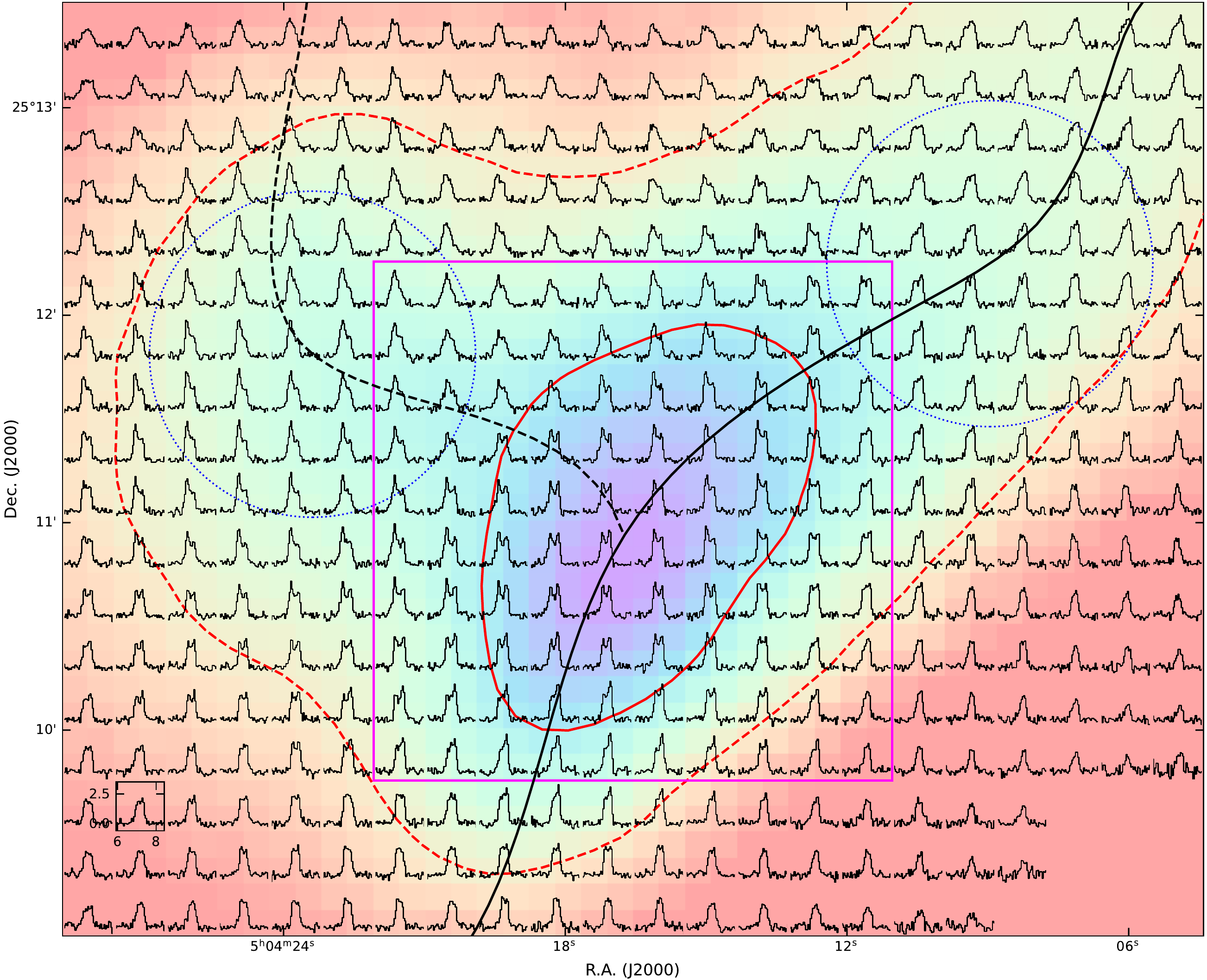}
    \caption{
    Line profile map of \mcot\ emission for L1544. 
    The background color is to show the \hmol\ column density distribution. 
    The black solid or dashed lines are the skeletons of the filamentary structure. 
    The red solid and dashed lines are contours with levels of $p_{40}$ and $p_{10}$, respectively. 
    The blue circles indicate the area where the averaged \mco\ and \sco\ line profiles of the filament region were obtained as presented in Figure \ref{fig:infall_profiles_fil}. 
    The magenta square is to indicate the area of \mco\ and \hcop\ line profile map in Figure \ref{fig:infall_profile_map}.
    }
    \label{fig:infall_profile_map_L1544_13CO}
\end{figure*}

\begin{figure*}
    \centering
    \includegraphics[width=7.1in]{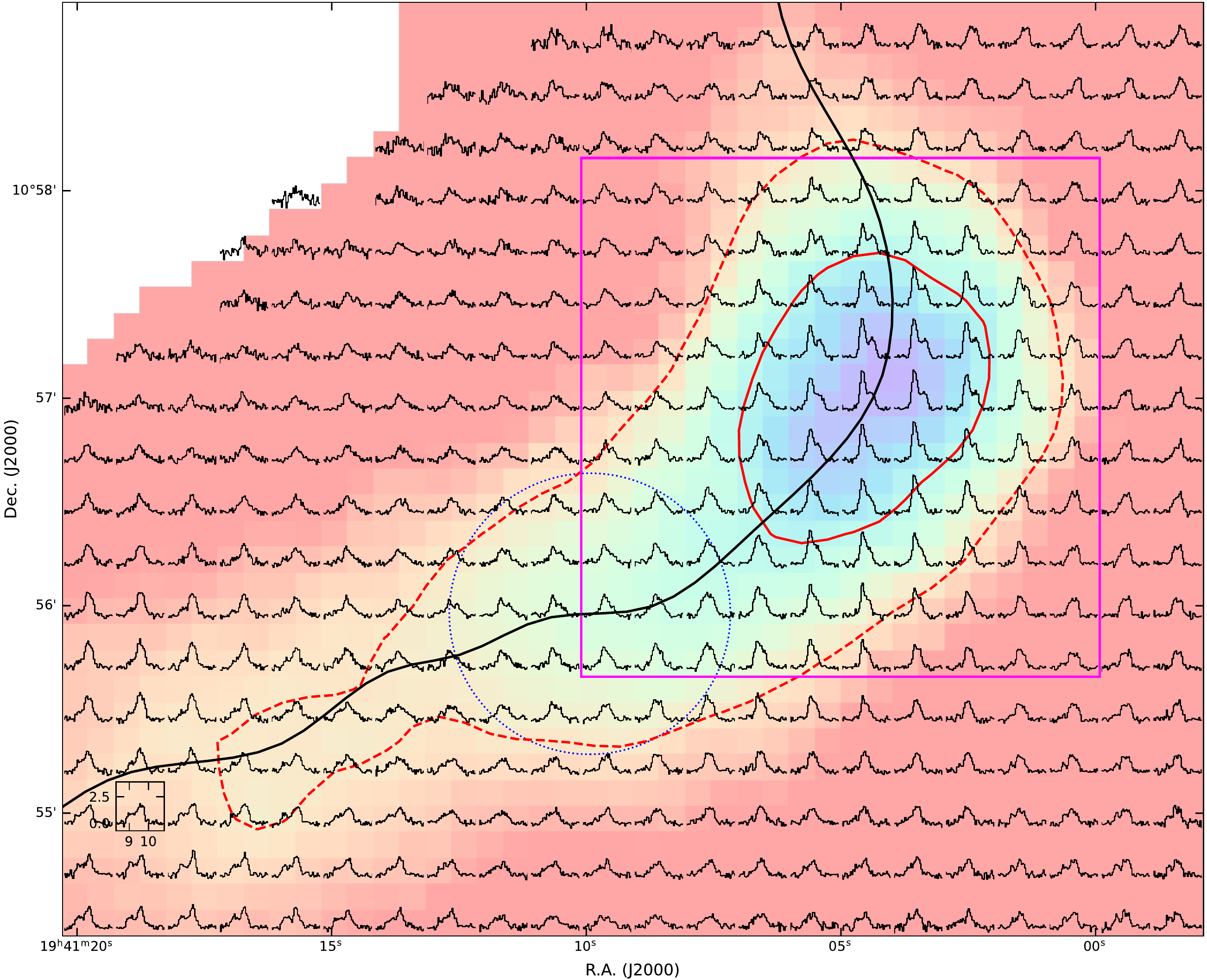}
    \caption{Same as Figure \ref{fig:infall_profile_map_L1544_13CO}, but for L694-2.}
    \label{fig:infall_profile_map_L6942_13CO}
\end{figure*}

Figures \ref{fig:infall_profile_map_L1544_13CO} and \ref{fig:infall_profile_map_L6942_13CO} display the \mco\ line profiles in the surrounding filament region outside the core area shown in Figure \ref{fig:infall_profile_map} which are the original data of the blue and red profile distributions and averaged line profiles shown in Figure \ref{fig:infall_profiles_fil}.

\begin{figure}
    \centering
    \includegraphics[width=3.39in]{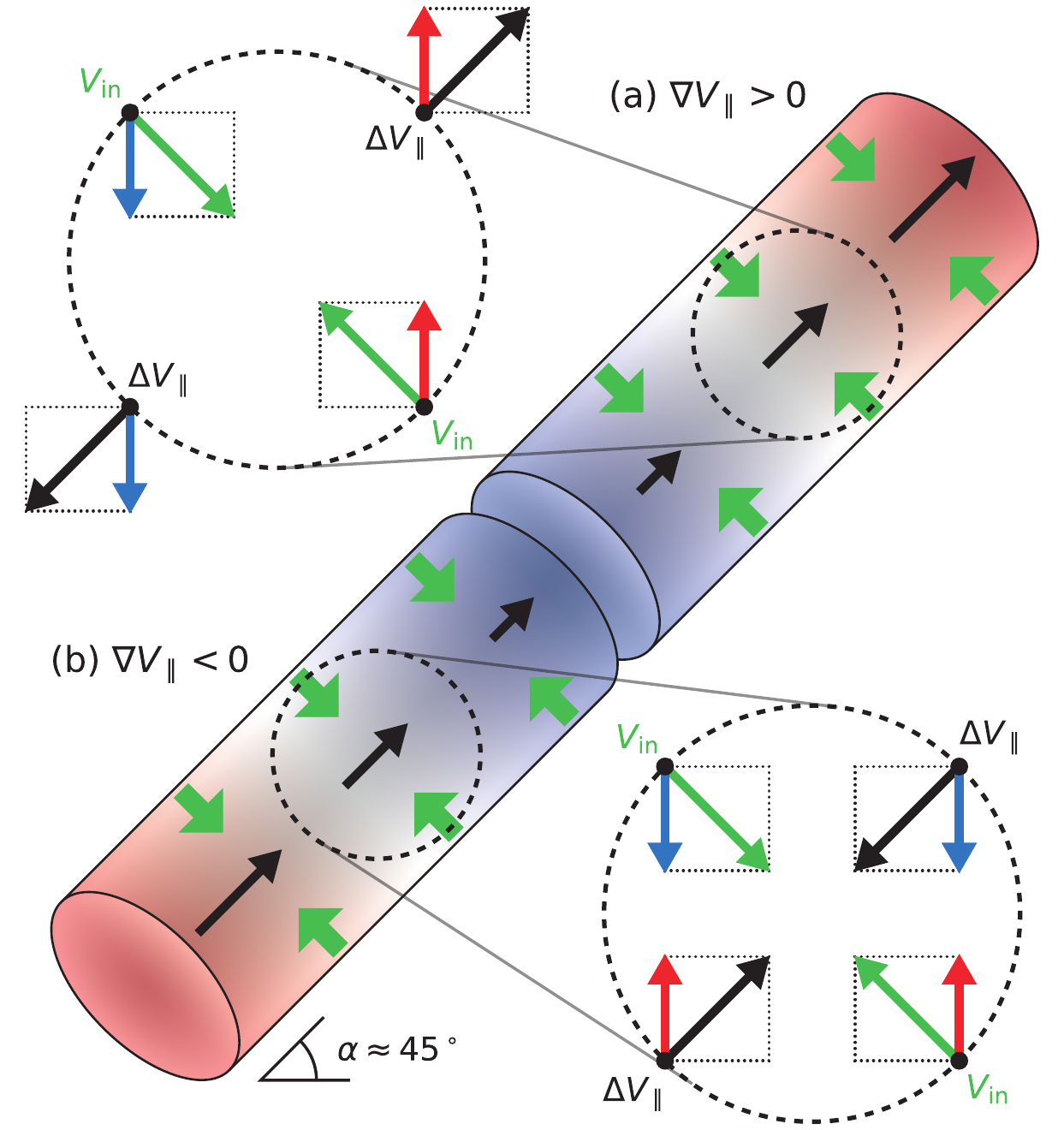}
    \caption{
    Simplified model on the origin of the blue- and red-shift motions detected in the filament with radial contracting motion and the large velocity gradient. 
    The black arrows inside the cylinder and in the enlarged area indicate the velocity gradient along the filament and the axial velocity difference compared with the average velocity, respectively. 
    The green arrows refer to the radial contraction motion of the filament. 
    The blue and red arrows mean blue- and red-shifted components, respectively, showing the observer's line-of-sight velocity components of axial motion and radial contraction in the area indicated by the dotted line. 
    The dotted circle indicates the area of interest where the asymmetric profiles are considered.
    }
    \label{fig:asymmetry_with_vgrad}
\end{figure}

We discussed the \mco\ line profile in the filament region in Section \ref{sec:infl-fil}.
The blue and red-shifted components of the infall asymmetry detected near the prestellar core trace the motion of the contracting gas towards the core center.
Similarly, we can also interpret the asymmetric line profile detected in low-density filaments as tracing the radial contraction motion towards the central axis of the filament.
However, as we discussed in Section \ref{sec:infl-fil}, there is a possibility that the longitudinal flow motion parallel to the filament axis, which is identified by the velocity gradient along the filament, could appear as such blue and red-shifted components in the line profile.
Figure \ref{fig:asymmetry_with_vgrad} displays a simplified model of a filament in the form of a linear cylinder. 
This is to help to understand the effects to the shapes of the asymmetric profiles due to the radial motions toward and longitudinal motions along the central axes of two types of filaments inclined in $45^\circ$ to the sky plane, (a) a filament with a positive velocity gradient in the longitudinal gas motions and (b) a filament with a negative velocity gradient in the longitudinal gas motions.

The main filament of L1544 is characterized by a sinusoidal velocity oscillation being related to the core-forming flow (Section \ref{sec:vels-cen} and Figure \ref{fig:velocity_profile}).
Accordingly, a decreasing velocity gradient appears near the positional offsets at about 0.1 and 0.3 pc along the filament, and Figure \ref{fig:asymmetry_with_vgrad} (b) depicts this situation.
In this case, both the radial infall motions onto the filament and the axial flow motions along the filament could appear as contracting gas motions to the line of sight and thus contribute on the shaping of an infall asymmetric profile.

On the other hand, Figure \ref{fig:asymmetry_with_vgrad} (a) is the other case where an increasing velocity gradient exists and the axial flow motions appear as expanding motions, as seen near the positional offset of about 0.2 pc along the L1544 filament.
This provides a possible explanation about the red profiles appearing near the position offset of 0.2 pc of the L1544 main filament (Figure \ref{fig:infall_profiles_fil}).

\bibliography{papers}{}
\bibliographystyle{aasjournal}

\end{document}